\documentclass[letterpaper,12pt]{thesis}
\usepackage{tabularx} 
\usepackage{amsmath}  
\usepackage{float}
\usepackage{CJKutf8}
\usepackage[toc,page]{appendix}
\usepackage{indentfirst}
\usepackage{graphicx} 
\usepackage[margin=1in,letterpaper]{geometry} 
\usepackage{caption}
\usepackage[utf8]{inputenc}
\usepackage[english]{babel}
\usepackage{fancyhdr}
\usepackage{cite} 
\usepackage[final]{hyperref} 
\usepackage{tensor}
\hypersetup{
	colorlinks=true,       
	linkcolor=blue,        
	citecolor=blue,        
	filecolor=magenta,     
	urlcolor=blue
}
\begin{document}
\frontmatter
\thispagestyle{empty}
\thispagestyle{empty}
\begin{center}

\begin{figure}
\begin{center}
\includegraphics[width=0.8\textwidth]{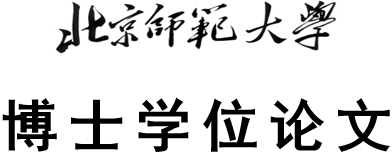}
\end{center}
\end{figure}
\vspace{1.5cm}
\begin{LARGE}

\textbf{Aspects of Black Hole Interior Volume and Entropy} \\
\vspace{3cm}
Shad Ali\\ 
\vspace{3cm}
Submitted to the Academic Degree Committee of Beijing Normal University for the Degree of Doctor of Philosophy
\\
\vspace{3cm}
Prof. Liu Wen Biao\\
Beijing Norma;l University\\
May. 2019
\end{LARGE}
\end{center}
\newpage
\clearpage
\thispagestyle{empty}
\hfill
\newpage
 \begin{titlepage}
 
  \begin{CJK}{UTF8}{bsmi}
  \begin{large}
 \begin{center}
 \textbf{北京师范大学研究生院}\\
 \vspace{1cm}
 北京师范大学学位论文原创性声明
\end{center}  
 
\vspace{1cm}

本人郑重声明： 所呈交的学位论文，是本人在导师的指导下，独立进行研究工作所取得的成果。除文中已经注明引用的内容外，本论文不含任何其他个人或集体已经发表或撰写过的作品成果。对本文的研究做出重要贡献的个人和集体，均已在文中以明确方式标明。本人完全意识到本声明的法律结果由本人承担。
\begin{center}
\vspace{1cm}
学位论文作者签名：　　　　　　　　　　日期：     年   月   日
\vspace{3cm}

学位论文使用授权书
\vspace{1cm}
\end{center}

学位论文作者完全了解北京师范大学有关保留和使用学位论文的规定，即：研究生在校攻读学位期间论文工作的知识产权单位属北京师范大学。学校有权保留并向国家有关部门或机构送交论文的复印件和电子版，允许学位论文被查阅和借阅；学校可以公布学位论文的全部或部分内容，可以允许采用影印、缩印或其它复制手段保存、汇编学位论文。保密的学位论文在解密后适用于本授权书。

\begin{center}
本人签名：\noindent\rule{4cm}{0.4pt}
    日期：\noindent\rule{4cm}{0.4pt}\\
 \vspace{0.75cm}       
导师签名：\noindent\rule{4cm}{0.4pt}
    日期：\noindent\rule{4cm}{0.4pt}

\end{center}
\end{large}
\end{CJK}
   
\end{titlepage}

\section*{Acknowledgments}
\addcontentsline{toc}{section}{Acknowledgement}
\begin{large}
At the end of this amazing journey toward Ph.D., I am thankful to Almighty Allah for giving me life, strength, and the opportunity to explore the universe in the way I wished. According to prophet Muhammad (PBUH); ‘‘He dieth not who giveth life to learning’’. With this spirit, I spent 30 years of my life as a student surrounded by noble souls, lovely people, and awesome family members. All of them were always on board to make me a better human being. However, the time I spent in China during my PhD studies is going to be an unforgettable chapter in my life. Coming to China was a new experience for me. First studying the Chinese language and understanding the Chinese culture was not less a surprise for me. Next, the completion of my PhD project was not easy at the start.

In this regard first of all, I would like to express my sincere gratitude to my advisor Prof. Liu Wen Biao for the continuous support of my Ph.D study and related research work, for his patience, motivation, and immense knowledge. His guidance helped me in all the time of research and writing of this thesis. I have to acknowledge that I could not have imagined having a best advisor and mentor for my Ph.D study. His patience and kindness as well as academic experience and expertise always have been precious to me.

Besides my advisor, I would like to thank to the Chinese Ministry of Education and China Scholarship Council (CSC) to provide us full bright scholarship, which covered the study and living expenses in Beijing Normal university. In fact, without the CSC support, it was not possible for me to stay and explore the china.

Next I would like to thank International Student Office (ISO) and specially Miss. Hu, for their continuous support and guidance. 

I am also Thank full to all staff of the department of Physics for their guidance, assistance and suggestions during my Ph.D  degree. 

I am also thankful to my all my lab mates and classmates who have been helped during my stay in Beijing Normal University. The academic committee members deserve special thank for evolving my work too. Also special thanks to Xin Yang Wang, Shan Han Zhang, Jian Zhe Yang, Yan Liu, Ming Zhang, Lan Shan Quan and Yan Haoping for their continuous support.

I would like to thank my parents, my brothers and sisters for sacrifice of time and provision of peaceful environment conducive for studies and accomplishment of my work.

\vspace{10mm}
\textbf{Shad Ali}
\end{large}
\newpage
\thispagestyle{empty}
\hfill
 \begin{CJK}{UTF8}{bsmi}
\section*{摘要}
\begin{large}
全文共分两部分和五章。 第一部分包括对我已经完成工作的简单介绍和文献研究，具体包括前两章。 第二部分是我们的研究工作删去！包括三章，对应三项具体的研究工作。 除了这三章之外，最后还有一些附录可参考。

这项工作的基本思想来源是从Marios Christodoulou和Carlo Rovelli (CR)最近在其题为''黑洞有多大？''的论文中进行的调查开始的。这项工作与Baocheng Zhang的黑洞内部体积和熵有关。在CR工作中， 考虑球对称的Schwarzschild黑洞并定义内部体积，因为球体$S$内的体积是由球体$S$包围的类似空间的球形对称三维超曲面的最大适当体积。使用此定义，他们发现黑洞的内部体积与超前Eddington时间成正比。这是该黑洞体积定义的主要特征，而这意味着黑洞可以存储大量的删去！信息。利用这种特殊性质，可以探测黑洞发射的霍金辐射的性质。人们还将结果扩展到带电的静态黑洞，并找到了一致的结果。删去！数值分析表明，这种公式的特殊性在于体积随着超前时间线性增加。

随后，张保成接着进行了CR工作，并研究了Schwarzschild黑洞内部标量量子模的熵。 他发现黑洞内部的标量量子模的熵与黑洞表面积成正比。 发现比例常数小于1，这意味着与Bekenstein Hawking 熵（视界面的熵） 相比内部熵较小。 请注意，这些分析仅适用于质量大于普朗克质量的黑洞。 如果质量小于普朗克质量，则需要了解不确定性关系。

在第2章中，我们按照上面对CR体积的讨论和Baocheng Zhang熵进行了讨论，并讨论了不同黑洞的内部熵。 我们调查的主要建议是基于内部体积和熵与提前时间之间的线性关系。 这意味着黑洞可能具有包含量子模式的大容量以存储丢失的信息。 这些量子模的熵与Bekenstein Hawking熵成正比，其比例常数小于1（即，正在进行的物体的大部分信息在黑洞的地平线上丢失）。 如果能够确定黑洞内部这些量子模的熵， 并利用霍金辐射下的Bekenstein Hawking熵构造内部量子模的熵之间的演化关系， 那么这可以为信息问题提供合理的解决方案悖论。

删去！在第3章中分析了d维带电黑洞以找到内部熵。我们回顾了第2章中使用的方法，并计算了d维带电黑洞的内部量子模式中的体积和熵。接下来，为了比较两个熵， 我们考虑使用黑洞发射率 作为准静态过程和黑洞辐射作为 黑体辐射的两个重要假设。 这使我们利用 Stefan Boltzmann定律得到黑洞内量子模的熵差分形式。接下来，使用 Bekenstein Hawking熵，找到了标量模熵与Bekenstein Hawking熵之间 的比例关系。 发现该比例关系随着的增加而减小。

第4章包括对上述讨论对改进的引力理论的扩展研究。为此，基于f（R）重力理论中的四维黑洞解与非线性麦克斯韦场理论相结合，我们使用Christodoulou和Rovelli提出的方法计算带电$f(R)$黑洞的内部体积。接下来，考虑到内部体积中的质量较少的标量场和仅携带能量的霍金辐射，我们计算带电$f(R)$黑洞内的标量场的熵，并研究霍金辐射下熵的演化。与此同时，还计算了霍金辐射下Hawking熵的演化。基于这些结果，得到了标量场熵内部演化与霍金辐射下Bekenstein-Hawking熵之间的比例关系。根据研究结果，我们研究和讨论了$f(R)$引力理论中修正的引力系数b如何影响两类熵之间的演化关系。这也表明来自带电$f(R)$黑洞的辐射率可随着修正系数$b$而增加。

在第五章中，考虑到$f$维$(R)-$AdS黑洞的相变，我们分析了与$f'(R)$相关的修正重力因子$b$对临界点参数$(P_c,T_c,\nu_c,G_c)$的影响。共存曲线，小黑洞和大黑洞的数密度差$(\frac {n_1-n_2} {n_c})$和配置熵$S_ {con}$。按照黑洞一阶相变的类比， 我们发现我们的结果与 Reissner Nordstrom达成一致，并对AdS黑洞进行充电。当穿过临界点的黑洞分子被发现具有消失的潜热并且变得难以与外部观察者区分时。在临界点，压力和温度与参数$b$无关，而自由能$G$取决于它。对于诸如比容，压力，温度和自由能等参数的减少，也发现了一致的结果。进一步的分析表明，参数$b$对共存曲线没有任何影响，这表明我们的结果与Reissner Nordstrom黑洞的结果一致，我们的诀窍是遵循Van der Waals阶段的条件。最后，我们还研究了数密度差异与配置熵维数$d$之间的关系。

关键词： 黑洞热力学，霍金辐射，内部体积和黑洞，黑洞蒸发， 相变，共存曲线和构型熵。

\end{large}
 \end{CJK}
\newpage

\section*{Abstract}
\addcontentsline{toc}{section}{Abstract}

Whole thesis is divided in to two parts and five chapters. The first part consist of introduction and literature study of my work. This section consists first two chapters. Second part consists of our research work and investigations that consists of three chapters. Besides these five chapters, there are appendices for exploring the points where it is needed.

The basic and conceptual notion of this work starting form the recent investigations of Marios Christodoulou and Carlo Rovelli (CR) in their paper entitled as ''How big is a black hole?''. This work is related to the black hole interior volume and the entropy by Baocheng Zhang. In CR work a spherically symmetric Schwarzschild black hole is considered and define the interior volume, as the volume inside a sphere $S$ is the maximal proper volume of a space-like spherically symmetric $3d$ hyper-surface bounded by the sphere $S$. Using this definition, they found the interior volume of black hole is proportional to the advance time. This is the main characteristic of this formulation. Which means that black hole could store a large amount of in-falling information. Using this special property, one can prob nature of Hawking radiation emitted by black hole. They also extend their result to the charged static black hole and found a consistent result. Their numerical analysis showed this special character of this formulation is that the volume linearly increases with advance time.

Latter the CR work is followed by Baocheng Zhang and investigated the entropy of scalar quantum modes in the interior of Schwarzschild Black hole. He found the entropy of scalar quantum modes in the interior of Black hole is proportional to black hole surface area. The proportionality constant is found to be less than unity, which means that the interior entropy is less as compared to the Bekenstein Hawking entropy (horizon entropy). Note that These analyses are only applicable for black hole with mass greater than Plank mass. If mass less than Planks mass, then one needed to understand the uncertainty relation.

In Second chapter, we followed the above discussion of CR volume along with Baocheng Zhang entropy and discussed the interior entropy of different black holes. The main proposal of our investigations is based on linear relation between the interior volume and entropy with advance time. This means that black hole could have a large volume containing quantum modes to store the lost information. The entropy of these quantum modes is directly proportional to the Bekenstein Hawking entropy with constant of proportionality less than one  (i.e. most of the information of an in-going objects are lost on the horizon of the black holes). If one could determine the entropy of these quantum modes in the interior of black hole and construct the evolution relation between the entropy of in the interior quantum modes with Bekenstein Hawking entropy under Hawking radiation, then this may give a reasonable solution for the problem of information paradox.

In same way, a $d$-dimensional charged black hole analyzed for finding the interior entropy in chapter 3. We recall the method used in chapter 2 and calculate volume and entropy in the interior quantum modes of a d-dimensional charged black hole and found a consistent result with CR work. Next, for comparing the two entropy's, we consider the two important assumptions of using the black hole emission rate as quasi-static process and black hole radiation as black body radiations. This led us to obtain the differential form of entropy of black hole interior quantum modes by using Stefan Boltzmann law. Next, using the Bekenstein Hawking entropy and found a proportional relation between the entropy of scalar modes and Bekenstein Hawking entropy. This proportional relation is found to decreases with the increase of dimensions.

Chapter 4 consists the extended study of the above discussion to the modified gravity theories. For this purpose, based on the 4-dimensional black hole solution in $f(R)$ gravity theory coupled to nonlinear Maxwell field theory, we calculate the interior volume of a charged $f(R)$ black hole using the method proposed by Christodoulou and Rovelli. Next, considering mass less scalar field in the interior volume and Hawking radiation carrying only energy, we calculate the entropy of the scalar field inside a charged $f(R)$ black hole and investigate the evolution of the entropy under Hawking radiation. In the meantime, the evolution of the Bekenstien-Hawking entropy under Hawking radiation has also been calculated. Based on these results, the proportional relation is obtained between the evolution in the interior of the scalar field entropy and Bekenstein-Hawking entropy under Hawking radiation. According to the result, we investigate and discuss how the modified gravity coefficient $b$ in $f(R)$ gravity theory affects the evolution relation between the two types of entropy. This also showed  that the radiation rate from a charged $f(R)$ black hole can increase with the modified coefficient $b$.

In chapter five, considering phase transition of a d-dimensional $f(R)$ AdS black holes, we analyzed the effects of modified gravity factor $b$ associated with $f'(R)$ on the critical point parameters $(P_c, T_c, \nu_c, G_c)$, coexistence curves, difference in number densities of small and large black holes $(\frac{n_1-n_2}{n_c})$ and configuration entropy $S_{con}$. Following the analogy of black hole first order phase transition, we found our results in agreement with Reissner Nordstr$\ddot{o}$m and charge AdS black hole. Where as the  black hole molecule crossing the critical point are found to have vanishing latent heat and become indistinguishable for outside observers. At critical point, the pressure and temperature are independent of parameter $b$, whereas the free energy $G$ is dependent on it. A consistent result is also found for reduced parameters like specific volume, pressure, temperature and free energy. Further analysis revealed that the parameter  $b$ doesn't have any effect on the coexistence curve, which shows that the our results are consistent with that of Reissner Nordstr$\ddot{o}$m black hole and our trick is according with the fact that obeys the condition of Van der Waals phase transition. Finally, we also investigate the relation between difference in number density and dimension $d$ of configuration entropy.

\textbf{Key words:} Black hole thermodynamics, Hawking radiation, Interior volume and of black hole, black hole evaporation, phase transition, Co-existence curves, and configuration entropy.
\hfill
\newpage
\clearpage
\thispagestyle{empty}
\hfill
\clearpage
\section*{Published Papers}
\begin{enumerate}
    \item Shad Ali, Xin-YangWang, Wen-Biao Liu*, "Entropy in a d-dimensional charged black
hole", Int.J.Mod.Phys. A, no. 27, 33(2018). 

    \item Shad Ali, X. Y. Wang, and W. B. Liu*, “Entropy Evolution in the Interior Volume of a
Charged f(R) Black Hole”, Commun. Theor. Phys., issue. 6, no.718 71(2019).

    \item Shad Ali*, Peng Wen, Wen-Biao Liu, “Entropy variation of a Charged (2 + 1)-
dimensional BTZ back hole under Hawking radiation”, International Journal of Theoretical
Physics, pp. 1 – 8, 59(2020).

    \item Shad Ali*, Misbah Ullah, and Johar Zeb, “Corrections to black hole entropy variation
under Hawking radiation”, BSM 2021, Online, "Andromeda Proceedings.

    \item Shad Ali*, Muhammad Arshad Kamran and Misbah Ullah Khan, “Entropy variation of
rotating BTZ black hole under Hawking radiation”, Physica Scripta, 97, 4 (2022).

    \item  Shad Ali*, “Configuration entropy and co-existence curves of f(R) AdS black holes: A
general approach”, High Energy Dens. Phys. 51 (2024).

    \item Shad Ali*, Tong Liu*, “Black hole entropy variation underHawking radiation: A review”,
New Astro. Rev., 101709, 99(2024).
\end{enumerate}

\newpage
\clearpage
\thispagestyle{empty}
\hfill
\clearpage
\tableofcontents
\listoffigures
\mainmatter

\newpage
\part{}
\chapter{Literature}

\section{Introduction}
The most striking feature of space-time is a black hole and most interesting object in the study of gravitational physics. Black hole is considered to be the final stage of a star's life, after losing all of its energy. Classically, a black hole is compact object with strong gravitational attraction even though light also can't come out from it, i.e. a place of no return for anything in it, due to its strong gravitational field. They are characterized by their boundary called event horizon. The existence of any in going object will vanishes at at a pint called the black hole singularity. This point can't be seen by the observer outside of black hole due to the existence of the event horizon. This is also a fact behind the incomplete study of black hole interior, or a secret to our physical world. Due to insufficient information about the interior of a black hole, a singularity may be considered as the cutoff in the geometrical structure of space and time. Inside a black hole the space-time is coupled to infinity. It is thought that singularity is a 1-dimensional point occurs at the center of the black hole, for detail see the schematic fig. ({\ref{image-1.1}}) shown below. 
\begin{figure}
\begin{center}
\includegraphics[width=0.5\textwidth]{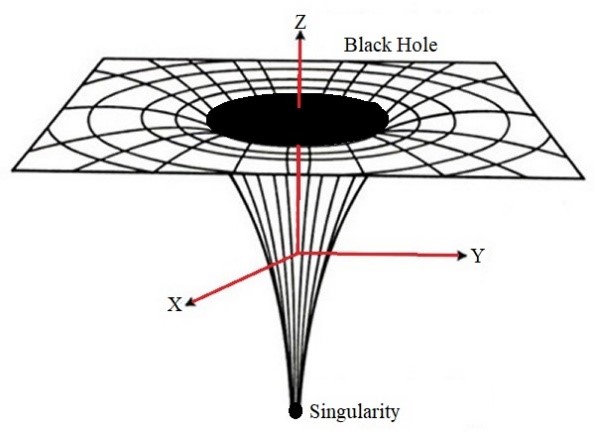}
\caption{Black hole singularity position}
\label{image-1.1}
\end{center}
\end{figure}

The theory of general relativity not only allows us to understand the point of singularity, but it is also unavoidable in our actual life  \cite{ruffini1971}. To completely understand the singularity, one need to understand the space and time in its accrual meanings. We will discuss some literature related to black hole for understanding the basic concept related to this study.

\section{Theory of relativity and black hole}

The special theory of relativity is proposed by Einstein in 1905 \cite{Weber, Einstein, John, rindler1991}. The bases of this theory are based on the Galilean principle of relativity, which states that the fundamental laws of physics are same for all reference frames moving with uniform motion. The characteristic feature of special theory of relativity is that it takes in account the speed of light $c$, structure of space-time and equivalence of acceleration and gravity. This theory is mainly summarized in two postulates as below.
\begin{itemize}
\item The laws of physics are same in all inertial frames reference (frames moving with uniform velocity). This postulate is also called the principle of relativity, and
\item The speed of light $c$ free space is universal constant.
\end{itemize}

The consequences of this theory includes dilation of time, length contraction, and equivalence of mass-energy, bending of light and the prediction of black holes, worm holes the birth of universe as result of Big Bang. This theory explains the concept of formulation for physical laws without gravity, which is also a failure of this theory, because one can't consider the zero gravity at any orientation or can't consider the relative motion of two bodies in different frames of reference. It also deals with the motion of bodies under the uniform speed relative to each other. The major problem for this theory is how to explain the bodies in two different frames of reference. This means that the notion of including only inertial frame is not sufficient for completing this theory. For example, twin paradox. This paradox is experimentally proved in 1971.

After realizing the cases of failure in the theory of special relativity, Einstein in 1915 urged himself to change this theory in General theory of relativity (simply GR), so that the above problems have solution in presenting the theory of general relativity. The GR is the generalization for the theory of special relativity. It gives a unified description of gravity, space-time particularly curvature space-time, which is directly related to the energy momentum of the body having the radiation and mass. It is experimentally verified theory by several test \cite{Wald:1984rga}. This theory is is proved experimentally by many ways e.g perihelion procession of Mercury's orbit,  the deflection of light rays by the Sun,  the gravitational red shift of light, equivalence principle, detection of gravitational waves and frame dragging phenomena. The first three tests of these are are based on the solution of Einstein field equation for spherically symmetric field. In theory of general relativity the mass is an important factor for the determination of gravitational effects. The relation is specified by a system of partial differential equation called Einstein Field Equation as given below,
\begin{equation}\label{fieldeq}
G_{\mu \nu }=R_{\mu \nu }-\frac{R g_{\mu \nu }}{2}=\frac{(8 \pi  G) T_{\mu \nu }}{c^2}
\end{equation}

Where $(\mu,\nu)=1,2,3,4$,.  $G_{\mu\nu} , R_{\mu\nu}$,  $g_{\mu\nu}$, $G$  and  $T_{\mu\nu}$ are the Einstein tensor, Ricci tensor, metric tensor, Newton gravitational constant and is the energy-momentum tensor respectively. This equation tells us about the relation of curvature tensor and energy momentum tensor, that is how the curvature tensor reacts in the presence of energy and momentum \cite{Wald:1984rga, Weber, Carroll:2004st}.  This equation is a counter piece of this theory of general relativity, General relativity theory gives a complete structure and properties of space-time and matter. Solving the Einstein equation completely is not possible, but it can be simplified in different specific form. The vacuum solution of can be obtained by using $R=0$. Consider a $4d$ Riemann space-time with Minkowski type metric signature $(-,+,+,+)$ see \cite{Weber, Carroll:2004st, Misner:1974qy}. Thus the metric takes the form
\begin{equation}\label{genmetric}
\text{ds}^2=\text{dx}^{\mu } \text{dx}^{\nu } g_{\mu \nu }
\end{equation}
using this relation, the Christoffel symbol can be calculated as 
\begin{equation}\label{chrstoffel}
\Gamma ^{\mu }_{\alpha \beta }\text{:=}\frac{1}{2} g^{\mu \sigma } \left(g_{\beta \sigma ,\alpha }-g_{\alpha \beta ,\sigma }+g_{\sigma \alpha ,\beta }\right)
\end{equation}
So, it is easy to define the geodesics of the Riemannian space 
\begin{equation}\label{geoeq}
\frac{\text{Dx}^2}{\text{D$\tau $}^2}\text{:=}\frac{d^2x^{\alpha }}{d\tau ^2}+\Gamma ^{\alpha }{}_{\mu \nu } dx^{\mu } dx^{\nu }=0
\end{equation}
By partial differentiation, the Riemannian tensor can be defined as
\begin{equation}\label{remtensor}
R^{\mu }_{\nu \alpha \beta }\text{:=}2\left(\partial _{[\alpha }\Gamma ^{\mu }{}_{|c}+\Gamma ^{\mu }{}_{\gamma [\alpha }\Gamma ^{\gamma }{}_{|\nu |\beta ]}\right)
\end{equation}
 for detail see also \cite{Wald:1984rga, Carroll:2004st, Stephani:2004ud}. The indices can be summarized as below:\\
Symmetric indices are written in parenthesis 
$$(\alpha \beta) \text{:=}\frac{\alpha \beta +\beta \alpha }{2}$$
Anti-symmetric indices are written in brackets,
$$[\alpha \beta ]\text{:=}\frac{\alpha \beta -\beta \alpha }{2}$$
The generalization of indices is written as,
$$(\alpha \beta \gamma) \text{:=}\frac{1}{3} (\alpha \beta \gamma +\beta \gamma \alpha +\gamma \beta \alpha )$$
The indices written in the vertical strokes $|\alpha\beta|$ are not included in anti-symmetric process. If the curvature is anti-symmetric, then it will have two pairs of commuting indices. The anti-symmetric terms will vanish, i.e.
$$R_{(\alpha \beta)  \mu \nu}=R_{(\mu \nu)  \alpha \beta }=0 \qquad R_{[\alpha \beta \mu \nu ]}=0 \qquad R_{\mu \nu \alpha \beta }=0$$
Let us take an example for explaining the symmetric and anti-symmetric indices. Consider the indices be $a,b,c$ etc. For two symmetric pairs we have $ab=ba$, whereas anti-symmetric pairs are $ab=-ba$. Note that here it is important to write that algebraic symmetries must be valid $R_{ab}=R_{ba}$. For a $6\times 6$ matrix there is over all $20$ independent components. Some main features of field equation are as follow
\begin{itemize}
\item Like all field equations, this equation is also a partial differential equation.
\item This equation is tensor equation, where the coordinate system is independent from the law of nature.
\item This equation satisfies the poison equation of Newtonian gravitational theory.
\item The mass-density analog in this equation is energy momentum tensor. If one need to find the simplest solution for the above field equation, then it can be conclude that $R=-8 \pi  G T_{\mu\nu}$ . For the flat space-time $T_{\mu \nu}\approx0$, then one gets:
\end{itemize}
\begin{equation}\label{ricciten}
R_{\mu \nu }=0
\end{equation}
Which is also called vacuum solution for Einstein field equation. To find the complete solution of Einstein's field equation is very complicated, so it will be easy to find the solutions in free space-time. The gravitational fields which are most important in our daily life, that are produced by sun, earth and moon. These gravitational fields are produced due to slowly rotating nearly spherical mass distribution, so these fields are spherically symmetric. Spherically symmetric fields are simple to understand. So, firstly, we will discuss these fields. In the following section, we will discuss the vacuum solution (also called Schwarzschild solution) of Einstein field equation and their effects on the space-time.

\section{Schwarzschild (Static) Black Hole Solution}

The metric (which is a tensor equation of second order under the coordinate transformation) is a field describing the nature of space-time and gravity. The simplest form of which was first computed by Karl Schwarzschild from Einstein field equation for a $4d$ space-time. Schwarzschild solution is also called the vacuum solution and it describes a spherically symmetric static black hole.  The line element of the Schwarzschild space-time in spherically coordinates (${{r,t,\theta,\phi}}$) is given by \cite{Weber, Wald:1984rga, Carroll:2004st, Stephani:2004ud, Gary}.
\begin{equation}\label{schmetric}
ds^2=-\left(1-\frac{2 M}{r}\right) dt^2+\left(1-\frac{2 M}{r}\right)^{-1} dr^2+r^2 \left(d\theta ^2+\theta  \sin ^2 d\phi ^2\right)
\end{equation}

Where $M$ is the mass of Schwarzschild black hole with horizon at $r_s=2 M$. The Schwarzschild black hole is characterized by single quantity $M$. For simplicity, we consider $c=\hbar=G=1$. This line element shows that the singular points are at $r=2M$ and $r=0$ i.e. the origin of flat space in polar coordinates. At these points, the coefficients become infinite. All rays of light travel parallel to each other, such rays make the event horizon (a null surface). All the light cone in the region of event horizon of black hole are bends towards the center $r=0$, instead of future time like infinity. A future directed light signal ends up at singularity before reaching the future null infinity. Which means that at any space-like slice with a black hole, there is always exists a future directed non-space-like curves, which vanishes at singularity before reaching future infinity. As result they remains in the interior of black hole. Thus, the black hole bifurcation (combining two) is possible to form a single one in late time, but the reverse is not possible. As for a single black hole do not bifurcate reversibly so, this property of black hole lead to the theorem called area theorem of the event horizon. According to which the area never decreases.
\begin{equation}\label{arealaw}
\delta A_{BH}>0
\end{equation}

In Schwarzschild solution, we will urge to know that actually what happens at $r=2M$. Here one can't ignore the effect for $r=2M$. In general relativity it is important to make difference between the curvature singularity (divergence of space-time) and coordinate singularity (which is place where the curvature is same but the metric component takes a bad choice). If we compute the square of Riemann tensor, then from metric Eq. (\ref{schmetric}), we gets
\begin{equation}\label{remtensor1}
R_{\mu \nu \rho \sigma } R^{\mu \nu \rho \sigma }=\frac{48 G^2 M^2}{r^6}
\end{equation}

Form this we see that for $r=0$ is not only coordinate singularity but also behaves as a curvature singularity, whereas $r=2M$ is just a coordinate singularity not a curvature singularity. A curvature singularity results in the break down in space-time. Such type of singularity can't be removed. In contrast the coordinate singularity is not so important, it can be removed by changing the coordinate by transformation, so that it will give a finite value at that point \cite{Heinicke:2015iva}.

\subsection{Tortoise Coordinates}

Schwarzschild metric describes the motion of particle outside the horizon  $(r>r_s)$ i.e in the exterior of black hole but can not describe for $(r<r_s)$ . It will be meaningful, if one could find the coordinates, which are not singular at the horizon but it can also be extended to the interior of the horizon. This is convenient by using the in going or outgoing radial null geodesics. A good choice for this is the use of tortoise coordinates. Let the outgoing and in going geodesics can be defined for Eq. (\ref{schmetric}) as $ds^2=0$, and $\theta=\phi=comstant$ so that, the above metric yields,
\begin{equation}\label{dszero}
\left(1-\frac{r_s}{r}\right)\left(\frac{dt}{d\lambda }\right)^2-\left(1-\frac{r_s}{r}\right)^{-1}\left(\frac{dr}{d\lambda }\right)^2=0
\end{equation}

\begin{figure}
\begin{center}
\includegraphics[width=0.5\textwidth]{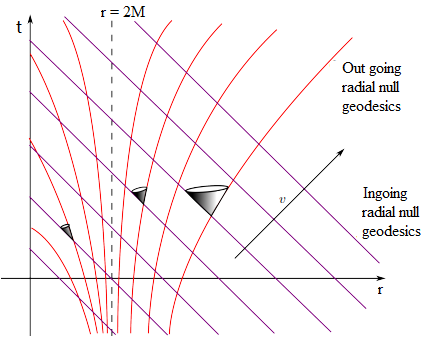}
\caption{In Schwarzschild coordinates the light cone appears to close up as $r\rightarrow 2M$ and for light cone opens when $r > 2M$}
\label{image-1.2}
\end{center}
\end{figure}

From this equation one can analyses the slope of the light cone in the direction of increasing or decreasing $r$. When $r\rightarrow r_s$ then the angle of light cone decreases and as the geodesic diverges from $r_s$  the angle of light cone increases as shown in the Fig. (\ref{image-1.2}) above.
Consider the Schwarzschild matrix given in Eq. (\ref{schmetric}), and define the transformation coordinates as

$$\nu\rightarrow t+r^* \qquad \mu\rightarrow t-r^*$$
\begin{equation}\label{metintor1}
ds^2=\left(1-\frac{r_s}{r}\right)\left(dt^2-d{r^*}^2\right)-r^2 d\Omega
\end{equation}
Where the tortoise coordinate is
\begin{equation}\label{r*}
r^*=r+2M ln|\frac{r}{2M}-1|\Rightarrow \frac{dr^*}{dr}=\left(1-\frac{2M}{r}\right)^{-1}
\end{equation}
In these coordinates for $r^*\rightarrow\infty$, when $r=r_s$. Let us consider a set of photon each of them is assigned to a number $\nu$. This number remain constant during the motion of photon. We choose the number as new coordinate. Then for radially in going and outgoing photons, we can write as
$$\nu= t+r^* (r)=t+r+2M ln|\frac{r}{2M}-1|$$
\begin{equation}\label{munu}
\mu= t-r^* (r)=t-r-2M ln|\frac{r}{2M}-1|
\end{equation}
As the photon moves inwards, the distance to black hole decreases with time, this time is termed as advanced time. $(t, r)\rightarrow(\nu, r)$. So from the above Eq. (\ref{metintor1}), we can write as,
\begin{equation}\label{tormet2}
ds^2=\left(1-\frac{2M}{r}\right)d\mu d\nu-r^2 d\Omega
\end{equation}
From first part of Eq. (\ref{munu}), we have
\begin{equation}\label{metintor2}
e^{\frac{r^*-r}{2 M}}=\left(\frac{r}{2 M}-1\right)
\end{equation}
$$r^*=\frac{\nu -\mu }{2}$$
$$e^{-\frac{r}{2 M}} e^{\frac{\nu -\mu }{4 M}}=\left(\frac{r}{2 M}-1\right)d\mu d\nu$$
\begin{equation}\label{14}
2M e^{-\frac{r}{2 M}} e^{\frac{\nu -\mu }{4 M}}=\left(1-\frac{r}{2 M}\right)d\mu d\nu
\end{equation}
Let us define
$$e^{\frac{\nu}{4 M}}=V \qquad e^{\frac{-\mu }{4 M}}=U$$
$$ e^{\frac{\nu }{4 M}}d\nu=4 M dV \qquad e^{-\frac{\mu }{4 M}} d\mu=-4 M dU$$
So, from above Eq. (\ref{14}), we gave
$$\left(\frac{2 M}{r}\right)e^{-\frac{r}{2 M}} e^{\frac{\nu -\mu }{4 M}}d\mu  d\nu =-\left(\frac{32 M^3}{r}\right)e^{-\frac{r}{2 M}} dU dV$$
Hence
\begin{equation}
\left(1-\frac{r}{2 M}\right)d\mu d\nu=-\left(\frac{32 M^3}{r}\right)e^{-\frac{r}{2 M}} dU dV
\end{equation}
From the above metric in Eq. (\ref{tormet2}), we get
\begin{equation}\label{32M^2}
ds^2=-\left(\frac{32 M^3}{r}\right)e^{-\frac{r}{2 M}} dU dV-r^2 d\Omega
\end{equation}
For $V>0$, it can be extended to $-\infty <U< \infty$, which describes both the exterior and interior of black hole. This means that using tortoise coordinates one can remove the coordinate singularity. Now consider Eq. (\ref{metintor1})
$$dt=d\nu -dr^*\Rightarrow dt^2-dr^{*2}=d\nu ^2-2d\nu dr^*$$
$$dt^2-dr^{*2}=d\nu ^2-2 d\nu  dr\left(\frac{dr^*}{dr}\right)$$
we get
\begin{equation}
dt^2-dr^{*2}=d\nu^2 -2\left({1-\frac{r}{2 M}}\right)^{-1}d\nu  dr
\end{equation}
Use this equation in Eq. (\ref{metintor1}) and simplifying, we get
\begin{equation}
ds^2=-d\nu ^2 \left(1-\frac{2 M}{r}\right)-2 d\nu  dr-r^2 d\Omega ^2
\end{equation}
These coordinates $(\nu,r,\theta, \phi)$ are called outgoing Eddington Finkelstein coordinates. Similarly, for in going photons, the resulting metric can be written as
\begin{equation}\label{collapsedmass}
ds^2=-d\nu ^2 \left(1-\frac{2 M}{r}\right)+2 d\nu  dr-r^2 d\Omega ^2
\end{equation}
Generally, we can write as
\begin{equation}
ds^2=-d\nu ^2 \left(1-\frac{2 M}{r}\right)\pm 2 d\nu  dr-r^2 d\Omega ^2
\end{equation}
As these coordinates are also defined only on the horizon, but they are regular on the horizon and can be extended to the origin of black hole. These two line elements are physically different from each other in every aspect. If we take radial null curves for which $d\theta=d\phi=0$, because $\theta$ and $\phi$ are constant and $ds^2=0$ then from Eq. (\ref{r*}) for in-going geodesics, we can write as
$$d\nu =0=r^*+t=\text{constant}$$
and 
\begin{equation}
\frac{d\nu }{dr}=\left\{\left(0,2 \left(1-\frac{2 M}{r}\right)\right)\right\}
\end{equation}
Form this solution, we see that when $r=2M$ falls towards the center $r=0$. Only massless particles $(M=0)$ moves on the horizon. Now, if we use these conditions for the second line element in Eq. (\ref{munu}) the result is of the form
$$d\mu =0=r^*-t=\text{constant}$$
\begin{equation}
\frac{d\mu }{dr}=\left\{\left(0,-2 \left(1-\frac{2 M}{r}\right)\right)\right\}
\end{equation}
What is meant by these two results? For the Eq. (\ref{metintor2}) everything goes away form the center at $r=2M$. This means that if the first part of Eq. (\ref{munu}) describes a black hole, then in contrast the second part shows a white hole, this issue is briefly discussed below in fig. \ref{image-1.3}
\subsection{Kruskal Coordinates}

Now consider the Kruskal coordinate, which covers the whole space-time manifold of maximally extended Schwarzschild solution and it is useful to find the coordinates which don't show the singularity beyond the event horizon. To do so the convenient way is to study the behavior of $r=r_s$ and choose the in going and outgoing radial null geodesics. Consider Eq. (\ref{metintor1}) and Eq. (\ref{munu}), we can write as 
\begin{equation}\label{metkrskl1}
ds^2=\left(1-\frac{r_s}{r}\right)dt^2+\left(1-\frac{r_s}{r}\right)^{-1}-r^2 d\Omega ^2
\end{equation}
For this equation the Kruskal coordinates as
$$T=\frac{1}{2}\mu +\nu )=\left(\frac{r}{r_s}-1\right)^{\frac{1}{2}}e^{\frac{r}{2 r_s}}sinh\left(\frac{t}{2 r_s}\right)$$
$$X=\frac{1}{2}(\mu -\nu )=\left(\frac{r}{r_s}-1\right)^{\frac{1}{2}}e^{\frac{r}{2 r_s}}cosh\left(\frac{t}{2 r_s}\right)$$
Using this, the metric along with the Eq.  (\ref{32M^2}) can be written as 
\begin{equation}\label{metTX}
ds^2= \frac{32r^3 _s}{r} e^{-\frac{r}{r_s}}\left ( -dT^2+dX^2 \right )+r^2 d\Omega ^2
\end{equation}
where the term r is defined from
$$\left( T^2-X^2\right) =\left(1-\frac{r}{r_s}\right)e^{-\frac{r}{r_s}}$$
These $(T, X,\theta, \phi)$ coordinates are called as Kruskal coordinates. They have many properties in flat and curved space time
\begin{equation}\label{TXandtheta}
T^2=\pm X^2+constant \qquad T^2-X^2=constant \qquad and \qquad \frac{T}{X}=tanh \left(\frac{t}{2r_s}\right)
\end{equation}
This equation represents the straight line through the origin. For $t\rightarrow\infty$then it shows a null surface, where $T=\pm X$, this surface is similar to that $r\rightarrow r_s$ in Eq. (\ref{metkrskl1}). The region for these coordinates ranges between $-\infty<X<\infty$ and $T^2<X^2+1$. These coordinates are known as Kruskal coordinates and can be drawn as above.
\begin{figure}
\begin{center}
\includegraphics[width=0.6\textwidth]{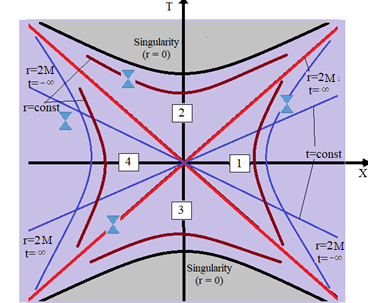}
\caption{Conformal diagram of Kruskal coordinates}
\label{image-1.3}
\end{center}
\end{figure}

Recall Schwarzschild solution, which describing the simplest black hole. We used it as flat space-time , where in the metric tensor, time coordinate has a different sign from space coordinates, this corresponds to the fact that you can move in different directions in space, but only in one direction in one time and time always grows. Schwarzschild solution shows that inside the event horizon of a black hole signs for time and space-time coordinates are interchanged, therefore you can move freely in both directions with time, but you can move only in one direction towards the singularity, this is what makes it inevitable. Even light is also bound to do so. It can't move a millimeter away from singularity. The best way to analyze different parts of a black hole is to work in Kruskal coordinates, which explains the Fig.  \ref{image-1.3}. In these coordinates, one can make four parts of the total space-time \cite{Carroll:2004st}.
\begin{itemize}
\item The space outside the black hole is in the right quadrant, which places equal r (radius) taking a form of hyperbolas going up and down. Here the event horizon looks like a straight diagonal line $(r=2M)$. This is the space in which our original coordinates are well defined.
\item Following the future directed null rays, we reach to the region $2$. Space inside the black hole is the upper quadrant, with constant $(r=2M)$ taking form of hyperbolas going from left to right. Singularity is at the zero radius, which is also looks like a hyperbola as above in region 2. Points with $t=const$. lie on straight lines crossing the center of the image. One good quality of these coordinates is that light cones always look the same here, light always follows diagonal lines as $45^0$. The future event horizon is the boundary of the region $2$.
\item Following the past null rays, we reach to the region $3$. Which is opposite to the region $2$. This region is simply the time reverse of region $1$. This is the part of space from which the bodies can come to our region $2$, but can't be get back to region $3$. This region can be thought as white hole. The past event horizon is the boundary of the region $3$.
\item The opposite region to our region is the region $4$. This region is connected to our region by a neck like configuration called the warm hole (as discussed by Einstein Rosen Bridge). This is the mirror image of our region $1$.
\end{itemize}
All these parts are shown in fig. (\ref{image-1.3}) in detail above.
\section{Charged static black hole}
Schwarzschild black hole is also known as static charge less black hole. Here in charged static black hole it is clear from the name, that any black hole having charge. Along with charge these black hole don't have the property of angular momentum. In general relativity the asymptotic final state of gravitationally collapsed charged object coupled to electromagnetism is completely described by three quantities, they are mass $(M)$ and charge $(Q)$. All the other degree of freedom evaporates at the time of collapse. According to Wheeler this expression of the static black hole is known as No Hair Theorem \cite{Wald:1984rga}. This theorem is no longer valid, if the non-abelian gauge field are not present, and the resulting black hole solutions are unstable see \cite{Novikov:1989sz}.

The solution for spherically symmetric charged black hole can be obtained in same way as that of Schwarzschild black hole with function
\begin{equation}\label{RNBH}
f(r)=1-\frac{2 M}{r}+\frac{Q^2}{r^2}
\end{equation}
Where $Q$ is the charge of the black hole. So, from Eq. (\ref{schmetric}), one gets a metric for static charged black hole. It is the unique solution of spherically symmetric and asymptotically flat space-time solution of Einstein-Maxwell equation.
The only non-zero component of electrically charged black hole is $T_{tr}=\frac{Q}{r^2}$ , and for magnetically charged black hole the non-zero component is the magnetic field $F_{\theta\phi}=Qsin\theta$. We will discuss only the electric charge case only. The location of the of the event horizon is the position where $g_{tt}\rightarrow0$, or we can write as 
$$f(r)=0$$
There are two solutions of this equation
\begin{equation}\label{RNBHradii}
r_{\pm}=M\pm \sqrt{M^2-Q^2}
\end{equation}
From this we see that a static charged black hole for there are two horizons, outer horizon (external horizon $r_+$) and inner horizon (Cauchy horizon $r_-$). The conformal diagram is given below, where we will examine three cases ($M > Q $	$M=Q$ and $M>Q$) \cite{Wald:1984rga, Carroll:2004st, Gary}. Here we will discuss these cases one by one as below;
\begin{itemize}
\item{Case 1. we consider the first case as $Q<M$ so, from above Eq. (\ref{RNBH}), we can write as
\begin{equation}
f(r)=\frac{\left(r-r_+\right) \left(r-r_-\right)}{r^2}
\end{equation}
This equation shows, that the metric component are singular at $r\rightarrow r_\pm$and $r\rightarrow 0$, but, it also shows curvature singularity at $r\rightarrow0$. In this case the singularity position is in space and can be avoid by some transformation of coordinates. Such as Eddington Finkelstein coordinates. The inner horizon is shown to be unstable at slight perturbation. The Fig. (\ref{image-1.4}(a)) shows maximally extended Reissner Nordstrom black hole with $Q<M$. If we examine the figure, it shows the complete space-time. It is clear that the infinity repeats in vertical direction in proper intervals. There are two singularities, but both of these have opposite signs. It is claimed, that the electric field between these two singularities is always pointing from left to right, which shows that the left singularity has a charge $Q>0$ and that on right has charge $Q<0$. 
\begin{figure}
\begin{center}
\includegraphics[width=0.5\textwidth]{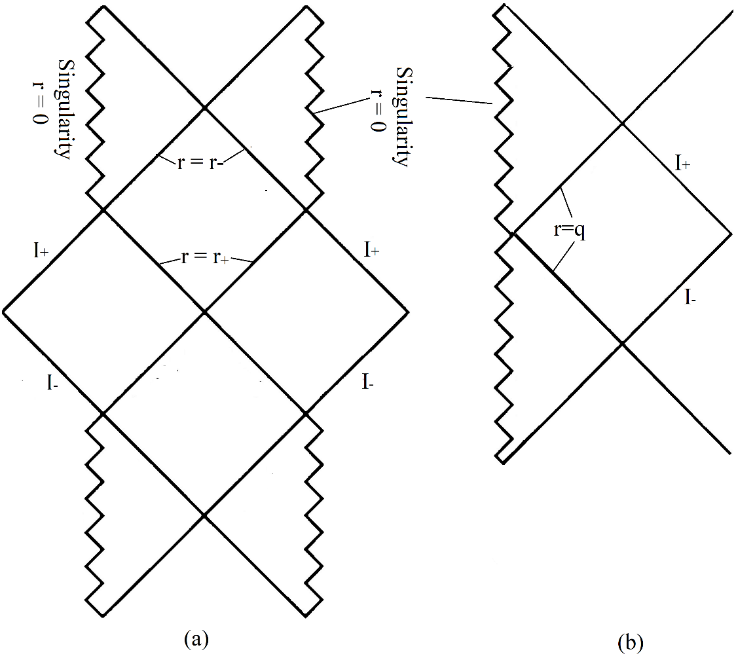}
\caption{Conformal diagram of maximally extended R-N black hole (a) $Q< M$ and (b). Extremal case $(M=Q$}
\label{image-1.4}
\end{center}
\end{figure}}
\item{Case 2. The second case is $Q=M$ as extremal limit, so from above Eq. (\ref{RNBH}), we can write as
\begin{equation}
[r]=1-\frac{2 M}{r}+\frac{Q^2}{r^2}=\left(1-\frac{Q}{r}\right)^2
\end{equation}
and the metric becomes
\begin{equation}
ds^2=\left(1-\frac{Q}{r}\right)^2dt^2+\left(1-\frac{Q}{r}\right)^{-2}dr^2+r^2d\Omega^2
\end{equation}
In this case, we have to assign a single charge. Let we take the charge is positive then the structure of conformal diagram is shown in Fig. (\ref{RNBH}(b)). Similarly, one can take negative charge for the opposite case.}

\item{Case 3. In this $q>M$ case there is no event horizon. As $r=0$ and $g_{tt}$ is negative, $r=0$ is time like singularity. The singularity in this case is called naked singularity i.e. an observer can see the singularity at all. Which is not possible, because if so, then the coulomb repulsion will dominant the gravitational attraction and it will need an infinite amount of energy to compress the mass zero volume.}
\end{itemize}
\section{Rotating black hole}
\begin{figure}
\begin{center}
\includegraphics[width=0.7\textwidth]{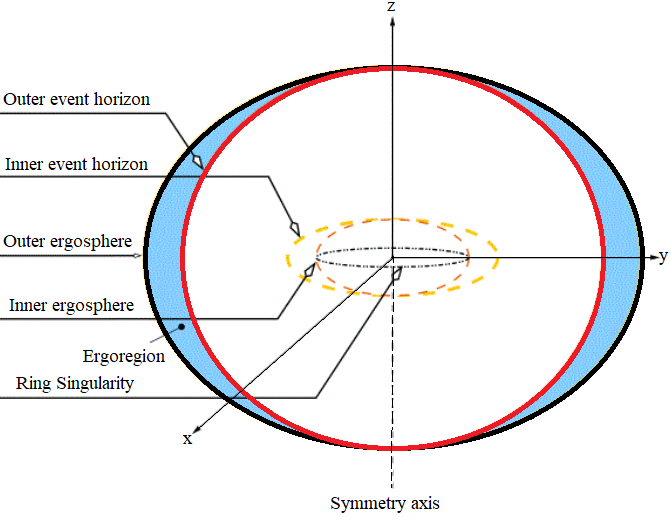}
\caption{A sketch of Kerr black hole with different parts.}
\label{image-1.5}
\end{center}
\end{figure}

The solution Einstein field equation for rotating uncharged black hole is known as Kerr solution. Astronomically, we know that stars are rotating and forms a weak field approximating to Einstein field equation. We know the approximate form of the metric at long distances from a static isolated body having mass m and angular momentum $J$ \cite{Carroll:2004st, Misner:1974qy}. The suitable coordinate system can be expressed as 
\begin{equation}
V= \left(1-\frac{2 M}{r}+O\left(\frac{1}{r^2}\right)\right)dt^2- \left(\frac{4 Jsin^2 {\theta}}{r}+O\left(\frac{1}{r^2}\right)\right)dt d\phi +\left(1-\frac{2 M}{r}+O\left(\frac{1}{r^2}\right)\right) \left(dr^2+r^2 d\Omega ^2\right)
\end{equation}
This coordinate system is valid for all solar system. If a rotating star having angular momentum $J$  undergoes a black hole during collapsing process, then it is expected that it will have angular momentum to retain its initial process.  Kerr black hole is the one, which possess the mass $M$ and angular momentum $J$, but don't have charge i.e. $Q=0$. So, Kerr black hole is characterized by two parameters i.e. mass and angular momentum. Unlike to Schwarzschild black hole, there exists two killing vectors $\partial_t$ and $\partial_\phi$. The Kerr solution is not spherically symmetric and static, but it has axis-symmetric property \cite{Visser:2007fj}. The line element for Kerr black hole is given by the Boyer Lindquist coordinates $(t,r, \theta,\phi)$:
\begin{equation}
ds^2=\left(1-\frac{2 M r}{\rho }\right)dt^2 -\frac{4 a M \text{rsin}^2\theta}{\rho ^2}dt d\phi +\frac{\rho ^2}{\Delta }dr^2+\rho ^2 d\theta ^2+\left( a^2+r^2+\frac{2 a^2 M \text{rsin}^2\theta}{\rho ^2}\right)\theta  \sin ^2 d\phi ^2
\end{equation}

$$\Delta (r)=a^2-2 M r+r^2, \qquad \rho ^2(r, \theta )=a^2 \theta  \cos ^2+r^2-2, \qquad J=Ma$$
Where $a$ is the Kerr parameter, $J=Ma$ is the angular momentum. Form the above line element, we can simply deduce the Schwarzschild black hole case by using $a=0$. The horizon is located at $\Delta(r)=0$. The external horizon and Cauchy horizons are
\begin{equation}
r_{\pm}=M\pm\sqrt{M^2-a^2}
\end{equation}
Similar to R-N black hole there are also two horizons, they are inner and outer event horizons. The space between the two horizons is called Ergo-sphere as shown in Fig.(\ref{image-1.5}) below
In the above metric equation, if $a\rightarrow 0$ then $r_\pm \rightarrow2M$, which shows that it has a coordinate singularity and coincides with Schwarzschild singularity. There is also exists the curvature singularity corresponding to $\rho\rightarrow 0$. As in $R-N$ black hole there is also three cases for comparring the angular momentum and mass as follow
\begin{itemize}
\item $a<M$
\item $a=M$
\item $a>M$
\end{itemize}
Here the second case is also called the extreme case. For this case, we can write $J=M^2$. The first and third cases are unstable and have no physical significance. We can construct the laws of black hole thermodynamics using static or rotating black hole solutions \cite{Davies:1978mf, Hawking:1976de, Hawking:1974rv, Hawking:1982dh}. This will be discussed in next chapter. In addition, we can extend the above discussion to the charged rotating black hole with related results, which is called Kerr Newman solution. The above discussion of black holes with mass $M$ can be summarized in the following table below.
\begin{figure}
\begin{center}
\includegraphics[width=0.7\textwidth]{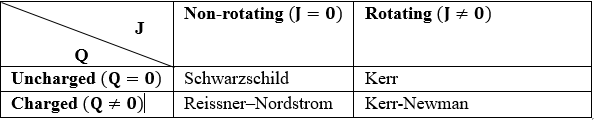}
\caption{Summarization of black hole types.}
\label{image-1.6}
\end{center}
\end{figure}

\section{Black hole thermodynamics}

Since after Einstein field equation in 1915, Schwarzschild was the first to discover the vacuum solution known as ''Schwarzschild metric''. Which consists of spherically symmetric space-time components. The distance from the center (where the mass of collapsing body is compressed) to the boundary of symmetric sphere is consider as the Schwarzschild radius. The boundary of the sphere is termed as ''Event horizon'' the object was named as "Black hole".  Latter in seventies Jacob Bekenstein and Stephen Hawking proposed that black hole emits radiations during quantum mechanical process near the horizon \cite{Hawking:1971tu, Hawking:1974rv, Li:2018bny}. They calculated the black hole temperature and entropy given below. 

\begin{equation}
T=\frac{\hbar}{8\pi G M} ,\quad  S=\frac{A}{4 \hbar G}
\end{equation}

They proposed that emission of radiation is actually a quantum mechanical vacuum polarization which is quasi-static in nature. It evaporates as time possess and eventually the black hole evaporates. At the first stages the question raised that from the second law of thermodynamics, we say that entropy of closed system never decreases, then what will be the entropy of our world, if black hole swallows a hot body having certain entropy? We can't ignore the effect because it will contradict the law of thermodynamics. So, it is supposed that to retain the situation for consistency of system's entropy, it need to introduce the entropy of black hole from one phase to the other during the swallowing of the an object. In such way that we can claim the decrease of entropy of the world due to its  transfer into black hole. This means the second law of thermodynamics requires that black hole has entropy, It can increase but never be decreased.

Hawking area theorem was the first step in this regard for the black hole thermodynamics, this theorem has direct relation to the second law of thermodynamics. See the references \cite{Bekenstein:1973ur, Bardeen:1973gs, Hawking:1976de, Wald:1999vt}. Similarly, we can say that as the radius of black hole horizon is proportional to black hole mass so, when something falls into black hole, it will be an increase in black hole mass, because nothing can come out from it. This clearly supports the notion of black hole has non-decreasing area \cite{Gary, Natsuume:2014sfa}. So, the main reason for the black hole thermodynamics is its area and entropy. It is claimed that if the black hole is formed under collapsed process, then initially it will be asymmetric (i.e. it will have no symmetry) and finally become symmetric to static black hole. All the physical quantities evaporate during the collapsing process and only few parameters can describe the black hole, like mass $M$, charge $Q$ and angular momentum $J$. This property of black is analogous to the thermodynamics.  
If we think for a while about the black hole radiations in addition with area theorem, then it is clear that area theorem is totally based on classical general relativity and there is scope for emission of particle from black hole like Hawking radiation and hence temperature can be assigned to black hole. \cite{Hawking:1974sw, Hawking:1974rv, jacobson1996}. From this fact J.M. Bardeen accomplished the laws of black hole thermodynamics that are analogous the laws of thermodynamics in our common world \cite{Bardeen:1973gs}. In the following section, we will explain the laws black hole thermodynamics.

\subsection{Laws of black hole thermodynamics}

In $1973$ Bardeen et. al proposed that the mass of black hole acts like as the energy $(M=E)$ of a thermodynamic system. Its surface gravity $\kappa$ acting like temperature $T$ and the horizon area $A$ acts like entropy $S$ as that of thermodynamic system. In classical general relativity, there are four laws of black hole thermodynamics. So, these laws are analogous to the law of thermodynamics \cite{Bardeen:1973gs, jacobson1996}. 

\subsubsection{Zeroth Law :}

The zeroth law states that for stationary black hole the surface gravity is constant over the horizon. This law is valid only in spherical symmetric cases. The spherical symmetry in this contest means that the gravitational force is constant on the horizon. So, to obey the zeroth law, a black hole must be in equilibrium  temperature with respect to its surroundings and gravity. The gravitational acceleration can be write as
\begin{equation}\label{acceleration}
a=\frac{M}{r_s ^2}
\end{equation}
And the surface gravity (acceleration per unit mass on the horizon of black hole is called the surface gravity) is
$$\kappa =a r_s=\frac{1}{4 M}$$
so the temperature $T$ of the black hole surface area is
\begin{equation}\label{temperature}
T=\frac{\kappa}{2\pi}
\end{equation}
 For simplicity, we used $c^2=G=1$. The analogous statement of zeroth law in thermodynamics is such that ''in equilibrium condition the temperature of the thermodynamic system is constant.''
\paragraph{\textbf{Theorem}: (Zeroth law of Black hole mechanics)}

\textit{ the surface gravity $\kappa$ is a constant on each consecutively connected component of future event horizon for a stationary black hole space-time, which satisfy the dominant energy condition $(-T_\nu ^\mu V^\nu$ is future directed time like or null vector, where energy momentum tensor $T_{\nu\mu}$ and future directed time like or null vectors $V^\nu$)}

\subsubsection{First Law :}

The first law of black hole thermodynamics is also called the law of energy conservation. This law states that any perturbation in the energy of static black hole will relate to the change in area, charge and angular momentum. Mathematically
\begin{equation}\label{firstlaw1}
dE=TdS+\phi dQ+\Omega dJ
\end{equation}
As the entropy $S$ and temperature $T$ of the black hole are proportional to the area and surface gravity respectively. So, if we can write the energy in term of mass the one gets

\begin{equation}\label{firstlaw2}
dM=\frac{\kappa}{8\pi} dA+\phi dQ+\Omega dJ
\end{equation}\paragraph{\textbf{Theorem:}}
\textit{In a stationary asymptotically vacuum solution of Einstein equation with bifurcate killing horizon, for a small amount of matter carrying energy  $\delta M$, angular momentum $\delta J$ crosses $\Lambda_H$ and the black hole finally become stationary, then then for Horizon $H_i=S_i{\bigcap} \Lambda_H$, (where  $H_i$ is the horizon and  $S_i$ is the surface.).They increases by an amount  $\delta{A_H}$ is given by Eq. (\ref{firstlaw2}).}
\subsubsection{Second Law :}

The entropy of black hole is always increases i.e.
\begin{equation}
dS\geq0 \Rightarrow\qquad dA\geq 0
\end{equation}

Bekenstein proposed that some multiple of black hole surface area (which can be measured in term of squared Planks length) can also be considered as entropy. This statement is conjectured with second law of black hole thermodynamics, also called generalized second law (GSL). It can be written as
\begin{equation}
\delta  \left(\frac{A c^3}{G \hbar }+S\right)\geq 0
\end{equation}

From this, we can see that the sum of black hole entropy plus the entropy of the thermodynamic object (which falls into black hole) is the total entropy and always greater than zero. This law also explains that when something goes into black hole, then we can't count for entropy.
\paragraph{\textbf{Theorem}}: (\textit{Second law of Black Hole Mechanics}) 
\textit{Hawking Area theorem \cite{jacobson1996, Hartman:2008b} Let $(M,g)$ is a asymptomatically predictable space-time obeying the null energy condition states that $T_{\mu \nu} V^{\nu} V^{\mu}\geq 0$, with $V_\mu$ is future directed time like or null vector. Let $S_1$ and $S_2$ are the Cauchy surfaces for $U$, such that $S_2\subset J^+ (S_1) $ then for Horizon $H_i=S_i{\bigcap} \Lambda_H$,}
$$Area(S_2)\geq Area(S_1)$$
\subsubsection{Third Law:}
The surface gravity of black hole can't be reduced to zero by any procedure and no matter how idealized. This was formulated by Bardeen, Carter and Hawking  \cite{Bardeen:1973gs}. For a Kerr black hole the surface gravity $\kappa=0$, if $\frac{J}{M^2}=1$, then it is possible to add mass with black hole and decrease its angular momentum. Another way is to make a naked singularity i.e. when one gets negative mass. In actual situations this case is not possible.

\section{Black hole evaporation}

According to Hawking prediction in \cite{Hawking:1974sw}, black hole emits Hawking radiations. He investigated that the emission of radiation is the result of particle and anti-particle pair creation in the vicinity of black hole horizon. They have opposite momenta. In this incident the particle with positive energy goes outward as a result black hole mass must be decreased at the rate at which the energy is radiated to infinity. For the maintenance of the energy conservation the particle with opposite momenta must act as back reaction \cite{Weber, Einstein}. This process is also known as Hawking effect. It is claimed that the emission of Hawking radiation is analogous to the radiation form a black body. Then the rate of mass loss by the black hole is given by Stefan Boltzmann law as:
\begin{equation}
\frac{dM}{d\nu}=-\sigma A T^4
\end{equation}
Where $\sigma$ is the Stefan Boltzmann constant. In simplest case the Hawking temperature is $T\propto \frac{1}{M}$ and $A\propto m^2$, so we can write as 
\begin{equation}
\frac{dM}{d\nu}=M^{-2}
\end{equation}

This shows that as the black hole losses mass, the temperature increases and at the final stage the black hole, it radiates all of its mass. The time of black hole mass evaporation is given by
\begin{equation}
\nu\approx M^3
\end{equation}
for which in cgs. system we can claim that 
\begin{equation}
\nu \approx 10^{74}\left(\frac{M}{M_0}\right)\sec.
\end{equation}

which show the life time of black hole to evaporate its total mass is more than the life time of universe $(10^{74})$ \cite{Wald:1984rga}. For a star with mass of sun collapsed into black hole $M\rightarrow M_0$, then the life time is approximately $10^{67}$years. It is proposed in \cite{Stephani:2004ud}, that if any small black hole has formed in the early universe then it would have final stage of evaporation in recent times. From which due to high temperature a large amount of $\gamma$-rays will be emitted into background, but since there is no such $\gamma$-rays so, it seems that in our universe there is no such a black hole. Recent analysis by using uncertainty principle reveals that during the evaporation the whole mass of black hole doesn't evaporates, there is always some minimum remnant mass in planks scale order, which show that black hole entropy never vanishes. 

Classically, in the last stages (when losses its mass down to the order of plank mass) of evaporation the effect on black hole will not be possible to study. For such description one need to wait for the theory of quantum gravity. Yet the original theory of quantum gravity is yet not discovered, but there are several ways, which can lead us to the desired position. one of them is the consideration of high dimensional black hole and other is the existence of plank length $l_p$ and mass $M_p$.
	
\section{Discussions and conclusions}

In this chapter, we discussed briefly the concept of black holes. Starting from the Einstein theory of general relativity and field equation, we discussed different geometries as result of solution from field equation. In classical physics, a black hole is compact object with strong gravitational attraction even though light also can't come out from it, i.e. a place of no return for anything in it, due to its strong gravitational field. Black hole is considered to be the final stage of a star's life after losing all of its energy. They are characterized by their boundary called event horizon. The existence of event horizon is the reason for incomplete study of black hole interior for an outsider. In any black hole (static or rotating) an in going object will vanish at 1-dimensional point in the center of black hole, called singularity. This was first proposed by Roger Penrose in $1965$. Due to insufficient information about the black hole interior, a singularity may be considered as the cutoff in the geometrical structure of space-time. Inside a black hole the space-time is coupled to infinity. In $1970$ Brandon Carter and Stephen Hawking proposed No-Hair theorem, which shows a black hole can be characterized by mass $M$, charge $Q$ and angular Momentum J. considering the black hole as thermodynamics object, they also investigated the four laws of black hole thermodynamics. Which is also revised here. Hawking proposed, a black hole can emit radiations due to quantum effects.

\clearpage
\thispagestyle{empty}
\hfill
\clearpage
\newpage
\chapter{Black Hole Interior volume and Entropy}

\section{Introduction}

The concept of black hole interior volume is yet not clearly investigated in the literature on black holes, there are several reasons behind this failure e.g. choice of space-like hyper-surface, the existence of black hole horizon, and also the interchange of space-time coordinates in the interior of a black hole. The choice of such a hyper-surface is itself a problem due to the lack of a mechanism. Parikh \cite{Parikh:2005qs} proposed that the interior volume of a black hole is independent of the choice of stationary time slicing, the killing vector may not be globally time-like. He also added that one can't construct a family of space-time, which has bound (constant) area but unbound volume. Anywhere, if this case is satisfied then there will be something wrong among one of these cases. $(i)$. There will be a cycle of vanishing length (Schwarzschild black holes), $(ii)$. There will be a conical singularity (de Sitter space), $(iii)$. The definition of volume became ambiguous (AdS branes and BTZ) this option is discussed in \cite{Zhang:2019pzd}, (iv). There was a conflict with symmetries (Rindler space), or (v) The area itself diverged (hyperbolic horizons). Christodoulou and Rovelli \cite{CR:2014} considered Eq. (\ref{collapsedmass}) black hole formed under collapsed process and showed, that for a space-like hyper-surface in the interior of a Schwarzschild black hole, the volume increases linearly with advance time as given by relation. 
\begin{equation}
V_{CR}=3\sqrt{3}\pi M^2 \nu
\end{equation}

Their investigations are recently extended to the static charged spherically symmetric RN black hole and rotating Kerr hole in \cite{Han:2018, Wang:2018dvo, Wang:2018txl, Ong:2015tua}. It is found that the volume in the interior is linearly growing with advanced time. In reference \cite{Yang:2018arj} the rate of mass change is taken into account, and it is found that volume increases toward evaporation, and in high-dimension cases, the increase in the rate of mass change decreases with the number of dimensions.

Similarly, the concept of entropy is an important term to discuss the hidden modes in the interior of black holes. It is assumed that the investigation of interior volume and entropy may be act as backbone for solving the problem of information paradox. The most important effort is recently made Baocheng Zhang using the statistical way in \cite{Zhang:2015} by calculating the entropy for the scalar massless field in the interior of Schwarzschild black hole by using the interior volume analogy as stated in \cite{CR:2014}. It is found by Baocheng Zhang, that the interior entropy of black hole is directly proportional to black hole horizon area.
\begin{equation}\label{baochengent}
S_{CR}=\frac{3 \sqrt{3} \gamma }{\left(90\times 8^4\right) \pi } A
\end{equation}
Where $A=16\pi M^2$. Note that, these analyses are only applicable if the black hole mass is greater than Plank mass. If the black hole mass is less or equal to the Plank mass, then it is needed to understand in the form of uncertainty relation \cite{dehghani:2009gu}. Actually, at the final stage, the rate of mass loss due to Hawking radiation is $\frac{dM}{d\nu}\approx M$. This mean that the black hole radiation are so slow that it can be ignored and the mass of black hole is approximately constant. But this situation is not consistent in actual situation of the CR volume and also to the entropy in the interior of CR volume, because the entropy is proportional to the interior volume. From the above Eq. (\ref{baochengent}), one can see that the proportionality constant is less than $\frac{1}{4}$, which shows that horizon entropy is much greater than $S_{CR}$. Similarly, it is also revealed, that at vanishing stage of mass the CR volume is non-vanishing and hence entropy. This means that there is always some amount of remnant mass after the black hole evaporation process. Such concept supports the non-vanishing entropy, which leaves the possibility of solving the problem of information paradox. Such a discussion for higher dimensional Schwarzschild black hole is presented in \cite{Yang:2018arj}, a semi-classical approach is followed by Majhi.et.al \cite{Majhi:2017tab}.

The main aim of this chapter is to revisit the interior volume and entropy of different types of black holes . We will also make comparison among them, in order to take out the effect on the interior entropy due to difference in geometry of black hole. Structure of our contents is such that, in next section we will evaluate the interior volume. In third section, we will consider the entropy calculation of the scalar modes as followed by Baocheng Zhang in \cite{Zhang:2015}. We will also make discussion on the entropy correction by following the semi-classical way as followed by \cite{Majhi:2017tab}. Finally, to discuss the black hole evaporation, we will discuss the differential form of different entropies and proportional relation between the entropy of the scalar modes and the Bekenstein Hawking Entropy.

\section{Volume in the Interior of Black Hole }
Recently, M. Christodoulou and C. Rovelli discussed the interior volume of a $3d$ spherically symmetric hypersurface and defined the interior volume of Schwarzschild black hole. Accordingly, the volume bound by a $3d$ space-like hyper-surface $\sum_\nu$ is the maximum volume bounded by a spherically symmetric sphere $S_\nu$ \cite{CR:2014, Christodoulou:2016tuua}. For a sphere $S$ in Minkowski space-time with radius $R$, it is very convenient to compute the volume $V$ as the volume inside a spherically symmetric $2$-sphere, which is given by $V=\frac{4}{3}\pi R^3$. This means that there is a space-like hyper-surface bounded by the sphere $S$ whose volume is equal to $V$ as shown in the Fig. (\ref{image-2.1}) below.  But, if there are many space-like hypersurfaces bounded by $S$, then what will be the volume? In such case, we need a hyper-surface that satisfy the two equivalent statements as given below
\begin{figure}
\begin{center}
\includegraphics[width=0.4\textwidth]{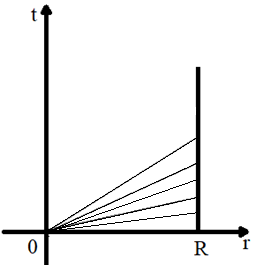}
\caption{The Volume of Space- like hyper-surface bounded by Minkowski space-time with $\theta<45^0$}
\label{image-2.1}
\end{center}
\end{figure}
In such case, we need a hyper-surface that satisfy the two equivalent statements as given below
\begin{itemize}
\item{The hyper-surface $\sum$ chosen must be lie on the same simultaneity surface as $S$}.
\item{The hyper-surface $\sum$ must be the largest spherically symmetric bounded surface by shphere $S$.}
\end{itemize}

If the hyper-surface chosen to satisfy the above two statements, then it will bound the largest possible volume as compared to other in the desired space-time. If $V_R$ is the volume of the largest hyper-surface then it must satisfy the conditions,
$$\delta V_R \geq 0 \qquad V_R>V_R'$$
Where $V_R'$ is the volume with respect to some other space like surface also bounded by $S$.

This means that if hyper-surface $\sum$ is the largest one, then the change in the volume must be zero and if we compare this volume $V_R$ of the largest hypersurface to the volume $V_R'$ bounded by some other hyper-surface $\sum'$, then it must satisfy the second condition given above. In the Fig. (\ref{image-2.1}) above the simultaneity surfaces of the inertial observer are straight lines in $t-r$ plane. From the above two conditions, let us consider $R$ is the radius of largest hyper-surface then, we can write the metric component as $g_{\mu\nu}=(R,0,0,0)$ and the metric as $ds^2=dR^2$. If, we take $R'$ as another hyper-surface, then the metric component $g'_{\mu\nu}=(T',-{\sum_{i=1}^3}R'_i)$, the metric is $ds'^2=-dT'^2+{\sum_{i=1}^3}d R'^2 _i  $. Which shows that $V_R>dV' _R$ in the chosen space-time. Note that any spherically symmetric sphere can reside only one largest hyper-surface inside it.  The inertial frame will be defined with respect to hyper-surface $\sum$ having the biggest volume (also called the proper volume).

\begin{figure}
\begin{center}
\includegraphics[width=0.4\textwidth]{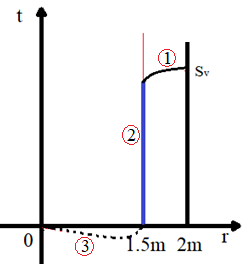}
\caption{$t-r$ diagram, showing the hyper-surface of infinite length}
\label{image-2.2}
\end{center}
\end{figure}

In case of curved geometry like Schwarzschild geometry, which have lacking of flatness. So, the statement of the simultaneity surfaces can't be considered, but one can extend the statement of largest hyper-surface to curved space-time as the volume bounded in curved space-time will be the volume due to the contribution of some largest hyper-surface. Let us consider $t-r$ plane in Schwarzschild type space time geometry as shown in Fig. (\ref{image-2.2}). the horizon is at $r=2M$. We consider a maximal sphere $S_\nu$ starting at the horizon and approaches to the asymptotically stretch surface at  $r=\frac{3}{2} M$. This surface is also called the limiting surface and can be calculated numerically. After this point the maximal surface either approaches to the singularity, or approaching back to horizon due to which the proper volume of maximal surface is infinite as shown in the Fig. (\ref{image-2.2}) below.
For a $2d$ object in flat geometry is, the interior volume is defined as
\begin{equation}
V=4\pi\int{r^2\sqrt{1-\left(\frac{dt(r)}{dr}\right)^2}}dr
\end{equation}
Which yields maximum volume of $\frac{4}{3}\pi r^3$ at $t(r)=0$

In case of non-eternal black hole formed under collapsing process, the maximal hyper-surface don't have infinite volume, because the maximal surface starting form horizon and approaching to the surface $r_\nu=\frac{3}{2} M$ don't extend to infinity. Actually, it is the surface that prevents the object to go into singularity. The reason is that the surface $r_\nu$ is the connection between the horizon and the center of collapsed object before going to the singularity point. For simplicity, see the fig (\ref{image-2.2}) carefully. Let us make sections of the maximal surface from horizon to the center of collapsing object as shown in the Fig. (\ref{image-2.3}) below. In this way, we can find the effective part of hyper-surface, which contributes to the maximal volume of hyper-surface. Starting from horizon, we see that part $1$ is a null surface so, its volume contribution of the black hole is equal to zero. The second part is the stretched part extending from the end of part $1$ to the boundary of collapsing object, the volume of this part is linearly increasing with the advance time, while the horizon remains constant due to curvature. This part also satisfy the Parikh statement for finding the bound area and unbound volume can led us to find the black hole interior volume. For the volume of largest hyper-surface, we will choose this part of hyper-surface which increases with the advance time as discussed above. This part is stretched and straight. The 3rd part is the collapsing part as shown by dotted line in Fig. (\ref{image-2.3}). Since the part of space-time in collapsing object acquires the time-like killing vector, which results in finite volume. This constant part of volume can also be neglected because the volume of part 2 linearly grows with advance time $\nu$ and this part have finite volume. So, if we want to find the bulk volume central part will be suitable for desired result as shown by part 2 in the figure below.
\begin{figure}
\begin{center}
\includegraphics[width=0.4\textwidth]{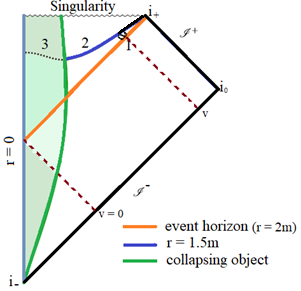}
\caption{The Penrose diagram, which shows the largest hyper-surface, which guarantees the interior volume of a static spherically symmetric  black hole formed under collapsed process (blue line Part-2). Part-1 is the null part and part-3 is the collapsed part with finite volume.}
\label{image-2.3}
\end{center}
\end{figure}
The above definition remains valid in the case of spherically symmetric black hole formed under collapsed process, like Schwarzschild black hole, RN-AdS black hole.
\subsection{Volume in the interior of Schwarzschild Black hole}

Consider the Schwarzschild geometry in Eddington Finkelstein coordinates $(\nu,r,\theta,\phi)$, the line element is given in Eq. (\ref{schmetric}) of Chap.1. For an uncharged spherically symmetric black hole $f(r)=1-\frac{2M}{r}$ and $2$-sphere form is $d\Omega^2=r^2(d\theta^2+sin^2 \theta d\phi)$ and the advance time is $\nu=t+r^*$.
\begin{equation}
r^*=r+2Mlog|r-2M|
\end{equation}

Here we consider the geometric units as $G=c=\hbar=k_B=1$ . The horizon $(r=2M)$ as foliated by the sphere $S_\nu$. The hyper-surface is defined as the product of anaffine parameter to the 2-sphere and can be written as $\sum=\gamma\times S^2$,  where $\gamma\rightarrow(\nu(\lambda),r(\lambda))$ is the affine parameter. So, the metric (\ref{schmetric}) in Chap.1 can be written as 
\begin{equation}\label{EDmet}
{ds}^2_{\sum}=\left(-f(r) \dot{\nu }^2+2 \dot{\nu } \dot{r}\right)d \lambda ^2+r^2d\Omega^2
\end{equation}

As considered only curved space times, so the contribution to the volume will not be same to $V=\frac{4}{3} \pi R^3$ as normal volume. Using the Eq. (\ref{EDmet}) the interior volume of a spherically symmetric black hole is given as
$$V_{CR}=\int_0 ^{2M} d\lambda \int d\Omega \sqrt{r^4\left(-f(r)\dot{\nu }^2+2 \dot{\nu } \dot{r}\right)sin^2 \theta}$$
\begin{equation}\label{intvolgenfor}
=4 \pi\int{\text{d$\lambda $}}\sqrt{r^4\left(-f(r)\dot{\nu }^2+2 \dot{\nu } \dot{r}\right)}
\end{equation}

This equation shows that the proper length of geodesic in auxiliary metric ($4\pi$-times) is precisely the volume of the hyper-surface $\sum$. An investigation for geodesic in an auxiliary manifold is given by maximization condition to find the maximal curve $r=\frac{3}{2}M$, which together with the metric of hyper-surface gives the interior volume. This equation also shows that finding the hyper-surface is same as solving the geodesic equation for the equation of motion with Lagrangian. So the auxiliary metric form the above equation can be written as 
\begin{equation}\label{effmetric}
{ds}^2_{eff}=r^4 \left(-f (r) \dot{\nu }^2+2 \dot{\nu } \dot{r}\right)
\end{equation}
So, calculating the volume of the hyper-surface $V_{\sum}$ is equivalent to calculating the volume from auxiliary metric with Lagrangian whose maximal value is

$$L\left(r,\nu, \dot{r}, \dot{\nu}\right)=1$$
i.e.
\begin{equation}\label{Eq.1}
\left(-f(r) \dot{\nu }^2+2 \dot{\nu } \dot{r}\right)=1
\end{equation}
using this Eq. (\ref{intvolgenfor}), we gets 
\begin{equation}\label{VCR}
V_{CR}=4\pi \lambda_f
\end{equation}

The metric $\tilde{g}_{\alpha\beta}$   has a Killing vector, $\zeta^\mu=(\partial_\mu)^\mu\propto (1, 0) $. Since $\gamma$ is an affinely parameterized geodesic in $M_{aux}$, the inner product of Killing vector $\zeta^\mu$ with its tangent $\dot{x}^\alpha=(\dot{\nu}, \dot{r})$ is conserved as
\begin{equation}\label{Eq.2}
\zeta\times \dot{x}^\alpha=r^4\left(-f(r) \dot{\nu }^2+2 \dot{r}\right)=A
\end{equation}
solving Eq. (\ref{Eq.1}) and (\ref{Eq.2}), they can be written in the form
\begin{equation}\label{Eq.3}
\dot{r}=-r^{-4}\sqrt{A^2+r^4f(r)}\qquad and  \qquad \dot{\nu}=\frac{1}{A+r^4\dot{r}}
\end{equation}

As for space-like nature $(-f(r) \dot{\nu }^2+2 \dot{\nu } \dot{r})>0$, then from Lagrangian, we can say that $L>0$, since $r$ is positive and the Lagrangian will vanish at $r=0$, which is the final point of the geodesic, as can be seen from (\ref{Eq.3}), $\dot{r}$ becomes infinite. Thus, $\gamma$ is a space-like geodesic in $M_{aux}$. A well-suited parametrization is to take $\lambda$ as the proper length in $M_{aux}$. So, the auxiliary metric given above can be written as,
\begin{equation}\label{auxmet}
ds_{M_{aux}}=-\sqrt{-r^4f(r)}d\nu=Ad\nu
\end{equation}

It can be easily seen the $A$ has to be negative for the geodesic to be space-like. Then, $\dot{\nu}$ and $\dot{r}$ are both negative and there are only positive terms in (\ref{Eq.1}). Integrating (\ref{Eq.2}), we get
\begin{equation}\label{VCR/4pi}
\frac{V_{\sum }}{4 \pi }=\lambda _f=\int _0 ^{2 M}\frac{r^4}{\sqrt{A^2+r^4f(r)}}
\end{equation}
Eq. (\ref{VCR/4pi}) shows that the restriction imposed on A.
$$A^2> -r_\nu^4f(r_\nu)>=0$$
\begin{equation}
\Rightarrow A^2=\frac{27}{16}M^4=A_c ^2
\end{equation}

This condition is obtained from the expression $(-f(r) \dot{\nu }^2+2 \dot{\nu } \dot{r})>0$ having the roots r=0 and $r=2M$ otherwise the position is maximum at $r_\nu=\frac{3}{2}M$. Since $(-f(r) \dot{\nu }^2+2 \dot{\nu } \dot{r})>0$ in the range $0<r<2M$. For every constant r there is a solution or we can say $r$ is constant the surface is space like geodesic of auxiliary manifold. For a stationary (maximal) point of the volume given in Eq. (\ref{Eq.3}(b))
$$\frac{d\nu}{d\lambda}=\frac{1}{A}\Rightarrow d\lambda=Ad\nu$$
Integrating we gets
\begin{equation}
\lambda_f=A(\nu_f-\nu)
\end{equation}
Thus, the $r = constant$ surface with the largest volume between two given values of advance time $\nu$ when $A$ is largest. This means that, $r=r_v$ which gives $A=A_c$.So, the volume will be the largest possible. These considerations provide the basis for the derivation of the asymptotic volume.
\begin{equation}\label{Acnu}
\lambda_f=A_c \nu
\end{equation}
Using Eq. (\ref{Acnu}) in Eq. (\ref{VCR}) one simply gets the interior volume as
\begin{equation} \label{VCR1}
V_{CR}=-4\pi\sqrt{-r^4f(r)}\nu=4\pi A_c \nu
\end{equation}
or we can write as
\begin{equation}\label{SCHVCR}
V_{CR}=3\sqrt{3}\pi M^2 \nu
\end{equation}

Where $A=-\sqrt{-r^4f(r)}$, Which shows that the volume is increases with advance time. This result can be extended to other cases by simply using metric Eq. (\ref{EDmet}) with function $f(r)$ from desired black hole metric and find the maximization of $A=A_c$, in order to get the asymptotic expression, use analogy to above equation, one gets the required black hole interior volume. Let use consider the outer cases as below.

In case of higher dimensional Schwarzschild black holes, following the above analogy, one can calculate the interior volume by considering a d-dimensional metric Eq. (\ref{EDmet}) as 
\begin{equation}\label{HDEDmet}
{ds}^2_{\sum}=\left(-f(r) \dot{\nu }^2+2 \dot{\nu } \dot{r}\right)d \lambda ^2+r^2d\Omega_{d-1} ^2
\end{equation}
With the radial function $f(r)=1-\left(\frac{r_h}{r}\right)^{d-2}$ and the horizon radius $r_h=\left(\frac{16 \pi  M}{(d-1) A_{d-1}}\right)^{\frac{1}{d-1}}$ and is $A_{d-1}=\frac{2 \pi ^{d/2}}{\Gamma \left(\frac{d}{2}\right)}$. The horizon area is $\textit{A}=A_{d-1} r_s^{d-1}$ \cite{Yang:2018arj, Bhaumik:2016sav}. Following the above calculations, the asymptotic expression of interior volume is 

$$V_{CR}=\int  \sqrt {-r_\nu^{2(d-1)} f(r_{\nu })}d\nu d\Omega_{d-1}=A_{d-1}\int{ \sqrt{r_{\nu }^{2 (d-1)} \left(\left(\frac{r_s}{r_{\nu }}\right)^{d-2}-1\right)}}d\nu$$

 This gives
\begin{equation}
V_{CR}=\sqrt{\frac{d-2}{d}}\left(\frac{d}{2 (d-1)}\right)^{\frac{d-1}{d-2}} \nu  A_{d-1} r_s^{d-1}
\end{equation}
Where $r_{\nu}=r_s\left(\frac{d}{2 (d-1)}\right)^{\frac{d}{d-2}}$ is used as the position of maximal surface. Same to the CR volume, this volume is also increases with advance time. but this volume also have relation with the dimensional of the black holes.
\subsection{Volume in the interior of Reissner Nordström black hole}

Let us consider a spherically symmetric charged black hole formed under collapsed process. The line element is given in Eq. (\ref{EDmet}) with function given in Eq. (\ref{RNBH}) \cite{Wang:2018dvo, Wang:2018txl}
Where the advance time is $\nu =r^*+t=r+t-\sqrt{Q^2-2 M r} tan^{-1} \left(\frac{r}{Q^2-2 M r}\right)$ with geometric units taken $G=c=\hbar=k_B=1$. At the horizon, we have $f(r)=0$ so, which yields the outer and inner (Cauchy) horizons as given in Eq. (\ref{RNBHradii}).
Where $r_+$ is the outer and $r_-$ is the inner horizons  of black hole. The interior volume for a spherically symmetric black hole is defined in Eq. (\ref{VCR1}). Where
\begin{equation}
A_c=\sqrt {-r_\nu^4 f(r_{\nu })}
\end{equation}
By maximization condition, we can get
\begin{equation}
r_\nu=\frac{3M+\sqrt{9M^2-8Q^2}}{4}
\end{equation}
which is same to Eq. (A3) of the reference \cite{Christodoulou:2016tuua}. As Eq. (\ref{SCHVCR}) of Schwarzschild black hole is calculated at large advance time $\nu$ i.e. $\nu>>M$. Which means that black hole is made during collapsing process. For complicated black hole solutions, one can't calculate $A_c$ analytically, but it is easy to calculate the function $(\sqrt{-r^4 f(r)})$ numerically, \cite{Wang:2017zfn}. The minimum value of $A_c$ for $r_\nu$ is obtained as,

$$=\left(-\left(\frac{3 M+\sqrt{9 M^2-8 Q^2}}{4}\right)^4 \left (\frac{2 \left(3 M^2-4 Q^2+M \sqrt{9 M^2-8 Q^2}\right)}{\sqrt{9 M^2-8 Q^2}+3 M}  \right )\right)^{\frac{1}{2}}$$
\begin{equation}
=\frac{1}{32} \sqrt{27 M^4-8 \text{MQ}^2 \sqrt{9 M^2-8 Q^2}-36 M^2 Q^2+9 M^3 \sqrt{9 M^2-8 Q^2}+8 Q^4}
\end{equation}
This minimum value of $A_c$ will give the maximum value of the interior volume of black hole. The calculated value of $A_c$ satisfies the CR volume result for $Q=0$ as
\begin{equation}
A_c=\frac{3\sqrt{3}}{4} M^2
\end{equation}
So, the interior volume of Reissner Nordstr$\ddot{o}$m black hole after using the above equation is 
\begin{equation}\label{RNBHvol}
V_{CR}=\pi{\frac{\sqrt{27 M^4-8 \text{MQ}^2 \sqrt{9 M^2-8 Q^2}-36 M^2 Q^2+9 M^3 \sqrt{9 M^2-8 Q^2}+8 Q^4}}{\sqrt{2}} }\nu
\end{equation}
This relation is consistent with the interior volume of Schwarzschild black hole given above in Eq. (\ref{SCHVCR}) at $Q=0$.

\subsection{interior Volume of Kerr Black hole }

Considering the analogy of the CR work, the interior volume in the interior of Kerr black hole is discussed in \cite{Wang:2018dvo, Bengtsson:2015zdaa} . In CR work, they used to calculate the interior volume of static symmetric black holes e.g. Schwarzschild black hole and RN black holes. Kerr black hole have axis-symmetric, so one can not use the maximization condition for calculating the maximal hyper-surface. However one can use the definition of vanishing surface as discussed by \cite{Zhang:2015}. Here we will give a short introduction to the interior volume of Kerr black hole. The Kerr metric in Eddington Finkelstein coordinates \cite{Kerr:1963ud} can be written as.

\begin{equation}\label{Kerrmet}
ds^2=-\frac{\Delta -a^2 \theta  \sin ^2}{\rho ^2}d\nu^2+\rho ^2 d\theta ^2+2 d\nu dr+\frac{A^2 \theta  \sin ^2}{\rho ^2}d\phi^2-2 a \theta  \sin ^2 dr d\phi -\frac{4 a \theta  M \text{rsin}^2}{\rho ^2}d\nu d\phi
\end{equation}
where
$$\Delta (r)=a^2-2 M r+r^2$$
$$\rho ^2(r,\theta )=r^2+a^2 \cos ^2\theta $$
$$A=(r^2+a^2)^2-a^2 sin^2\theta$$
and
$$J=a M$$
Here $J$ is the angular momentum of Kerr black hole. The interior and exterior horizons can be calculated for $\Delta=0$,
\begin{equation}\label{Kerrrad}
r_\pm=M\pm \sqrt{M^2-a^2}
\end{equation}

Kerr black hole has the property of spinning, so the horizons of a Kerr black hole may not be spherically symmetric. The character may affect the physical properties of Kerr's black hole. As discussed in CR work for a Schwarzschild black hole, the interior volume is mainly the contribution of the largest spherically symmetric stretched part of a $3d$ hyper-surface. The largest hyper-surface can be calculated by using the condition of vanishing curvature, which doesn't extend to singularity, so one can say that the spinning character don't affect the interior volume of a Kerr black hole. So, without loss of generality, the  interior volume can be defined as
\begin{equation}
V_{CR}=\int{\sqrt{-\Delta}\rho sin\theta d\nu d\theta d\phi}
\end{equation}
which gives
\begin{equation}\label{KerrBHvol}
V_{\text{CR}}=2 \sqrt{-\Delta } \pi \nu \left(\sqrt{r^2 + a^2} + \frac{r^2}{2 a} log\left(\frac{\sqrt{a^2+r^2}+a}{\sqrt{a^2+r^2}-a}\right)\right)
\end{equation}

This result is also investigated in \cite{Bengtsson:2015zdaa}. Form this equation we can say that if $r=r_s$, then the hyper-surface will be largest and the volume calculated also has its peak value.  Generally, this volume equation can be written as
\begin{equation}\label{Kerrvol}
V_{CR}=2\pi F(r,a)\nu =2\pi M^2 F_{max}\left(\frac{a}{M}\right)\nu
\end{equation}
where 
$$F(r,a)=\sqrt{-\Delta } \left(\sqrt{r^2 + a^2} + \frac{r^2}{2 a} log\left(\frac{\sqrt{a^2+r^2}+a}{\sqrt{a^2+r^2}-a}\right)\right)$$
and 
$$F_{max}\left(\frac{r}{M}\right)=F\left(\frac{r_\nu}{M},\frac{a}{M}\right)$$
If we plot the function $F_{max}\left(\frac{r}{M}\right)$ vs $\frac{r}{M}$, it gives 

\begin{figure}
\begin{center}
\includegraphics[width=0.6\textwidth]{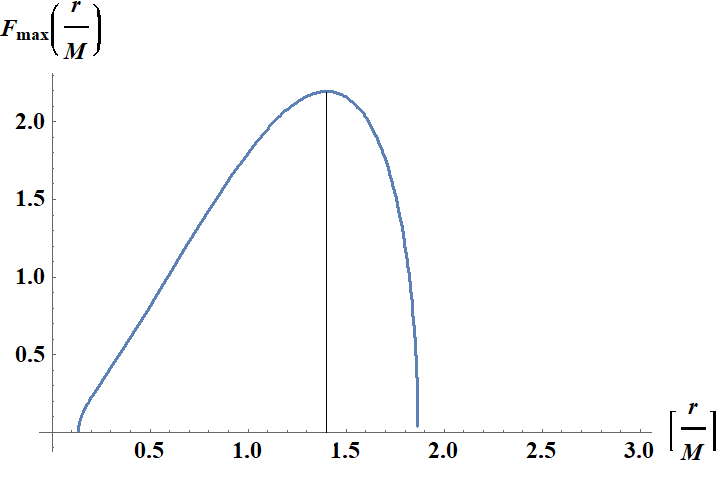}
\caption{The plot of $F_{max}\left(\frac{r}{M}\right)$ vs $\frac{r}{M}$. The function is maximum at $(\frac{r}{M})=1.4039$ and $(\frac{a}{M})=0.2$}
\label{image-2.4}
\end{center}
\end{figure}

This plot shows that the position of the hyper-surface is at $r_\nu=1.4039$ when $a=0.2$. For the detail see \cite{Bengtsson:2015zdaa}. Considering Kerr metric given in Eq. (\ref{Kerrmet}) the interior volume is calculated as in Eq. (\ref{Kerrvol}). The extrinsic curvature for a Kerr black hole is calculated as
\begin{equation}
K=\frac{\sqrt{2} \left(a^2 cos2\theta (M-r)+a^2 (M-3 r)+2 r^2 (3 M-2 r)\right)}{a^2 \left(cos2\theta+2 r^2\right)^2 \sqrt{\frac{a^2+(3 M) r}{a^2 cos2\theta+a^2+2 r^2}}}
\end{equation}

From this equation, if $a=0$ the curvature of a Kerr black hole diverges. According to the reference \cite{Ong:2015tua}, the position of the largest hyper-surface $r_\nu$ is calculated by the maximization of the extrinsic curvature, so if the trace of the extrinsic curvature is zero the volume will also be automatically maximum. It means that at the vanishing curvature the hyper-surface will be largest for the maximum volume. The hyper-surface is largest at $r_\nu$ for value of $1.4$ as $\theta$ varies from $0$ to $\pi$. For which the volume is calculated is above.  Using the above volume discussions, we can calculate the entropy in the interior of black hole using Baocheng Zhang method.

\section{Entropy in the Interior of Black Hole}

The CR volume in the interior of black hole depend on how the space-time is sliced into the space and time in it, i.e. the CR volume gives us the information about the slice dependent of the CR volume. Unlike to the surface area of the black hole (which is same for all observer), a Schwarzschild black hole has the volume given above in Eq. (\ref{SCHVCR}). This interior volume is found directly proportional to the advanced time $\nu$. It is supposed that if one could make a connection between the interior entropy and horizon entropy then it may solve the problem of information paradox. So, this interior volume $V_{CR}$ is investigated for understanding and solving the problem of black hole information paradox. This investigation is also important to understand the black hole radiations and corresponding thermodynamics. This will be discussed in next section. for this, let us first consider the Schwarzschild black hole for calculating the interior entropy

\subsection{Interior entropy of Schwarzschild Black hole}
Consider the Schwarzschild black hole formed under collapse process as discussed in reference \cite{CR:2014}. Eq. (\ref{EDmet}). The interior volume $V_{CR}$ is discussed in section 2.2 above. We have  seen from Eq. (\ref{SCHVCR}), (\ref{RNBHvol}) and (\ref{Kerrvol}) that the black hole interior is different from normal volume so, for finding the interior entropy, it is important that how many modes of quantum field are existed there in it. Baocheng Zhang discussed the entropy of the quantum scalar modes in the interior of Schwarzschild black hole. Following the quantum statistical method the number of quantum states in the volume with phase space labeled by $(\lambda, \theta, \phi, p_{\lambda}, p_{\theta}, p_{\phi})$ is calculated. According to  uncertainty principle $\delta x \delta p \approx 2\pi$. 

So, quantum mode corresponds to the cell of volume $(2\pi)^3$ in that phase space. The number of quantum modes in in the cell is 
\begin{equation}
\frac{phase{\quad}space}{volume{\quad}of{\quad}the{\quad} cell}=\frac{d\lambda d\theta d\phi dp_{\lambda}dp_{\theta} dp_{\phi}}{(2\pi)^3},\qquad \hbar=1
\end{equation}
This phase space consists of many quantum modes so, for the total interior quantum modes one can take the integral form of the above statement as
\begin{equation}\label{freeEnergy}
g(E)=\frac{1}{(2 \pi )^3}\int{d\lambda d\theta d\phi dp_{\lambda}dp_{\theta} dp_{\phi}}
\end{equation}
This equation gives the number of quantum modes with energy less than $E$. By taking the transformation of $(\nu,r)$ such that $\nu \rightarrow \nu(t,\lambda)$ and $r\rightarrow r(t,\lambda)$ with
\begin{equation}
-f(r)\left(\frac{\partial \nu }{\partial T}\right)^2+\frac{\partial \nu }{\partial T}\frac{\partial r}{\partial T}=-1
\end{equation}
we can write the metric in transformed form. The purpose of this substitution is to make the possibility of constant time variation as vanishing factor in the interior of black hole. so that the modified metric can be written as
\begin{equation}\label{EDmet+T}
{ds}^2=-dT^2+\left(-f(r) \dot{\nu }^2+2 \dot{\nu } \dot{r}\right)d \lambda ^2+r^2d\Omega^2
\end{equation}
Here 
$$g_{TT}=-1,\quad g_{\lambda \lambda }=\left(-f(r)\dot{\nu }^2 +2 \dot{\nu } \dot{r}\right), \quad g_{\theta \theta }=r^2, \quad g_{\phi \phi }=\theta  r^2 \sin ^2$$

are the metric components. The general form of the metric is ${ds}^2=-dT^2+H(T)d \lambda ^2+r(T)^2d\theta^2+r(T)^2d\phi^2$.
This shows that the interior of the black hole is not static, i.e. for constant radial component the hypersurface is dynamical at given time $T$ . As, these investigations are made for $\nu>>M$ and $r=\frac{3}{2}M$, near the maximal hyper-surface, the proper time between two neighbouring hyper-surfaces tends to zero as $t$ increases so, no evolution happening there. Thus our statistical calculations are not affected by the non-static character of the metric. It is calculated on approximately $T=$constant which is the hyper-surface that leads to the CR volume. Now, the entropy of the scalar modes in the interior of Schwarzschild black hole can be obtained by following the statistical way.

From the WKB approximation, the scalar field in the interior of black hole consisting quantum modes is
\begin{equation}\label{Scalarfield}
\Phi =e^{\text{-iET}}e^{\text{iI}(\lambda ,\phi )}
\end{equation}
Using this in Klein Gordon equation 
\begin{equation}
\frac{1}{\sqrt{-g}}{{\partial_ \mu }\left(\sqrt{-g} g^{\mu \nu } \partial_\nu \Phi \right)}=0
\end{equation}
we can get the equation of motion. Expanding and solving the above equation in components $(T, \lambda,\phi)$ and solving, gives (See Appendix \ref{app.A})
\begin{equation}\label{eqofenergy}
E^2-\frac{p_{\lambda }^2}{-f(r) \dot{\nu }^2 +2 \dot{\nu } \dot{r}}-\frac{p_{\theta }^2}{r^2}-\frac{p_{\phi }}{ r^2 \sin ^2 \theta }=0
\end{equation}
Where, we used
$$\frac{\partial I}{\partial \lambda }=p_{\lambda }, \qquad \frac{\partial I}{\partial \theta }=p_{\theta }, \qquad \frac{\partial I}{\partial \phi }=p_{\phi }$$
as eigen states of the diagonal elements of the  metric. Solving Eq. (\ref{eqofenergy}) for $p_\lambda$, we gets
\begin{equation}\label{P:lambda}
p_{\lambda }=\sqrt{-\dot{\nu }^2 f(r)+2 \dot{\nu } \dot{r}} \sqrt{E^2-\frac{p_{\theta }^2}{r^2}-\frac{p_{\phi }}{r^2 \sin ^2\theta}}
\end{equation}
Use Eq. (\ref{P:lambda}) in Eq. (\ref{freeEnergy}) and simplifying, we gets 
$$g(E)=\int{d\theta  d\lambda  d\phi\sqrt{- f(r)\dot{\nu }^2+2 \dot{\nu } \dot{r}}\int{\sqrt{E^2-\frac{p_{\theta }^2}{r^2}-\frac{p_{\phi }}{r^2 \sin ^2\theta}}}dp_{\theta } dp_{\phi }}$$
$$=\frac{1}{(2 \pi )^3}\int d\theta  d\lambda  d\phi\sqrt{- f(r)\dot{\nu }^2+2 \dot{\nu } \dot{r}} \left (\frac{2 \pi }{3} E^2 r^2 sin^2 \theta  \right )$$
$$=\frac{E^3}{(2 \pi )^3}\frac{2 \pi }{3}\int{d\lambda  \sqrt{r^4 \left(2 \dot{\nu } \dot{r}-\dot{\nu }^2 f(r)\right)}}{\int{\sin^2 \theta d\theta} \int{d\phi}}$$
$$\frac{E^3}{12 \pi ^2}\left ( 4\pi\int{d\lambda  \sqrt{r^4 \left(2 \dot{\nu } \dot{r}-\dot{\nu }^2 f(r)\right)}} \right )$$
\begin{equation}\label{Qstate}
g(E)={\frac{E^3}{12 \pi ^2}{V_{CR}}}
\end{equation}

For the integration in the first step, we used the general formula $ \sqrt{1-\frac{x^2}{a^2}+\frac{y^2}{b^2}} dx dy=\frac{2 \pi }{3} a b$ From Eq. (\ref{Qstate}), we see that $g(E)\propto V_{CR}$ but it is still have similarity with normal space time, but the physical interpretation is not need to be same because the volume in general relativity is the result of the curved space time. The main reason for physical difference can be considered as the volume is bound within the closed hyper-surface that is increasing with advanced time $\nu$. So, the number of quantum states inside the black hole must also be increased withe time. This statement can be considered as the effective factor for volume in curved space-time. Consider this difference, we can calculate the free energy as
\begin{equation}\label{BfreeE}
F(\beta )=\frac{1}{\beta }\int {dg(E)} ln(1-e^{-\beta (E)})   
\end{equation}
$$ F(\beta)=-\int{\frac{dg(E)}{e^{-\beta(E)} -1}}$$
$$F(\beta)=-\frac{V_{\text{CR}}}{12 \pi ^2}\int{\frac{E^2 dE}{e^{-\beta (E)}-1}}$$
Solving the integral, we get the result as
\begin{equation}\label{FreeEnergy}
F(\beta)=-\frac{\pi^2 V_{CR}}{180\beta^4}
\end{equation}
Finally, the entropy is
\begin{equation}\label{Ent1}
S_{CR}=\beta^2 \frac{\partial F}{\partial \beta}=\frac{\pi^2 V_{CR}}{45\beta^3}
\end{equation}

Which looks like the entropy in normal volume. Using the value of interior volume $V_{CR}$ from Eq. (\ref{SCHVCR}) and inverse temperature, we can get the entropy of black hole interior quantum modes. Since a black hole have the property of emitting the radiation, the emission is claimed to be quasi-static and increases with time due to which the black hole temperature not constant. Treating the black hole radiation as black body radiations then one can defined by Stefan Boltzmann law. Accordingly, the rate of mass loss from Schwarzschild black hole due to Hawking radiation can be written as
\begin{equation}\label{Boltzmannlaw}
\frac{dM}{d\nu}=-\frac{1}{\gamma M^2}
\end{equation}

This equation sates that the time in which the radiation lasts form a black hole is proportional to the triple power of mass M and can be written as
$$\nu\approx\gamma M^3$$
This also satisfies the condition of reference \cite{CR:2014} and $\nu>>M$.
Consider Schwarzschild black hole inverse temperature as $\beta=T^{-1}= 8\pi M$, then the entropy equation becomes 
\begin{equation}
S_{CR}=\frac{3 \sqrt{3} \gamma  M}{45\times 8^3}=\frac{3 \sqrt{3}\gamma A}{(45\times 8^4) \pi }
\end{equation}
Where $ A=16\pi M^2$ is the surface area of Schwarzschild black hole. As the radiation from black hole are in the Planks scale order and also the final evaporation stage has yet not discovered so, this means that the mass loss of black hole is so small that during the emission of radiation, we can take $\frac{dM}{d\nu}\approx M$, which is inconsistent with the requirement for the calculation of the CR volume, because in such case one can not take the increase in volume of black.  The above equation also confirms the statement that the entropy of quantum field in the interior of CR volume is proportional to the area of the black hole horizon and the coefficient of $A$ is much smaller than $\frac{1}{4}$. This means it don't satisfy the first law of black hole satisfied by the Bekenstein Hawking relation. From this entropy relation, we can say that there are more information loss on the black hole horizon. Here we can rise the question that how to fit the above relation in first law of black hole thermodynamics.  some of these problems were discussed in Ref. \cite{Zhang:2015}.

For calculation of entropy in high dimensional black hole, one need to consider the metric Eq. (\ref{HDEDmet}) and the phase space with coordinates labeled as$\lambda, \theta_1,\theta_2,.....\theta_{d-1},p_{\lambda},p_{\theta_1 },p_{\theta_2 },....p_{\theta_{d-1} }$. Following the above analogy the number of quantum modes in a cell of volume ${(2\pi})^d$ is given by 
\begin{equation}
\frac{d\lambda d\theta_1 d\theta_2.....d\theta_{d-1} d p_{\lambda} dp_{\theta_1 } dp_{\theta_2 }....dp_{\theta_{d-1}}}{(2\pi)^d}
\end{equation}
The total number of quantum modes in the phase space over the volume of a black hole with energy less than $E$  are given by 
\begin{equation}\label{HDQstate}
g(E)=\frac{1}{(2\pi)^d}\int{d\lambda d\theta_1 d\theta_2.....d\theta_{d-1} d p_{\lambda} dp_{\theta_1 } dp_{\theta_2 }....dp_{\theta_{d-1}}}
\end{equation}
using the Klein Gordon equation for massless scalar field with same analogy as in \ref{app.A}, we can get the equation of motion as
\begin{equation}
E^2-\frac{1}{2 \dot{\nu } \dot{r}-\dot{\nu }^2 f(r)}p_{\lambda }^2-\frac{1}{r^2}p_{\theta _1}^2-\frac{1}{ r^2 \sin ^2\theta _1}p_{\theta _1}^2-.....\frac{1}{r^2 \left(\sin ^2\theta _1 \text{...}\sin ^2 \theta _{d-2} \right)}p_{\theta _{d-1}}^2=0
\end{equation}
It can write as 
\begin{equation}
p_{\lambda }={\sqrt{\dot{\nu } \dot{r}-\dot{\nu }^2 f(r)}}\sqrt{E^2-\frac{1}{r^2}p_{\theta _1}^2-\frac{1}{ r^2 \sin ^2\theta _1}p_{\theta _1}^2-.....\frac{1}{r^2 \left(\sin ^2\theta _1 \text{...}\sin ^2 \theta _{d-2} \right)}p_{\theta _{d-1}}^2}
\end{equation}
Using this equation in Eq. (\ref{HDQstate}) the total number of quantum modes are calculated as 

\begin{equation}\label{HTDQstate}
g(E)=\frac{E^d}{(2 \pi )^d}\frac{A_{d-1}}{2 d}V_{CR}
\end{equation}

using this Eq. (\ref{HTDQstate}), the free energy can be calculated as 

\begin{equation}
F(\beta )=\frac{1}{\beta }\int{dg(E) log\left(1-e^{-\beta E}\right)}
\end{equation}
Here $\beta$ is the inverse temperature. 
$$\beta=\frac{4\pi}{f'[r]}=\frac{16\pi^2 M}{(d-1)(d-2)A_{d-1}}$$
Using the Riemann Zeta function, the general form of free energy can be written as   
\begin{equation}\label{HDent}
F(\beta )=\frac{A_{d-1} V_{\text{CR}}}{(2 \pi )^d \beta ^{d+1}}(d+1) \zeta  \Gamma (d+1)
\end{equation}
Finally, the entropy can be calculated by 
\begin{equation}\label{entropyformula}
S_{CR}=\beta^2 \frac{\partial F}{\partial \beta}
\end{equation}
gives the entropy of Schwarzschild black hole in same way as in Eq. (\ref{Ent1}). This entropy is same as that in normal volume given in Eq. (\ref{Ent1}). 
\subsection{A Semi-classical approach for Entropy calculation}

Baocheng Zhang \cite{Zhang:2015} followed the statistical way for entropy of scalar field in the interior of Schwarzschild black hole as discussed previous section, in contrast of Baocheng Zhang work a semi-classical approach is followed by Majhi and Samanta in \cite{Majhi:2017tab} for calculation of entropy in the interior of Schwarzschild black hole. There exist several errors due to which they were un able to obtain the correct result as calculated by Baocheng Zhang. In reference \cite{Zhang:2017aqf}, It is claimed by Baocheng Zhang that the two methods of Baocheng Zhang and Majhi are similar, because WKB approximation is consistent with constraints analysis. If so, then why the two results are not consistent with each other? what is the main error between the two methods?  To check this, we analyze the Majhi work and get the following results

\begin{itemize}
\item{The metric equation considered by Majhi and Samanta is either not related to the interior of black hole or it is not static, because if we have to calculate the time $t$ then it depend on $\nu$ and $r$. As both $\nu$ and $t$ represent time component.}
\item{Ignoring spherical coordinates of the metric clearly shows that, we can't calculate the volume from the mentioned metric. Which shows that metric is not related to the interior of black hole.}
\item{The term $\nu$ is taken twice in their metric Eq. (5) in their paper \cite{Majhi:2017tab}, which is also not related to the solution needed.}
\end{itemize}
After making these possible modifications, it could give the same result as that of Baocheng Zhang investigations. In our calculations, we will calculate the interior entropy by using their approach for charged static black hole with possible modifications in order to get a correct result as that of Baocheng Zhang. Consider ingoing Eddington Finkelstein coordinates as in Eq. (\ref{EDmet}) as in chapter 1, the interior volume can be obtained by following the CR method. They used the Lagrangian concept, where the position and momenta can be equally represented. The volume of phase-space is defined as
\begin{equation}\label{hypvol}
V_{\sum}=\int{d \nu dr d P_\nu d P_r}
\end{equation}
To find this integration, the Hamiltonian of the particle is taken as required quantity. For this purpose, the ansatz metric becomes
\begin{equation}\label{ansatzmet}
ds^2=g_{ab}dx^a dx^b=(-f(r) \dot{\nu }^2 +2 \dot{\nu } \dot{r})d\lambda^2
\end{equation}

The metric coefficients are independent of $t$ coordinate so the corresponding conserved quantity is identified as the energy (Hamiltonian). The action of the particle is,
\begin{equation}\label{action}
I=-M\int_{1}^{2} ds_{\text{ansatz}}=-M\int_{1}^{2}( g_{ab}dx^a dx^b)^{\frac{1}{2}}
\end{equation}

This action has symmetry of parametrization. The velocity of the particle is $u^a=\frac{dx^a}{d\tau}$ , where $\tau$ is the arbitrary parameter. Which can define the path of the particle. $x^a=x^a(\tau)$, let $\tau$ be the integration variable, then the action is
\begin{equation}
I=M\int_{1}^{2}Ld\tau=M\int_{1}^{2}( g_{ab}{\frac{dx^a}{d\tau}} {\frac{dx^b}{d\tau}})^{\frac{1}{2}}d\tau
\end{equation}
Where the Lagrangian is $L=M( g_{ab}{\frac{dx^a}{d\tau}} {\frac{dx^b}{d\tau}})^{\frac{1}{2}}$. So, the equation of motion can be obtained by using the Euler-Lagrangian formula,
\begin{equation}
\frac{d}{d\tau }\left(\frac{\partial L}{\partial \dot{x}^{\mu }}\right)-\frac{\partial L}{\partial \dot{x}^{\nu }}=0
\end{equation}
Which can be solved for the geodesic of the desired particle as
\begin{equation}\label{Geodesic}
\dot{u}^{\nu }+\Gamma _{\text{ab}}^{\nu } u^a u^b =0
\end{equation}

In reference \cite{Majhi:2017tab}, it is stated that the Christoffel symbol is symmetric in upper and lower indices, which gives the vanishing value of the Christoffel symbol, so the geodesic equation obtained is true for any space-time. From this geodesic equation, we can write as
\begin{equation}
\frac{dx^0}{d\tau}=0
\end{equation}

Following these relations, the canonical Hamiltonian will also have vanishing value. Here the non-zero Christoffel symbols are
\begin{equation}\label{christoffel}
\Gamma _{\phi \phi }^{\theta }=\frac{1}{2}(\sin ^2)'[\theta ], \qquad \Gamma _{\phi \theta }^{\phi }=\frac{1}{2}\frac{\left(\sin ^2\right)'\left [\theta  \right ]}{\sin ^2\theta}
\end{equation}

Form which we can write as $\Gamma_{r\theta}^{0}=0$ and hence from Eq. (\ref{Geodesic}) we have $\dot{u_0}=0$. So, we can't identify the Hamiltonian for representation of dynamical system. Here the canonical momenta are
\begin{equation}
P_a=\frac{\partial L}{\partial \dot{x}^a}=\frac{\partial }{\partial \dot{x}^a}\left ( M g_{ab} {\frac{dx^a}{d\tau}}{\frac{dx^b}{d\tau}}\right )=\frac{M^2}{L}g_{ab}{\frac{dx^b}{d\tau}}
\end{equation}
Which yields vanishing canonical Hamiltonian as $H_c=P_a \frac{{dx}^a}{d\tau}-L=0$ by using the suitable values of $P_a$. This vanishing Hamiltonian is the guaranteed of re-parametrization of invariant theory, which is also occurs in Minkowski space-time. The four conserved momenta are given by
\begin{equation}
p^2= g_{ab} P^a P^b=M^2
\end{equation}

We need Hamiltonian for above metric in Eq. (\ref{ansatzmet}). This metric is not same as that of Majhi.et.al. as they used constraint analysis to evaluate the Hamiltonian as energy and also their metric were found inappropriate. So, we will follow their method by using the charged black hole by using these the modification. The two types of constraints used are is "On-shell constraint" which guarantees the particle as of virtual character given below:
\begin{equation}\label{const}
\psi =P^2-M^2\approx 0 \qquad\phi={\frac{\dot{P}^0} \tau }{M}-x^0
\end{equation}
The total Hamiltonian is proportional to primary constraints.
\begin{equation}\label{H1}
H_T=\psi  \zeta (\tau )=\zeta (\tau ) \left(P^2-M^2\right)
\end{equation}
Where $\zeta (\tau )$ is the proportionality constant. Generally, the Hamiltonian equation can be calculated as 
$$\frac{\partial H}{\partial p}=\dot{q} \qquad \frac{\partial H}{\partial q}=-\dot{p}$$
So, in case of the above metric and its Hamiltonian, The Hamiltonian equations for quantities $x$ and $p$  can be written as (See also Appendix \ref{app.B})
\begin{equation}\label{dot{x1}}
\dot{x}^a=\frac{dx^a}{d\tau}=2 \zeta {\{x^a, P_a}\} P^a=2 \zeta P^a
\end{equation}
Similarly
\begin{equation}
\dot{P}^a=-\zeta \frac{\partial g^{\text{bc}}}{\partial x^a}P_b P_c
\end{equation}
\begin{equation}\label{dot{x2}}
\dot{x}^a=2 \zeta P^a=2 \zeta\frac{M^2}{L}g^{ab}\frac{{dx}^b}{d\tau }=2 \zeta\frac{M^2}{L}g^{ab}u_b=2 \zeta\frac{M^2}{L}g^{ab}u_b=2 \zeta\frac{M^2}{L}u^a
\end{equation}
For Eq. (\ref{dot{x1}}) and Eq. (\ref{dot{x2}}), we gets
\begin{equation}\label{zeta1}
\zeta =\frac{L}{2 M^2}
\end{equation}
So the total Hamiltonian form Eq. (\ref{H1}) can be written as 
\begin{equation}\label{H2}
H_T=\frac{L}{2 M^2} \left(P^2-M^2\right)
\end{equation}
Here $\frac{L}{2 M^2}$ is the proportionality constant. The system is said to be first class with gauge freedom, which can be removed by imposition of gauge fixing condition. Let us consider the secondary constraint as 
\begin{equation}\label{const1}
\phi _2=\frac{\dot{P}^0}{M}\tau -x^0
\end{equation}
And the primary constraint can be written as
\begin{equation}\label{const2}
\phi_1=P^2-M^2\approx 0
\end{equation}
This makes the system as a second class of constraint. Consider
\begin{equation}
\dot{\phi }_2=\frac{\partial }{\partial \tau }\left ( \frac{P_0}{M}\tau-x^0 \right )=\frac{P^0}{M}+\frac{\dot{P}_0}{M}\tau-\dot{x}^0
\end{equation}
or we can write as 
\begin{equation}\label{const3}
\dot{\phi }_2=\frac{P^0}{M}+\frac{\tau}{M}\left\{p^0,H_T\right\}-\left\{x^0,H_T\right\}
\end{equation}
Use the values of $\dot{p}^0=\left\{p^0,H_T\right\}$ and $\dot{x}^0=\left\{x^0,H_T\right\}$, See aliso Appendix \ref{app.C}.
From the last two terms, we can calculate 
\begin{equation}\label{EOM1}
 \dot{x}^0=\left\{x^0,H_T\right\}=2\zeta P^0, \qquad \zeta \left(2 P^a\frac{\partial g^{0 b}}{\partial x^a}P_b-g^{0 b}\frac{\partial g^{bc}}{\partial x^a}P_aP_b \right )
\end{equation}
This equation vanishes because the terms $g^{0a}$ and $g^{0b}$ are zero for our metric. Using Eq. (\ref{EOM1}) in Eq. (\ref{const3}), we gets
\begin{equation}\label{zeta2}
\zeta =\frac{1}{2 M}
\end{equation}
Comparing the two $\zeta$ Eq. (\ref{zeta1}) and (\ref{zeta2}), we gets $L=M$ or we can write as
\begin{equation}
L=\sqrt{ g_{ab} \frac{dx^a}{d\tau} \frac{dx^b}{d\tau}}
\end{equation}
Which shows the work with proper gauge and $\tau$ is the proper time. Now the equation of motion obtained are as:
\begin{center}
\begin{multline}\label{EOM2}
\dot{x}^a=\left\{x^a,H_T\right\}=\zeta\left\{x^a,P^2\right\}=\frac{1}{2M}\left\{x^a,P^2\right\}=\frac{P^a}{M}  
\\ 
\dot{P}^a=\left\{P^a,H_T\right\}=\zeta  \left\{P^a,P^2\right\}=\frac{1}{2 M}\left ( 2\frac{\partial g^{\text{ab}}}{\partial x^c}g^{\text{cd}}-\frac{\partial g^{\text{db}}}{\partial x^c}g^{\text{ca}} \right )P_b P_d
\end{multline}
\end{center}
using the identity $\partial_c g^{ab}=-g^{ad} g^{be} {\partial_c g_{de}}$ in above equation, We can also define the Christoffel symbol as
\begin{equation}\label{dotPinchrst}
\dot{P}^a=-\frac{1}{m}\Gamma ^a _{\text{bc}}P^b P^c 
\end{equation}
Finally, using the $\dot{x}^a$ from Eq. (\ref{EOM2}) in Eq. (\ref{dotPinchrst}), we can recall the geodesic Eq. (\ref{Geodesic}). Now to find the effective Hamiltonian, the constraint metric can be calculated by defining the two constraints as 
$$\phi _1=P^2-m^2 \qquad \phi _2=\frac{P^o}{m}\tau -x^o$$
generally the constraint metric can be written as 
\begin{equation}\label{constmetgenrleq}
C_{AB}=\left\{\phi _A,\phi _B\right\}, \qquad (A,B)=(1,2)
\end{equation}
Solving this equation for $(A,B)=(1,2)$  (for detail See Appendix \ref{app.D}), from the above, we get the constraint matrix as,
\begin{equation}\label{constmet}
C_{AB}=\left\{\phi _A,\phi _B\right\}=2 P^o
\begin{pmatrix}
0 & 1\\ 
-1 & 0
\end{pmatrix}
\end{equation}
and the inverse constraint metric is
\begin{equation}\label{insconstmet}
(C_{AB}) ^{-1}=\frac{1}{2 P^o}
\begin{pmatrix}
0 & -1\\ 
1 & 0
\end{pmatrix}
\end{equation}
Knowing the inverse constraints, we can calculate Dirac bracket between two dynamical variables of the Hamiltonian using the equation as
\begin{equation}
H^*=H-\Sigma_{c_j} \phi _j
\end{equation}
Using values in this equation, we can write as 
$$\{f,g\}^*=\{f,g\}-{\Sigma  \left\{f,\phi _A\right\}{C_{AB}^{-1}} \left\{\phi _B,g\right\}}$$
or 
$$\{f,g\}^*=\{f,g\}+{ \left\{f,\phi _A\right\}{\begin{pmatrix}
0 & -1\\ 
1 & 0
\end{pmatrix}} \left\{\phi _B,g\right\}}$$
If $\phi_1=P^2$ and $\phi_B=\phi_2$ the we can write as 
$$\{f,g\}^*=\{f,g\}+{ \left\{f,P^2\right\}\frac{1}{2P^o} {\begin{pmatrix}
0 & -1\\ 
1 & 0
\end{pmatrix}} \left\{\phi _2,g\right\}}$$
$$=\{f,g\}-\frac{1}{2 p^0}\left ({\left\{f,P^2\right\} \left\{\phi _2,g\right\}-\left\{f,\phi _2\right\} \left\{P^2,g\right\}}  \right )$$
From this The equation of the dynamical variable can be find as
\begin{equation}\label{fhog}
\dot{f}=\{f,H\}^*
\end{equation}

This equation gives the effective Hamiltonian as non-vanishing quantity. Notice that $H$ must be selected in such a way that the equation of motion (\ref{EOM2}) along with (\ref{dotPinchrst}) can be generated from Eq. (\ref{fhog}). In fact the considering $H=P^0$, Now we can calculate the Hamiltonian equation as 
$$\dot{x^a}=\left\{x^a,P^o\right\}^*=g^{\text{ao}}+\frac{1}{2 P^o}\left [ \left \{ x^a, P^2 \right \}\left \{ \phi_2, P_0 \right \} -\left \{ x^a, \phi_2 \right \}\left \{ P^2,P^0 \right \}\right ]$$
To do so, solving all parts of this equation we get,
\begin{equation}
\dot{x^a}=\left\{x^a,P^o\right\}^*=g^{ao}+\frac{1}{2 P^o}\left [ 2 P^a+\frac{2}{m}\Gamma _{bc} ^oP^b P^c  \right ]
\end{equation}
and
$$\dot{P}^a=\left\{P^a,P^o\right\}^*=\left\{P^a,P^o\right\}+\frac{1}{2 P^o}\left [ \left\{P^a,P^2\right\} \left\{\phi _2,P^o\right\}-\left\{P^a,\phi _2\right\}\left\{P^2,P^o\right\} \right ]$$
solving all of its parts, it gives
\begin{equation}
\dot{P}^a=-\frac{1}{P^o}\left [ P^b P^c \Gamma _{\text{bc}}^o+P^b P^c g^{\text{oa}} \Gamma _{\text{bc}}^a \right ]
\end{equation}
Where $a=0$ can be eliminated by gauge fixing and the space components for $a=\mu$ can give the required equation of motion.
\begin{equation}
x^{\mu }=\frac{P^{\mu }}{P^o}, \qquad P^{\mu }=\frac{1}{P^o}\Gamma ^{\mu}_{ab}P^a P^b
\end{equation}
$P^o$ can be eliminated to get the geodesic equation. for $P^o=-P_o$
\begin{equation}
g^{ab} P_a P_b=-(P^o)^2+2r^{-4}P_{\nu } P_r+r^{-4}f(r)P_r^2=M^2
\end{equation}
$$-(P^o)^2+h^{ij}P_iP_j=M^2$$
For massless particles $M=0$, we get
\begin{equation}
E^2=P_o ^2=h^{ij}P_iP_j
\end{equation}
On expansion, it can be written as 
\begin{equation}
E^2-\frac{p_{\lambda }^2}{-f(r) \dot{\nu }^2 +2 \dot{\nu } \dot{r}}-\frac{p_{\theta }^2}{r^2}-\frac{p_{\phi }}{\theta  r^2 \sin ^2}=0
\end{equation}
which is same as Eq. (\ref{eqofenergy}). Where we used
$$\partial_{\lambda} I=p_{\lambda }, \qquad \partial_{\theta} I=p_{\theta }, \qquad \partial_{\phi} I=p_{\phi }$$

Next, following statistical way to calculate the entropy in the interior of d-dimensional charged black hole. The result found is exactly same as that of Baocheng Zhang result. Solving for $p_\lambda$ the total number of quantum states are 
We can calculate the free energy as in Eq. (\ref{BfreeE})
$$F(\beta )=\frac{1}{\beta }\int {dg(E)} ln(1-e^{-\beta (E)})$$
The entropy in the interior of black hole
$$S_{CR}=\frac{\pi^2 V_{CR}}{45\beta^3}$$
If we use Eq. (\ref{RNBH}) we will get the same result as the above. We will try to get the same result for charged static black hole. Let, we use value of $V_{CR}$ from Eq. (\ref{RNBHvol}) as previously discussed can be easily calculated. Using the Baocheng Zhang analogy the statistical entropy  for scalar modes in the interior of charged black hole is also discussed in the next section  below.
\subsection{Entropy in the interior of Reissner Nordström black hole}

The problem of black hole information paradox is very interesting in studying the black hole. Until now, there are many in investigations made in this regard, which shows that there is a large space in the interior of black hole for restoring the lost information as discussed earlier. Following the CR work, the interior volume of scalar field in a charged static black hole is calculated in Eq. (\ref{RNBHvol}) as discussed in \cite{Ong:2015, Han:2018}. In order to calculate the entropy, following the  Baocheng Zhang analogy the total number of quantum states in scalar mode of static charged black hole were calculated as given in Eq. (\ref{Qstate}), where $V_{CR}$ is given in Eq. (\ref{RNBHvol}). The free energy is evolved in Eq. (\ref{FreeEnergy}). Where $\beta$ is the inverse temperature of black hole. Further the entropy is
\begin{equation}\label{RNbeta}
\beta=T^{-1}=\frac{2\pi\left(M+\sqrt{ M^2- Q^2}\right)^2}{\sqrt{ M^2- Q^2}}
\end{equation}
Using the values of interior volume and $\beta$ from Eq. (\ref{RNBHvol}) and (\ref{RNbeta}) , the entropy in the interior of a charged Reissnor Nordstr$\ddot{o}$m black hole from the general formula (\ref{entropyformula}) can be calculated as
\begin{equation}
S_{CR}={\frac{\pi^3\sqrt{27 M^4-8 \text{MQ}^2 \sqrt{9 M^2-8 Q^2}-36 M^2 Q^2+9 M^3 \sqrt{9 M^2-8 Q^2}+8 Q^4}  }{45\sqrt{2}\beta^3}}\nu
\end{equation}
This entropy is a function of mass and charge like the interior volume of a charged black hole. This entropy has the special character due to interior volume of black hole, which has  linear relation with advance time. Using this relationship, we can further prob the black hole interior information by using the notion of radiation.

According to the references \cite{Hawking:1974sw, Page:1976ki}, black hole emits radiations due to Hawking effect. It is claimed in \cite{Wald:1984rga}, the emission of Hawking radiation is analogous to the radiation form a black body. If so, then the rate of mass loss by the black hole is given by Stefan Boltzmann law as: 
\begin{equation}\label{BLTZMNlaw}
\frac{dM}{d\nu}=-\sigma A T^4
\end{equation}
Where
$$A=4\pi r_+ ^2=4\pi \left(M+\sqrt{ M^2- Q^2}\right)^2$$
is the black hole area. Using the values of T and A in Eq. (\ref{BLTZMNlaw}), we get
\begin{equation}
d\nu =\frac{\left(\sqrt{M^2-Q^2}+M\right)^6}{\left(M^2-Q^2\right)^2}dM
\end{equation}

Using these values, we can get the entropy of scalar modes in the interior of black hole in differential form as.
\begin{equation}\label{intRNSRC}
dS_{CR}=-{\frac{\gamma\sqrt{27 M^4-8 \text{MQ}^2 \sqrt{9 M^2-8 Q^2}-36 M^2 Q^2+9 M^3 \sqrt{9 M^2-8 Q^2}+8 Q^4}  }{360\sqrt{2}\sqrt{M^2-Q^2}}}dM
\end{equation}
This differential entropy is independent of advance time due to similarity between black hole radiation and black body radiation. Now a bound can be made between the interior entropy scalar modes and the Bekenstein entropy termed as "the proportional relation between the entropy of the entropy of scalar field and the horizon entropy". According to the Bekenstein Hawking definition of entropy
\begin{equation}
S_{BH}=\frac{A}{4}=\pi (M+\sqrt{ M^2- Q^2})^2
\end{equation}
Whose differential form is given by 
\begin{equation}\label{RNBHent1}
dS_{BH}=\frac{2\pi(M+\sqrt{ M^2- Q^2})^2}{\sqrt{ M^2- Q^2}}dM
\end{equation}
Using the Eq. (\ref{intRNSRC}) and (\ref{RNBHent1}), the proportional relation between the two type of entropy is calculated as 
\begin{equation}\label{Entproprel}
dS_{CR}=-{\frac{\gamma\sqrt{27 M^4-8 \text{MQ}^2 \sqrt{9 M^2-8 Q^2}-36 M^2 Q^2+9 M^3 \sqrt{9 M^2-8 Q^2}+8 Q^4}  }{720\sqrt{2}\pi\left (M+\sqrt{M^2-Q^2}  \right )^2}}dS_{BH}
\end{equation}
Let us simply it can be written as
\begin{equation}\label{Entpropftn}
dS_{CR}=-F(M,Q)dS_{BH}
\end{equation}
where 
$$F(M,Q)={\frac{\sqrt{27 M^4-8 \text{MQ}^2 \sqrt{9 M^2-8 Q^2}-36 M^2 Q^2+9 M^3 \sqrt{9 M^2-8 Q^2}+8 Q^4}  }{720\sqrt{2}\pi\left (M+\sqrt{M^2-Q^2}  \right )^2}}$$

\begin{footnotesize}
$$dS_{CR}=-{\frac{\gamma\sqrt{27 M^4-8 \text{MQ}^2 \sqrt{9 M^2-8 Q^2}-36 M^2 Q^2+9 M^3 \sqrt{9 M^2-8 Q^2}+8 Q^4}  }{360\sqrt{2}\sqrt{M^2-Q^2}}}dM$$
\end{footnotesize}

\begin{figure}
\begin{center}
\includegraphics[width=0.5\textwidth]{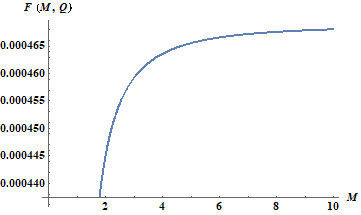}
\caption{Plot between $F(M,Q)$ and M for $( Q=0.5)$, which shows the proportional relation of the two entropies during the evaporation}
\label{image-2.5}
\end{center}
\end{figure}

The above equation is direct confirmation the Baocheng Zhang’s result given in \cite{Zhang:2015} at $Q=0$. The above result is also the validation of the area law in quantum field theory and also suggesting a link between the between the black hole interior and horizon entropy. This equation shows that the entropy of scalar field in the interior of black hole is proportional to the Bekenstein Hawking entropy, where the proportionality constant is $F(M,Q)$.  The plot shown below, show the relation of the two types entropy from the early stage of black hole to the final stage of evaporation, where the mass become the size of plank mass and quantum mechanical effects stops. This figure shows that in the initial stages the evaporation is so slow that the mass remains constant, in the central part the evaporation proportional to mass decrease. When enough mass is evaporated form black hole then at Plank level evaporation stops. As shown by the final straight part of the Fig. (\ref{image-2.5}) above.
\subsection{Entropy in the interior of Kerr Black hole}

Following the Baocheng Zhang scenarios, in reference \cite{Wang:2018dvo}, the entropy of static black holes (either charged or uncharged) are discussed in the above sections. It is also verified that the entropy in the interior of massless scalar modes always increases with advance time $\nu$. So, this character will affect the interior volume of black hole. To calculate the interior entropy, it is necessary to evaluate the number of quantum states in the interior of black hole. Considering the metric for a maximal hyper-surface (with $r=r_\nu$), the volume is determined in Eq. (\ref{Kerrvol}). Next, considering the massless scalar field in the interior of a Kerr black hole and solving the equation of motion, one obtains Eq. (\ref{eqofenergy}) in general form as 
\begin{equation}
-E^2+h^{ab}P_a P_b=0
\end{equation}

Where $h^{ab}$ is the auxiliary metric on the maximal hyper-surface. So, the total number of quantum sates contained in the interior of scalar field are given in Eq. (\ref{Scalarfield}). Where $V_{CR}$ is given in Eq. (\ref{KerrBHvol}). The free energy calculated is given in Eq. (\ref{FreeEnergy}) and finally the entropy in Eq. (\ref{Ent1}). Next, using two conditions: the black hole emission rate as quasi-static process (it needs the emission rate must satisfy the condition $\nu>>M$ and the radiation as black body radiation as black body radiations. The former condition guarantees the horizon temperature as Hawking temperature given by,
\begin{equation}\label{insvtemp}
\beta =\frac{1}{T}=\frac{4\pi \left(r_{\pm }+a^2\right)}{r_+-r_-}
\end{equation}

Where $r_\pm$ is given in Eq. (\ref{Kerrrad}), the first condition guarantees that the temperature of the scalar field and the horizon will be in equilibrium, if the process is considered at an infinitesimal level. So, if the temperature of the scalar field inside the black hole is considered equal to the temperature of the horizon, then the rate of mass loss can be seen by the Stefan Boltzmann law in Eq. (\ref{Boltzmannlaw}). Using the values of $\beta$ and $A$, we can get Eq. (\ref{Boltzmannlaw}) as
\begin{equation}\label{BoltzmannlawKerr}
\frac{dM}{d\nu }=\frac{\left(r_+-r_-\right){}^4}{32 \gamma \pi ^3 M^3 r_+^3}=\frac{\left(M^2-a^2\right)^2}{32 \gamma \pi ^3 M^3 \left(\sqrt{M^2-a^2}+M\right)^3}
\end{equation}
Where $\gamma$ is a positive constant, it depends on quantum modes coupling with gravity. Its value doesn't affect our discussion. As the volume of black hole changes with advance time so, using the above equation, we can introduce the volume in terms of differential mass form Eq. (\ref{Kerrvol}) as
\begin{equation}\label{KerrvolBoltzmannlaw}
dV_{CR}=2 \pi M^2   F\left(\frac{r}{M},\frac{a}{M}\right)d\nu
\end{equation}
In addition of Eq. (\ref{BoltzmannlawKerr}), Eq. (\ref{KerrvolBoltzmannlaw}) gives
\begin{equation}
\dot{V}_{CR}=-64 \gamma \pi ^4 M^4 F_{\max }\left(\frac{r_{\nu }}{M},\frac{a}{M}\right)\frac{\left(\sqrt{M^2-a^2}+M\right)^3}{\left(M^2-a^2\right)^2}\dot{M}
\end{equation}
Finally, the differential entropy is
\begin{equation}
\dot{S}_{\text{CR}}=-\frac{\pi ^2}{45 \beta }\dot{V}_{CR}=-\frac{8 \pi^3  \gamma  M^2}{45} F_{\max }\left(\frac{r_{\nu }}{M},\frac{a}{M}\right)\frac{\left(\sqrt{M^2-a^2}+M\right)^3}{\sqrt{M^2-a^2} \left(\sqrt{M^2-a^2}+a^2+M\right)^3} \dot{M} 
\end{equation}

Now to calculate the proportional relation between the horizon entropy and the entropy of massless scalar field, we calculate the variation of Bekenstein Hawking entropy \cite{Bekenstein:1972tm, Bekenstein:1973ur, Bardeen:1973gs} defined as
$$S_{BH}=\frac{A}{4}=\pi(r^2 _+ -a^2)$$
Where $A=4\pi (r_+^2+a^2 )$ the area of the event horizon for Kerr black hole. The differential form is 
\begin{equation}\label{DRNBHent}
\dot{S}_{BH}=\frac{2\pi(1+\sqrt{ M^2- a^2})^2}{\sqrt{ M^2- a^2}}M\dot{M}
\end{equation}
Here $\dot{M}<0$. The proportional relation can be made between the two types of entropy can be written as 
\begin{equation}
\dot{S}=-\frac{\pi ^2}{180}\gamma f\left(\frac{r_{\nu }}{M},\frac{a}{M}\right)\dot{S}_{BH}
\end{equation}
Where
$$f\left(\frac{r_{\nu }}{M},\frac{a}{M}\right)=f_{\max }\left(\frac{a}{M}\right)=F_{\max }\left(\frac{r_{\nu }}{M},\frac{a}{M}\right)\left (1-\sqrt{1-\left(\frac{a}{M}\right)^2}  \right )\left ( \frac{a}{M} \right )^{-2}$$
The plot of this function $f_{max} (\frac{a}{M})$ vs $(\frac{a}{M})$ is given above.
\begin{figure}
\begin{center}
\includegraphics[width=0.7\textwidth]{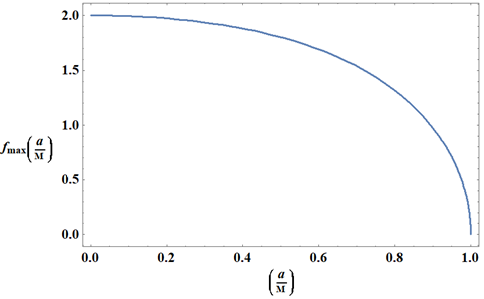}
\caption{Plot of $f_{max} (\frac{a}{M})$ vs $(\frac{a}{M})$}
\label{image-2.6}
\end{center}
\end{figure}

This figure shows the evaporation process form initial mass to the final stage. At the initial stages, it is approximately constant because of slow evaporation and low temperature, but as the time passes the process no longer remain constant. It increases up to its final stage and at the final stage no information can be retrieved from black hole having mass of planks order. At this level, one need to consider uncertainty instead of considering the proportional relation.
\section{Discussions and conclusions}

Classically, the volume enclosed by the black hole horizon is thought to be a challenge due to the space-time in the interior of black hole infinity coupling and also the existence of event horizon. An event horizon can be considered as the boundary between the two space-times. Parikh firstly proposed that interior volume of black hole is independent form the choice of space-time slicing \cite{Parikh:2005qs}. Recently, M. Christodoulou and C. Rovelli \cite{CR:2014, Christodoulou:2016tuua}, consider a spherically symmetric Schwarzschild black hole in Eddington Finkelstein coordinates and define the interior volume as "the volume of a space-like hyper-surface in the interior of Schwarzschild black hole increases linearly with advance time $\nu$". They also extend the discussion to static charged black hole. Form their analysis, maximum contribution in black hole volume is due to stretch central part of the largest hyper-surface with constant radius $r_\nu$. This mean that this part of the hyper-surface has bound area and unbound volume. For other black hole see \cite{Wang:2018dvo, Majhi:2017tab, Christodoulou:2016tuua}. Following the CR work, Baocheng Zhang \cite{Zhang:2015, Zhang:2017aqf} investigate the entropy of scalar modes in the interior of Schwarzschild black hole and found, the interior entropy  is also found to be proportional to $\nu$. The proportionality constant is found to be less than factor of $\frac{1}{4}$, which shows the horizon bear maximum entropy as compared to the interior entropy of the scalar mode.

Further, considering the emission of Hawking radiation form black hole, a proportional relation is calculated between entropy's of black hole interior quantum modes and the Bekenstein Hawking entropy. It is found that the entropy's of the quantum field in the black hole is always proportional to the Bekenstein Hawking entropy during the radiation emission process. This shows, that in the process of Hawking radiation and the horizon of the black hole is closely related to its interior. Considering the above discussion, one can make a more general idea for the loss of information inside the black hole. A more reliable statement was given by Page \cite{Page:1976ki}, accordingly there is an upper bound between the black hole horizon and Bekenstein Hawking entropy. The relation between the entropy of interior scalar modes and horizon entropy of a black hole shows, that the information stored in black hole are directly related to quantum modes and are not loss by black hole during the evaporation process. These analyses of Baocheng Zhang showed that there is a large room in the interior of black hole for storing the information, because the interior volume and entropy are directly related to advance time. Note that these analysis are only applicable for black hole with $M>M_{pl}$. If $M\leq M_{pl}$, then it is needed to consider the uncertainty relation.

Finally, we discussed the black hole evaporation under Hawking radiations by considering two assumption tocalculate the proportional relation. First it is claimed that the emission process is slow enough, so that one can consider the temperature in the interior as Hawking temperature.  Secondly, the emission of Hawking radiation is analogous to the radiation form a black body. So, the rate of mass loss by the black hole is given by Stefan Boltzmann law. Using Stefan Boltzmann law, one can calculate the differential form of entropy. The advantage of this differential form is that one can consider the black hole temperature as varying quantity. Using the differential form of Bekenstein Hawking entropy, we can get the proportional relation, which explains the process of evaporation. See Appendix (\ref{app.A})
\newpage

\newpage
\part{}
\clearpage
\thispagestyle{empty}
\hfill
\clearpage
\chapter{Entropy in a d-dimensional charged black hole}

\section{Introduction}
In general relativity, it is difficult to define the interior volume of a black hole, because it depends on how to select the reasonable space-like hyper-surfaces inside a black hole due to existance of event horizon and the interchange of space-time coordinates. At first time Parikh \cite{Parikh:2005qs} proposed that a reasonable definition of the volume inside a black hole should be slicing-invariant, which has been discussed by many other peoples, for detail see \cite{Grumiller:2005zk, Ballik:2010rx, Ballik:2013uia, DiNunno:2008id, Finch:2012fq, Cvetic:2010jb, Gibbons:2012ac}. Recently, Christodoulou $et. \, al$ \cite{CR:2014} proposed a special method to define the interior volume of a Schwarzschild black hole, which is analogues to the method of calculating geodesic equation for a particles in space-time. Using this method, they demonstrate that the maximal hyper-surface at $r=\frac{3}{2}$, which corresponds the interior volume of a Schwarzschild black hole. Meanwhile, they also showed that the interior volume of a Schwarzschild black hole grows linearly with advance time.  The interior volume of hyper-surface was calculated as in equation below,
\begin{equation}
V_\Sigma=3\sqrt{3} \pi m^2 v.
\end{equation}

Moreover, in CR paper the interior volume of a charged black hole was also calculated by using the same method. These results showed, that the interior volume corresponds to the maximization of hyper-surface at constant $r$.

It is supposed that, the special property of the interior volume could be  a candidate to resolve the problem of information paradox. Since the volume increases with advanced time so, a black hole could have a large interior volume containing many modes of quantum fields and the entropy of these modes may relate to the Bekenstein-Hawking entropy. It means that the loss of information on the horizon due to Hawking radiation may be stored in interior of black hole in the form of quantum states. Therefore, it is necessary to investigate the black hole interior entropy in the form of these quantum modes. For a Schwarzschild black hole, the interior entropy of the scalar field has been investigated in Ref. \cite{Zhang:2015, Yang:2018arj}. The results shows that considering Hawking radiation, the entropy is proportional to the Bekenstein-Hawking entropy.

In this paper, we consider the scalar field inside a $d$-dimensional charged black hole and calculate its entropy. Using our previous method discussed in \cite{Wang:2018dvo}, a reasonable result can be obtained by considering Hawking radiation. The proportional relation can be found between the entropy of the scalar field and Bekenstein-Hawking entropy. Finally, we investigate the relation between the dimensions and proportionality coefficients for the two types of entropy.

The organization of this chapter is as follows. In the next section, we calculate the interior volume of the $d$-dimensional charged black hole. In section 3, we calculate the entropy of the scalar field using the standard statistical method and discuss the relation between the entropy and the Bekenstein-Hawking entropy. In section 4, we use the Hamiltonian analysis to calculate the entropy of the scalar field, and compare the results to that in section 3. Finally, discussion and conclusions are given in section 5.

\section{The Interior Volume of a d-dimensional charged Black Hole}
To calculate the volume of  $d$-dimensional charged black hole, first we need to review the definition of interior volume inside a spherically symmetric black hole \cite{CR:2014}. In flat space-time, the maximal volume can be defined as a special space-like hyper-surface $\Sigma$, enclosed by the two-dimensional sphere $S$. But, in curved space-time, we can't use the same case as in flat space-time but a special space-like hyper-surface must satisfy the following two equivalent conditions:

(a) $\Sigma$ lies on the same simultaneity surface as $S$; 

(b) $\Sigma$ is the largest spherically symmetric surface bounded by $S$. 

\noindent
From the above statements, condition (a) can't be extended to the curved space-time, because in general, the simultaneity surface has no special significance and an event in general relativity can't be defined in two simultaneity surfaces at same time. In contrast, condition (b) can be extended to the curved case immediately. The Ref. \cite{CR:2014} has shown that the maximal hyper-surface is obtained at $r=\frac{3}{2}m$, which corresponds to the interior volume of a Schwarzschild black hole when $v \gg m$. 

Next, we will calculate the interior volume of a $d$-dimensional charged black hole. The metric of a $d$-dimensional charged black hole in the Eddington-Finkelstein coordinates is \cite{Gunasekaran:2012dq}
\begin{equation}\begin{split}
ds^2 = -f(r) dv^2 + 2 dv dr +r^2 d \Omega_{d-2}^2,
\end{split}
\end{equation}
where
\begin{equation}
\begin{split}
&f(r) =1-\frac{m}{r^{d-3}}+ \frac{q^2}{r^{2(d-3)}},\\
&m = \frac{16 \pi M}{(d-2) \Omega_{d-2}}, \\
&q = \frac{8 \pi Q}{\Omega_{d-2} \sqrt{2(d-2)(d-3)}}, \\
&\Omega_{d-2} =\frac{2 \pi^{\frac{d-1}{2}}}{\Gamma(\frac{d-1}{2})}.
\end{split}
\end{equation}
A $(d-1)$-dimensional spherically symmetry hyper-surface $\Sigma$ can be regarded as the direct product of a $(d-2)$-sphere and a curve $\gamma$ in the $v-r$ plane
\begin{equation}
\Sigma \equiv \gamma \times S^{d-2},
\end{equation}

\begin{equation}
\gamma \mapsto (v(\lambda), r(\lambda)).
\end{equation}
The curve $\gamma$ is parametrized in $\nu$ and $r$ by $\lambda$. Therefore, the hyper-surface $\Sigma$ is coordinated by $\lambda, \theta, \phi$. The line element of the induce metric on $\Sigma$ is 
\begin{equation}
ds_{\Sigma}^2= (-f(r) \dot{v}^2 + 2 \dot{v} \dot{r}) d \lambda^2 + r^2 d \Omega_{d-2}^2
\end{equation}
where the dot $(\textbf{.})$  indicates differentiation by $\lambda$ and $-f(r) \dot{v}^2 + 2 \dot{v} \dot{r} > 0$ to satisfy the space-like condition of the hyper-surface.

Finally, using the condition of $v \gg m$ and $\dot{r}=0$, the proper volume of this hyper-surface has been obtained as 
\begin{equation}\label{volume1}
V_{\Sigma}= \Omega_{d-2} \sqrt{-f(r) r^{2(d-2)}} v,
\end{equation}
where
\begin{equation}
r=(4 \pi )^{\frac{1}{d-3}} \left[\frac{(d-2)  \sqrt{\frac{(d-1)^2 M^2}{(d-2)^2}-\frac{2 Q^2}{d-3}}+(d-1) M}{(d-2)^2 \Omega_{d-2} }\right]^{\frac{1}{d-3}}.
\end{equation}

This is proper volume of the hyper-surface at constant $r$, which corresponds to the interior volume of a $d$-dimensional charged black hole.

\section{Entropy in the Volume of a d-dimensional charged Black Hole}

According to above statements, the interior volume of a $d$-dimensional charged black hole grows linearly with advanced time $v$. This special character of the interior volume may influence the statistical quantities of the quantum field inside the black hole and it may propose a solution to the problem information paradox. Hence, it is significant to investigate the statistical properties of quantum fields inside the black hole. Here we only involve the massless scalar field inside the black hole. 

According to our previous paper \cite{Wang:2018dvo}, for any point $p$ on the maximal hyper-surface at constant $r$, there must exist a Gaussian normal coordinate system $\{T, x_p^i \}$, $i=1,2,3$ defined by a family of geodesics, where $T$ is the affine parameter of the geodesic and $x_p^i$ denotes the spatial coordinate of point $p$. Thus, the line element can be expressed as
\begin{equation}\label{stmh}
ds^2 = -dT^2 + h_{ij} dx^i dx^j,
\end{equation}
where $h_{ij}$ is the induced metric on the hyper-surface.

This line element is equivalent to $ds^2 = -dT^2 + h_{ij}(T) dx^i dx^j$, which means that the hyper-surface in the black hole is dynamical for the defined time $T$. In our previous work, we have shown that the standard statistical method can be used to investigate the statistical properties of quantum fields on the hyper-surface, even if the hyper-surface is dynamical.

Next, we will use a common method in curved space-time to discuss the motion of mass less scalar field in the interior of black hole. The equation of motion for the massless scalar field is given by
\begin{equation}\label{os}
P^\mu P_\mu = g^{\mu \nu} P_\mu P_\nu = g^{00}E^2 + h^{ij} P_i P_j=-E^2 +  h^{ij} P_i P_j=0,
\end{equation}
Where $g^{\mu \nu}$ is the inverse metric of space-time on the hyper-surface and $h^{ij}$ is the inverse induced metric on the hyper-surface. Form both the equation of motion and the space-time metric, we can obtain 
\begin{equation}
P_{\lambda} = \left(-f(r) \dot{v}^2 + 2 \dot{v} \dot{r} \right)^{\frac{1}{2}} \left(E^2 - \frac{1}{r^2} P_{\phi_1}^2 - \cdots - \frac{1}{r^2 sin\phi_1 \cdots sin\phi_{d-3}}P_{\phi_{d-2}}^2 \right)^{\frac{1}{2}}.
\end{equation}
Thus integrating the phase cell in the phase space, the number of quantum states with energy less than $E$ can be obtained as
\begin{equation}
\begin{aligned}\label{g(E)}
g(E)&=\frac{1}{(2 \pi)^3} \int d\lambda d\phi_1 \cdots d\phi_{d-2} dP_{\phi_1} \cdots dP_{\phi_{d-2}}\\
&=\frac{E^{d-1} \Omega_{d-2}}{2 (2 \pi)^{d-1} (d-1)} V_\Sigma,
\end{aligned}
\end{equation}
From Eq. (\ref{g(E)}), we can see that the number of quantum states increases gradually with advanced time due to its direct relation with the interior volume of black hole. 

Hence, we continue to calculate the free energy of scalar field at some inverse temperature $\beta$ as
\begin{equation}
\begin{aligned}
 F(\beta)&=\frac{1}{\beta} \int dg(E) \ln(1 - e^{-\beta E})\\
&=- \frac{\Omega_{d-2} \zeta(d) \Gamma(d) }{2(d-1) (2 \pi)^{d-1} \beta^d} V_{\Sigma}.
\end{aligned}
\end{equation}
Hence, the entropy is obtained as
\begin{equation}\label{entropy}
 S_\Sigma=\beta^2 \frac{\partial F}{\partial \beta}=\frac{d \Omega_{d-2} \zeta(d) \Gamma(d) }{2(d-1) (2 \pi)^{d-1} \beta^{d-1}} V_\Sigma,
\end{equation}
where $\beta$ is the inverse temperature of scalar filed. This entropy is similar to that in the flat phase space, because, of advanced time $v$ as discussed above.

In our previous work, we consider two assumptions, one of which is about Hawking radiation as black-body radiation and the other is about the evaporation of black hole as quasi-static process. Based on these two assumptions, the temperature of the scalar field inside the black hole can be regarded as the Hawking temperature. Therefore, the inverse of the temperature $\beta$ can be expressed as
\begin{equation}\label{beta}
\beta = \frac{1}{T} = \frac{(d-2) \Omega_{d-2} ^2 \left[(4 \pi )^{\frac{1}{d-3}} C^{\frac{1}{d-3}}\right]^{2 (d-1)}}{4 \left\{(d-3) M \Omega_{d-2}  \left[(4 \pi )^{\frac{1}{d-3}} C^{\frac{1}{d-3}}\right]^d-(4 \pi )^{\frac{d}{d-3}} Q^2 C^{\frac{3}{d-3}}\right \}},
\end{equation}
where 
\begin{equation}\label{C}
  C = \frac{(d-2) \sqrt{\frac{(d-1)^2 M^2}{(d-2)^2}-\frac{2 Q^2}{d-3}}+(d-1) M}{(d-2)^2 \Omega_{d-2} }.
\end{equation}
Now consider the Hawking radiation, the rate of mass loss due to Hawking radiation from a black hole is given by the Stefan-Boltzmann law \cite{Montvay:1981jj}
\begin{equation}\label{sbl}
\frac{dM}{dv} = -\sigma_d A_{d-2} T^{d}, 
\end{equation}
where $\sigma_d$ is 
\begin{equation}\label{sigmad}
  \sigma_d = \frac{2 \pi^{\frac{d-2}{2}}}{\Gamma(\frac{d}{2})} \Gamma(d) \zeta(d),
\end{equation}
and $A_{d-2}$ is the area of the event horizon which is 
\begin{equation}\label{Ad-2}
  A_{d-2} = (4 \pi )^{\frac{d-2}{d-3}} \left[\frac{(d-2) \sqrt{\frac{(d-1)^2 M^2}{(d-2)^2}-\frac{2 Q^2}{d-3}}+(d-1) M}{(d-2)^2  }\right]^{\frac{d-2}{d-3}} \Omega_{d-2}^{-\frac{1}{d-3}}.
\end{equation}
Thus, substituting both the Eq. (\ref{sigmad}) and Eq. (\ref{Ad-2}) into the Eq. (\ref{sbl}), we have
\begin{equation}\label{dv}
  \begin{split}
    dv = &-\frac{2^{\frac{-2 d^2+3 d+7}{d-3}} \pi ^{\frac{d^2-3 d+2}{6-2 d}} C^{\frac{2-d}{d-3}} \Gamma \left(\frac{d}{2}\right)}{\Omega_{d-2}  \zeta (d) \Gamma (d)} \\
    & \left[\frac{(d-2) \Omega_{d-2} ^2 \left((4 \pi )^{\frac{1}{d-3}} C^{\frac{1}{d-3}}\right)^{2 (d-1)}}{(d-3) M \Omega_{d-2}  \left((4 \pi )^{\frac{1}{d-3}} C^{\frac{1}{d-3}}\right)^d-(4 \pi )^{\frac{d}{d-3}} Q^2 C^{\frac{3}{d-3}}}\right]^d dM,
  \end{split}
\end{equation}
where C is given in Eq. (\ref{C}).

Now, we will calculate the relation between the entropy of scalar field and the Bekenstein-Hawking entropy along with the emission of Hawking radiation form assumed black hole. As we have discussed earlier, that a reasonable proportional relation between the two types of entropy can be obtained by using statistical method in equilibrium for infinitesimal process. In other words, we can say, that proportionality relation can be investigated by using the differential form. As well as, the variation of both interior volume and the entropy depends on the variation of advanced time $v$ so, the variation of mass can be ignored the final stage of the Hawking radiation.

Differentiating two sides of the Eq.(\ref{volume1}) and substitute the Eq.(\ref{dv}) into it, then the differential form of the interior volume is expressed as
\begin{eqnarray}\label{vga}
  \begin{aligned}
    \dot{V}_\Sigma = &-\frac{2^{\frac{-2 d^2+3 d+7}{d-3}} \pi ^{\frac{d^2-3 d+2}{6-2 d}} c^{\frac{2-d}{d-3}} \Gamma \left(\frac{d}{2}\right)}{\zeta (d) \Gamma (d)} \left[\frac{(d-2) \Omega_{d-2} ^2 \left((4 \pi )^{\frac{1}{d-3}} c^{\frac{1}{d-3}}\right)^{2 (d-1)}}{(d-3) M \Omega_{d-2}  \left((4 \pi )^{\frac{1}{d-3}} c^{\frac{1}{d-3}}\right)^d-(4 \pi )^{\frac{d}{d-3}} Q^2 c^{\frac{3}{d-3}}}\right]^d \\
    & \left \{ \frac{(4 \pi )^{-\frac{4}{d-3}} c^{-\frac{4}{d-3}}}{(d-3) (d-2) \Omega_{d-2} ^2}\left[-\left(d^2-5 d+6\right) \Omega_{d-2} ^2 \left((4 \pi )^{\frac{1}{d-3}} c^{\frac{1}{d-3}}\right)^{2 d} \right. \right. \\
    &\left. \left. +16 \pi  (d-3) M \Omega_{d-2}  \left((4 \pi )^{\frac{1}{d-3}} c^{\frac{1}{d-3}}\right)^{d+3}-2^{\frac{12}{d-3}+5} \pi ^{\frac{2 d}{d-3}} Q^2 c^{\frac{6}{d-3}}\right] \right \}^{\frac{1}{2}} \dot{M},
  \end{aligned}
\end{eqnarray}
 Here $\dot{V}_\Sigma$ and $\dot{M}$ represent $\frac{dV_\Sigma}{dv}$ and $\frac{dM}{dv}$ respectively. Substituting Eq.(\ref{beta}) and Eq.(\ref{vga}) in Eq.(\ref{entropy}), we gets the differential form of entropy for infinitesimal quasi-static process
\begin{equation}\label{fentropy}
  \begin{split}
    \dot{S}_{\Sigma} = &- \frac{2^{-\frac{d^2+2 d-9}{d-3}} d (d-2)  \pi ^{\frac{3 d^2-11 d+12}{6-2 d}} \Omega_{d-2} ^3 C^{-\frac{d}{d-3}} \left((4 \pi )^{\frac{1}{d-3}} C^{\frac{1}{d-3}}\right)^{2 d} \Gamma \left(\frac{d}{2}\right)}{(d-1) \left[(4 \pi )^{\frac{d}{d-3}} Q^2 C^{\frac{3}{d-3}}-(d-3) M \Omega_{d-2}  \left((4 \pi )^{\frac{1}{d-3}} C^{\frac{1}{d-3}}\right)^d\right]} \\
    &\left \{\frac{(4 \pi )^{-\frac{4}{d-3}} C^{-\frac{4}{d-3}}}{(d-3) (d-2) \Omega_{d-2} ^2} \left[-\left(d^2-5 d+6\right) \Omega_{d-2} ^2 \left((4 \pi )^{\frac{1}{d-3}} C^{\frac{1}{d-3}}\right)^{2 d} \right. \right. \\ 
    &\left. \left. +16 \pi  (d-3) M \Omega_{d-2}  \left((4 \pi )^{\frac{1}{d-3}} C^{\frac{1}{d-3}}\right)^{d+3}-2^{\frac{12}{d-3}+5} \pi ^{\frac{2 d}{d-3}} Q^2 C^{\frac{6}{d-3}}\right] \right\}^{\frac{1}{2}} \dot{M},
  \end{split}
\end{equation}
$\dot{M}<0$.

Finally, we have to calculate the variation of the Bekenstein-Hawking entropy in order to compare it with Eq.(\ref{fentropy}) given above. The Bekenstein-Hawking entropy is defined as \cite{Hawking:1974sw, Bekenstein:1972tm, Bekenstein:1973ur}
\begin{equation}\label{BHS}
 S_{BH} = \frac{A}{4},
\end{equation}
where $A = \Omega_{d-2} r_+^{d-2}$ is the area of black hole horizon. Differentiating two side of the Eq. (\ref{BHS}), the differential form of the entropy can be obtained as
\begin{equation}\label{dsbh}
\dot{S}_{BH} =  \frac{4^{\frac{1}{d-3}} (d-1) \pi ^{\frac{d-2}{d-3}} \Omega_{d-2}}{(d-3) \sqrt{\frac{(d-1)^2 M^2}{(d-2)^2}-\frac{2 Q^2}{d-3}}} \left[\frac{(d-2) \sqrt{\frac{(d-1)^2 M^2}{(d-2)^2}-\frac{2 Q^2}{d-3}}+(d-1) M}{(d-2)^2 \Omega_{d-2} }\right]^{\frac{d-2}{d-3}} \dot{M},
\end{equation}
Where $\dot{M}$ is also negative. Using Eq.(\ref{dsbh}) and Eq.(\ref{fentropy}), the differential relation between the entropy of the scalar field and the Bekenstein-Hawking entropy can be expressed as
\begin{equation}\label{two types entropy}
\dot{S}_\Sigma = - F \left(\frac{Q}{M} \right)   \dot{S}_{BH}.
\end{equation}
The function $F \left(\frac{Q}{M} \right)$ is given as 
\begin{equation}\label{F}
F \left(\frac{Q}{M} \right) = -\frac{2^{-\frac{d^2+2 d-11}{d-3}} d (d-3) (d-2)  \pi ^{\frac{3 d^2-9 d+4}{6-2 d}} \Gamma \left(\frac{d}{2}\right) \Omega_{d-2} ^2 \sqrt{\frac{(d-1)^2}{(d-2)^2}-\frac{2}{(d-3)}(\frac{Q}{M})^2} A}{B},
\end{equation}
where 
\begin{equation}
  \begin{split}
    A=&\left\{(4 \pi )^{\frac{1}{d-3}} \left[\frac{(d-2) \sqrt{\frac{(d-1)^2}{(d-2)^2}-\frac{2}{(d-3)}(\frac{Q}{M})^2}+d-1}{(d-2)^2 \Omega_{d-2} }\right]^{\frac{1}{d-3}}\right\}^{2 (d-1)}\\
    &\left[\frac{(d-2) \sqrt{\frac{(d-1)^2}{(d-2)^2}-\frac{2}{(d-3)}(\frac{Q}{M})^2}+d-1}{(d-2)^2 \Omega_{d-2} }\right]^{-\frac{2 (d-2)}{d-3}} \\
    &\left \{\frac{(4 \pi )^{-\frac{4}{d-3}} }{(d-3) (d-2) \Omega_{d-2} ^2} \left[\frac{(d-2) \sqrt{\frac{(d-1)^2}{(d-2)^2}-\frac{2}{(d-3)}(\frac{Q}{M})^2}+d-1}{(d-2)^2 \Omega_{d-2} }\right]^{-\frac{4}{d-3}} \right. \\
    &\left. \left\{-\left(d^2-5 d+6\right) \Omega_{d-2} ^2 \left\{(4 \pi )^{\frac{1}{d-3}} \left[\frac{(d-2) \sqrt{\frac{(d-1)^2}{(d-2)^2}-\frac{2}{(d-3)}(\frac{Q}{M})^2}+d-1}{(d-2)^2 \Omega_{d-2} }\right]^{\frac{1}{d-3}}\right\}^{2 d} \right. \right. \\
    &\left. \left. -2^{\frac{12}{d-3}+5} \pi ^{\frac{2 d}{d-3}} \left( \frac{Q}{M} \right)^2 \left[\frac{(d-2) \sqrt{\frac{(d-1)^2}{(d-2)^2}-\frac{2}{(d-3)}(\frac{Q}{M})^2}+d-1}{(d-2)^2 \Omega_{d-2} }\right]^{\frac{6}{d-3}} \right. \right. \\
    & \left. \left. +16 \pi  (d-3) \Omega_{d-2}  \left\{(4 \pi )^{\frac{1}{d-3}} \left[\frac{(d-2) \sqrt{\frac{(d-1)^2}{(d-2)^2}-\frac{2}{(d-3)}(\frac{Q}{M})^2}+d-1}{(d-2)^2 \Omega_{d-2} }\right]^{\frac{1}{d-3}}\right\}^{d+3}\right \} \right \}^{\frac{1}{2}}, \\
    B = &(d-1)^2 \left\{(d-3) \Omega_{d-2}  \left\{(4 \pi )^{\frac{1}{d-3}} \left[\frac{(d-2) \sqrt{\frac{(d-1)^2}{(d-2)^2}-\frac{2}{(d-3)}(\frac{Q}{M})^2}+d-1}{(d-2)^2 \Omega_{d-2} }\right]^{\frac{1}{d-3}}\right\}^d \right. \\
    &\left. -(4 \pi )^{\frac{d}{d-3}} \left(\frac{Q}{M} \right)^2  \left[\frac{(d-2) \sqrt{\frac{(d-1)^2}{(d-2)^2}-\frac{2}{(d-3)}(\frac{Q}{M})^2}+d-1}{(d-2)^2 \Omega_{d-2} }\right]^{\frac{3}{d-3}}\right \}.
  \end{split}
  \end{equation}

\section{Hamiltonian Analysis of Entropy}
Recently, Bibhas Ranjan Majhi $et, \, al$ \cite{Majhi:2017tab} calculate the entropy of the scalar field in the interior of Schwarzschild black hole using constraint Hamiltonian analysis.The reason for using the constraints analysis was space-time metric (\ref{stmh}) and the vanishing value of canonical Hamiltonian for the system. They used two type of constraints, which are the particle on-shell constraint and the gauge fixing constraint. The first type constraint guarantees the non-virtual character of scalar particles and can be expressed as
\begin{equation}
  \Psi = P_\mu P^\mu - m^2 \approx 0,
\end{equation}
and the latter type constraint guarantees, that the parameter of the particle's path can be interpret as proper time, which can be expressed as 
\begin{equation}
  \Phi = \frac{P^0}{m} \tau - x^0 \approx 0.
\end{equation}
Using the two constraint, they derive the equation of motion from Hamiltonian equation as 
\begin{equation}
  \dot{x}^i = \frac{P^i}{P^0}, \, \dot{P}^i =\frac{1}{P^0} \Gamma_{\rho \sigma}^i P^\rho P^\sigma.
\end{equation}
Since $P^0=-P_0$, the on-shell constraint can be written as
\begin{equation}\label{Hos}
  P^\mu P_\mu = g^{\mu \nu} P_\mu P_\nu = -(P_0)^2 + h^{ij} P_i P_j = m^2.
\end{equation}
Using massless scalar field, $m=0$, the Eq. (\ref{Hos}) is exactly the same as the Eq. (\ref{os}). Therefore, the Hamiltonian or the energy of the scalar particles is given by
\begin{equation}
  E^2 = P_0^2 = h^{ij} P_i P_j.
\end{equation}

Subsequent, the Gibbs' free energy can be calculated as 
\begin{equation}
  \begin{split}
    G&=\frac{1}{(2 \pi)^3 \beta} \int d\lambda d\phi_1 \cdots d\phi_{d-2} dP_{\phi_1} \cdots dP_{\phi_{d-2}} ln(1-e^{- \beta E})\\
    &= - \frac{\Omega_{d-2} V_\Sigma}{2 (2 \pi)^d (d-1) \beta^d} \zeta(d) \Gamma(d).
  \end{split}
\end{equation}
Therefore, the entropy can be expressed as 
\begin{equation}\label{Hentropy}
  S_\Sigma = \beta^2 \frac{\partial G}{\partial \beta} = \frac{d \Omega_{d-2} V_\Sigma}{2 (d-1) (2 \pi)^{d-1} \beta^{d-1}} \zeta(d) \Gamma(d).
\end{equation}
Form Eq. (\ref{Hentropy}), we can see that it is exactly the same as the Eq. (\ref{entropy}). Which means that, the method of Hamiltonian analysis is same as that of standard statistical method. Therefore, it is convenient to calculate the proportionality relation between two types of entropy, if we consider the Hawking radiation as discussed in section 3.

\section{Discussions and Conclusions}
In this section, we will discuss the proportionality relation between the two kinds of entropy. The relation between the  proportionality coefficient $F\left(\frac{Q}{M}\right)$ and the dimension as shown in Figure \ref{image-3.1} below.

\begin{figure}
\begin{center}
\includegraphics[width=0.6\textwidth]{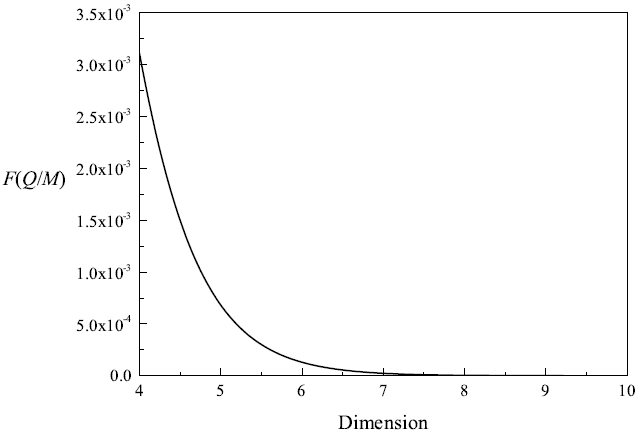}
\caption{Plots of the dimension versus $F\left(\frac{Q}{M}\right)$, where $\left(\frac{Q}{M}\right)=0.5$.}
\label{image-3.1}
\end{center}
\end{figure}

Form Figure \ref{image-3.1}, we can see that $F\left(\frac{Q}{M}\right)$ is gradually decreased with increasing of the dimension. It also illustrates that the variation of the entropy for a scalar field inside the black hole. For high dimension of space-time, this variation is much smaller as compared to the variation of Bekenstein-Hawking entropy. This means that, Bekenstein-Hawking entropy decreases rapidly with the  increase in dimensions of space-time as compare to the entropy of the scalar field. In other words, the Hawking radiation can getting faster with the increase in dimensions of space-time.

Summarizing this work, we calculated the interior volume of $d$-dimensional black hole. Involving massless scalar field inside the black hole, the proportionality relation of the entropy of the scalar field and the Bekenstein-Hawking entropy is calculated by using the standard statistical method and differential forms. Meanwhile, we recalculated the relation using the Hamiltonan analysis and found that the two different methods have same result. Finally, we found that the proportionality coefficient is gradually decreased with increasing the dimension. It can be interpreted as the Hawking radiation can getting faster when the dimension of space-time increases.

\clearpage
\thispagestyle{empty}
\hfill
\clearpage
\newpage

\chapter{Entropy evolution in the interior volume of a charged f(R) black hole}

\section{Introduction}
The definition of the interior volume of a spherically symmetric black hole is proposed by Christodoulou and Rovelli \cite{CR:2014}. This definition can be expressed as that the volume of the largest hyper-surface inside the black hole bounded by two-sphere in the event horizon is the interior volume. Based on this definition, the interior volume of a Schwarzschild black hole has been calculated. It is shown that the contribution to the largest hyper-surface mainly comes from the hyper-surface at constant radius with $r = \frac{3}{2}M$ when $v \gg M$. For a Schwarzschild black hole the the volume at this constant hyper-surface can be regarded as its largest interior volume, which can be expressed as, 
\begin{equation}
  V_{\Sigma} \sim 3\sqrt{3} \pi M^2 v.
\end{equation}
Since the interior volume has linear relation with advance time $v$ so, it can be regarded as that the interior volume of a Schwarzschild black hole increases with the advanced time. Subsequently, the definition for the interior volume has been extended from a spherically symmetric black hole to an axially symmetric black hole, and then the interior volume of a Kerr black hole has been calculated \cite{Bengtsson:2015zdaa}. In a Kerr black hole, the hyper-surface at $r = r_v$ can also be regarded as the largest hyper-surface approximately, where $r_v$ takes a special value which corresponds to the maximal value of the volume expression of the hyper-surface at $r =$ constant. Therefore, the interior volume of a Kerr black hole is similar to the Schwarzschild case, and it also increases with the advanced time $\nu$.

The special property of the interior volume may be a candidate to resolve the information paradox problem \cite{Ong:2015}. Since the interior volume increases with advanced time, a black hole can have a large volume containing many modes of quantum field and the entropy of these modes may relate to the Bekenstein-Hawking entropy. It means that the lost information on the horizon claimed in Hawking radiation may be stored in the interior of the black hole as quantum states. Moreover, the information on the horizon is associated with Bekenstien-Hawking entropy \cite{Page:1993, Marolf:2017, Jacob:2005}. So, if we obtain the entropy of the quantum field modes inside the black hole and construct the evolution relation between the entropy of the quantum field modes and Bekenstien-Hawking entropy under Hawking radiation, it may provide a way to solve the information paradox. Therefore, it is necessary to investigate the entropy of the quantum modes inside the black hole. The entropy of a scalar field in the interior volume of a Schwarzschild black hole has been investigated in Ref. \cite{Zhang:2015}. Subsequently, the method has been extended from a Schwarzschild black hole to other kinds of black holes including a Kerr black hole \cite{Yang:2018arj, Han:2018, Ali:2018, Wang:2018dvo}. It is shown that the entropy is also increases linearly with the advanced time, because it is proportional to the interior volume. Moreover, for all kinds of black holes, the entropy of the scalar field in the interior volume is proportional to Bekenstein-Hawking entropy under Hawking radiation approximately except the late stage of Hawking radiation.

Besides, the modified gravity is focused again in recent 20 years because our universe is experiencing an accelerated expansion. Some types of the modified gravity approaches are adding higher powers of the scalar curvature $R$, the Riemann and Ricci tensors, or their derivatives in the Lagrangian formulation \cite{Sheykhi:2012}. Among these attempts are Lovelock gravity, braneworld cosmology, scalar-tensor theory and $f(R)$ theory. In these modified gravity theories, $f(R)$ theory is proved to be able to mimic the whole cosmology history, from inflation to the actual accelerated expansion era \cite{Nojiri:2003, Atazadeh:2008}. So, it is interesting and meaningful to extend $f(R)$ gravity from cosmology to the black hole, which could provide us some features of black hole different from Einstein's gravity. Therefore, using the method proposed by Christodoulou and Rovelli, we calculated the interior volume of a charged $f(R)$ black hole. After that, based on the interior volume, we calculate the entropy of the quantum field modes in it, and attempt to construct the evolution relation between the entropy of the quantum field and Bekenstein-Hawking entropy under Hawking radiation. According to the result, we investigate how the modified coefficient $b$ in the $f(R)$ gravity theory affects the evolution relation of the two types of entropy.

The organization of the chapter is as follows. In section 2, we calculate the interior volume of a charged $f(R)$ black hole. In section 3, we calculate the entropy of a scalar field in a $f(R)$ black hole, and construct the evolution relation between this entropy and Bekenstein-Hawking entropy under Hawking radiation. In section 4, some discussions and conclusions are given.

\section{The interior volume of a charged f(R) black hole}
Christodoulou and Rovelli proposed the definition of the interior volume of a spherically symmetric black hole \cite{CR:2014}. This definition can be expressed as that the volume of the largest hyper-surface inside the black hole bounded by two-sphere in the event horizon is the interior volume. Based on the definition, the interior volume of a Schwarzschild black hole has been investigated. In a Schwarzschild black hole, the contribution to the largest hyper-surface mainly comes from the hyper-surface at $r = \frac{3}{2} M$ at the late advanced time. So, the interior volume of a Schwarzschild black hole can be regarded as the volume of the hyper-surface at $r = \frac{3}{2} M$. In the meantime, the general expression for the interior volume of an arbitrary spherically symmetric black hole has been proposed as
\begin{equation}\label{ge}
  V = 4 \pi \sqrt{-r_v^4 N(r_v)}v,
\end{equation}
where $r_v$ corresponds to the particular hyper-surface at constant radius $r$, which is the largest hyper-surface in a spherically symmetric black hole.

The action of $R+f(R)$ gravity in 4-dimensional spacetime coupled to a nonlinear Maxwell field is
\begin{equation}
  S = \int_\mathcal{M} d^4 x \sqrt{-g} \left[R + f(R) - F_{\mu \nu} F^{\mu \nu} \right],
\end{equation}
where $R$ is the Ricci scalar curvature, $f(R)$ is an arbitrary function of $R$, $F_{\mu \nu}$ is the electromagnetic field tensor, which is related to the electromagnetic potential $A_\mu$ by $F_{\mu \nu} = \partial_\mu A_\nu -\partial_\nu A_\mu$. Considering the constant scalar curvature  $R = R_0$, the metric for a 4-dimenaional charged $f(R)$ black hole in the Eddington-Finkelstein coordinates is \cite{Sheykhi:2012, Moon:2011}
\begin{equation}
ds^2 = -N(r) dv^2 + 2 dv dr +r^2 (d\theta + \text{sin}^2 \theta d \phi^2),
\end{equation}
where
\begin{equation}
N(r) =1-\frac{2 m}{r}+ \frac{q^2}{b r^2} - \frac{R_0}{12}r^2.
\end{equation}
In this metric, the modified coefficient is $b = 1 + f' (R_0)$, and the parameter $m$ and $q$ are related to the ADM mass $M$ and the electric charge $Q$ of the black hole as
\begin{equation}
  M = m b, \, \, \, \, Q = \frac{q}{\sqrt{b}}.
\end{equation}
It illustrates that the modified coefficient $b$ in $f(R)$ gravity changes the mass $M$, the electric charge $Q$ of black hole and the corresponding thermodynamic potential. Treating the Ricci scalar curvature as $R_0 = -\frac{12}{l^2} = -4 \Lambda$, the metric is used to describe a asymptotical AdS spacetime. The Hawking temperature $T$ with the outer event horizon $r = r_+$ is
\begin{equation}\label{HT}
    T = \left. \frac{N'(r)}{4 \pi} \right|_{r = r_+} = \frac{1}{4 \pi } \left[\frac{4 M}{b \left(\sqrt{\frac{A}{R_0}}+\sqrt{\frac{B}{R_0}}\right)^2}-\frac{4 \sqrt{2} Q^2}{\left(\sqrt{\frac{A}{R_0}}+\sqrt{\frac{B}{R_0}}\right)^3}-\frac{R_0 \left(\sqrt{\frac{A}{R_0}}+\sqrt{\frac{B}{R_0}}\right)}{6 \sqrt{2}} \right],
\end{equation}
where
\begin{equation*}
  \begin{split}
    A = & 8 -\frac{12 \sqrt{2} M}{b \sqrt{\frac{\frac{2 \sqrt[3]{2} b \left(Q^2 R_0-1\right)}{\sqrt[3]{C}}-\frac{2^{2/3} \sqrt[3]{C}}{b}+4}{R_0}}}-\frac{2 \sqrt[3]{2} b \left(Q^2 R_0 -1\right)}{\sqrt[3]{C}}+\frac{2^{2/3} \sqrt[3]{C}}{b}, \\
    B =& \frac{2 \sqrt[3]{2} b \left(Q^2 R_0-1\right)}{\sqrt[3]{C}}-\frac{2^{2/3} \sqrt[3]{C}}{b}+4, \\
    C =& \sqrt{b^2 R_0 \left(4 b^4 Q^2 \left(Q^2 R_0+3\right)^2-36 b^2 M^2 \left(3 Q^2 R_0+1\right)+81 M^4 R_0 \right)}\\
    &+b^3 \left(6 Q^2 R_0+2\right) - 9 b M^2 R_0.
  \end{split}
\end{equation*}
Moreover, Bekenstein-Hawking entropy of a charged $f(R)$ black hole can be expressed as
\begin{equation}\label{SBH}
  S_{BH} = \pi r_+^2 b,
\end{equation}
where $r_+$ can be written as
\begin{equation}
  r_+ = \frac{\sqrt{2}}{2} \left(\sqrt{D}+\sqrt{\frac{E}{R_0}-\frac{12 \sqrt{2} M}{R_0 b \sqrt{\frac{D}{R_0}}}}\right),
\end{equation}
and
\begin{equation*}
  \begin{split}
    D =& \frac{1}{R_0} \left[\frac{2 \sqrt[3]{2} b \left(Q^2 R_0-1\right)}{\sqrt[3]{C}}-\frac{2^{2/3} \sqrt[3]{C}}{b}+4 \right], \\
    E =& -\frac{2 \sqrt[3]{2} b \left(Q^2 R_0-1\right)}{\sqrt[3]{C}}+\frac{2^{2/3} \sqrt[3]{C}}{b}+8, \\
    C = & \sqrt{b^2 R_0 \left(4 b^4 Q^2 \left(Q^2 R_0+3\right)^2-36 b^2 M^2 \left(3 Q^2 R_0 +1\right)+81 M^4 R_0\right)} \\
    & + b^3 \left(6 Q^2 R_0+2\right) -9 b M^2 R_0.
  \end{split}
\end{equation*}

Since a charged $f(R)$ black hole is a spherically symmetric black hole, we can directly use the general expression Eq. (\ref{ge}) to calculate the interior volume. Analogous to the method in Ref. \cite{CR:2014}, the value of $r_v$ which is the position of the largest hyper-surface in the black hole should satisfy
\begin{equation}\label{crv}
\frac{d}{dr} \left(\sqrt{-r^4 N(r)} \middle) \right|_{r = r_v} = 0.
\end{equation}
From Eq. (\ref{crv}), $r_v$ can be obtained as
\begin{equation}
  r_v = \sqrt{6} \left[ \sqrt{\frac{G}{R_0}-\frac{18 \sqrt{6} M}{R_0 b \sqrt{\frac{F}{R_0}}}}+\sqrt{\frac{F}{R_0}} \right],
\end{equation}
where
\begin{equation*}
  \begin{split}
    F =& -\frac{4 b \left(3 Q^2 R_0-4\right)}{\sqrt[3]{Z}}+\frac{\sqrt[3]{Z}}{b}+8, \\
    G = & \frac{4 b \left(3 Q^2 R_0-4\right)}{\sqrt[3]{Z}}-\frac{\sqrt[3]{Z}}{b}+16, \\
    Z = & 3 \sqrt{3} \sqrt{b^2 R_0 \left(64 b^4 Q^2 \left(Q^2 R+4\right)^2-288 b^2 M^2 \left(9 Q^2 R_0+4\right)+2187 M^4 R_0\right)} \\
    & +243 b M^2 R_0 -16 b^3 \left(9 Q^2 R_0+4\right).
  \end{split}
\end{equation*}
Substituting the value of $r_v$ into Eq. (\ref{ge}), the interior volume of a charged $f(R)$ black hole can be obtained as
\begin{equation}\label{volume}
  \begin{split}
    V_\Sigma =& \frac{2}{3} \pi  v \left[ \frac{2 \sqrt{6} M}{b } \left(\sqrt{\frac{F}{R_0}}+\sqrt{\frac{Y}{R_0}}\right)^3- 6 Q^2 \left(\sqrt{\frac{F}{R_0}}+\sqrt{\frac{Y}{R_0}}\right)^2 \right. \\
    & \left. + \frac{1}{72} R_0 \left(\sqrt{\frac{F}{R_0}}+\sqrt{\frac{Y}{R_0}}\right)^6+ \left(\sqrt{\frac{F}{R_0}}-\sqrt{\frac{Y}{R_0}}\right)^4\right]^{\frac{1}{2}},
  \end{split}
\end{equation}
where
\begin{equation*}
  \begin{split}
    F = & -\frac{4 b \left(3 Q^2 R_0-4\right)}{\sqrt[3]{Z}}+\frac{\sqrt[3]{Z}}{b}+8, \\
    Y = & -\frac{18 \sqrt{6} M}{b \sqrt{\frac{-\frac{4 b \left(3 Q^2 R_0-4\right)}{\sqrt[3]{Z}}+\frac{\sqrt[3]{Z}}{b}+8}{R_0}}}+\frac{4 b \left(3 Q^2 R-4\right)}{\sqrt[3]{Z}}-\frac{\sqrt[3]{Z}}{b}+16, \\
    Z = & 3 \sqrt{3} \sqrt{b^2 R_0 \left(64 b^4 Q^2 \left(Q^2 R_0+4\right)^2-288 b^2 M^2 \left(9 Q^2 R_0+4\right)+2187 M^4 R_0\right)} \\
    & +243 b M^2 R_0 -16 b^3 \left(9 Q^2 R_0+4\right).
  \end{split}
\end{equation*}

\section{Interior and exterior entropy evolution under Hawking radiation}
Eq. (\ref{volume}) indicates that the interior volume of a charged $f(R)$ black hole is proportional to the advanced time $v$. It means that the interior volume of a charged $f(R)$ black hole, like a Schwarzschild black hole, also increases with the advanced time. The particular character of the interior volume can influence the statistical properties of the quantum field modes in the volume. One quite important quantity for the distribution of the quantum field modes is entropy. So, the entropy of the quantum field can reflect the features of the interior volume and the quantum field modes. Moreover, the entropy of the quantum field in the black hole may be a candidate to resolve the information paradox. The lost information of the black hole during Hawking radiation may be related to Bekenstein-Hawking entropy. Actually, the interior volume of the black hole continues to increase under Hawking radiation, and the entropy of the quantum field inside the black hole varies with the interior volume. If we can find the evolution relation between the entropy of the quantum field and Bekenstein-Hawking entropy under Hawking radiation, the entropy evolution of the interior (volume) and Bekenstien-Hawking entropy (event horizon) of the black hole can be connected. Therefore, it is significant to investigate the entropy of the quantum field in the interior volume and find the connection between this entropy and Bekenstein-Hawking entropy under Hawking radiation. In the following, we only consider the massless scalar field inside a charged $f(R)$ black hole, and Hawking radiation carrying only energy from the black hole which means that the electric charge $Q$ of the black hole is regarded as a constant.

According to Ref. \cite{Zhang:2015}, although the interior volume of a Schwarzschild black hole increases linearly with the advanced time, the equilibrium statistical method can also be used to calculate the statistical properties of the scalar field in the interior volume. It is mainly because the largest hyper-surface which corresponds to the interior volume accumulates on the hyper-surface at $r = \frac{3}{2} M$ at late advanced time. Therefore, near the hyper-surface at $r = \frac{3}{2} M$, the proper time between two adjacent largest hypersurfaces tends to zero as the advanced time increases. It means that there is no evolution between these two adjacent largest hypersurfaces. In other words, calculating the statistical properties of the scalar field can be seen as on the approximate simultaneity hyper-surface, and this simultaneity hyper-surface just corresponds to the interior volume of a Schwarzschild black hole. Hence, the equilibrium statistical method can be used on the largest hyper-surface to calculate the statistical properties of the scalar field. In Ref. \cite{Zhang:2015}, the entropy of the scalar field in the interior volume of a Schwarzschild black hole can be given as
\begin{equation}\label{sentropy}
  S_\Sigma = \frac{\pi^2}{45 \beta^3} V_\Sigma,
\end{equation}

where $\beta$ is the inverse temperature and $V_\Sigma$ is the interior volume of the black hole. Following this idea, we can also use the equilibrium statistical method to calculate the entropy of the scalar field in a charged $f(R)$ black hole. Since a charged $f(R)$ black hole is a spherically symmetric black hole, the largest hyper-surface can also be regarded as accumulating on the hyper-surface at $r = r_v$. Therefore, analogues to the Schwarzschild case, near the hyper-surface at  $r = r_v$, there is still no evolution between two adjacent largest hypersurfaces as the advanced time increases. It means that the equilibrium statistical method can also be used to calculate the statistical properties of the scalar field in a charged $f(R)$ black hole. In addition, according to Ref. \cite{Wang:2018dvo}, the entropy of the scalar field in the interior volume has been generally demonstrated that it can always be expressed as Eq. (\ref{sentropy}). Therefore, we can directly use Eq. (\ref{sentropy}) to investigate the evolution of the scalar field entropy under Hawking radiation.

Now, we want to find the connection between the evolution of the entropy of the scalar field in the interior volume and Bekenstein-Hawking entropy under Hawking radiation. According to Ref. \cite{Wang:2018dvo}, in order to construct the connection, two important assumptions as the black-body radiation and quasi-static process should be proposed. Based on these two assumptions, the temperature of the scalar field in the interior volume can be regarded as Hawking temperature in an infinitesimal evaporation process. Moreover, according to the above statement, the equilibrium statistical method can be used to calculate the statistical properties of the scalar field in the interior volume of the black hole in an infinitesimal process. Therefore, we should find the connection between the evolution of the two types of entropy in an infinitesimal evaporation process.

Substituting Hawking temperature Eq. (\ref{HT}) and the interior volume of a charged $f(R)$ black hole Eq. (\ref{volume}) into Eq. (\ref{sentropy}), the entropy of the scalar field in a charged $f(R)$ black hole can be expressed as
\begin{equation}\label{ssigma}
  S_{\Sigma} = \frac{1}{4320} \alpha(M, Q; b) v.
\end{equation}

In an infinitesimal evaporation process, the evolution of the scalar field entropy can be expressed as the differential form as
\begin{equation}\label{dSSigma}
  dS_\Sigma = \frac{1}{4320} \left[\alpha(M, Q; b) dv + v \frac{\partial \alpha}{\partial M} dM \right].
\end{equation}

Next, we want to construct the evolution relation between the entropy of the scalar field and Bekenstein-Hawking entropy under Hawking radiation. Taking the differentiation of Eq. (\ref{SBH}), we have
\begin{equation}\label{dSBH1}
    dS_{BH} = \pi \, \theta_1 (M, Q; b) \, dM.
\end{equation}
According to the black body assumption, the radiation process should satisfy the Stefan-Boltzmann law as
\begin{equation}\label{SBL}
  \frac{dM}{dv} = - \sigma T^4 A,
\end{equation}
where $\sigma$ is a positive constant related to the number of quantized matter fields coupling with gravity \cite{LEP:2009}, $A$ is the area of the event horizon, and $T$ is Hawking temperature. Substituting Hawking temperature $T$ and the area $A$ into Eq. (\ref{SBL}), we have
\begin{equation}\label{dM}
  dM =  -\frac{\sigma}{128 \pi ^3} \, \theta_2 (M, Q; b) \, dv.
\end{equation}

Substituting Eq. (\ref{dM}) into Eq. (\ref{dSBH1}), the differential relation between the Bekenstein-Hawking entropy and the advanced time can be given as
\begin{equation}\label{dSBH}
   dS_{BH} = -\frac{\sigma}{128 \pi ^2} \, \theta (M, Q; b) \, dv.
\end{equation}
Finally, combining Eq. (\ref{dSSigma}) and Eq. (\ref{dSBH}), the proportional relation between the variation of the scalar field entropy and the variation of Bekenstein-Hawking entropy in an infinitesimal evaporation process can be expressed as
\begin{equation}
  \frac{dS_\Sigma}{dS_{BH}} = -\frac{4 \pi^2}{135 \sigma} \left[\frac{\alpha(M, Q; b)}{\theta(M, Q; b)} + \frac{v}{\theta} \frac{\partial \alpha}{\partial M} \frac{dM}{dv} \right].
\end{equation}
According to the quasi-static assumption, we have $\frac{dM}{dv} \ll 1$, and then the second term can be ignored. Therefore, writing the differential form as the derivative form, the evolution relation between the entropy of the scalar field in a charged $f(R)$ black hole and Bekenstein-Hawking entropy under Hawking radiation can be expressed as
\begin{equation}
\begin{split}
    \dot{S}_\Sigma =& -\frac{4 \pi^2}{135 \sigma}\, \gamma(M, Q; b) \, \dot{S}_{BH},
\end{split}
\end{equation}

Where the dot indicates derivative by advanced time $v$ and $\gamma (M, Q; b) = \frac{\alpha(M, Q; b)}{\theta(M, Q; b)}$. Since the expansion of the function $\gamma(M, Q; b)$ is too long and very complicated, the concrete expression is not given here.

\section{Discussions and conclusions}
Based on the detailed numerical calculation, the relation between the proportionality coefficient $\gamma \left(M, Q; b\right)$ and the modified coefficient $b$ in the $f(R)$ gravity theory is shown in Figure 

\begin{figure}
\begin{center}
\includegraphics[width=0.6\textwidth]{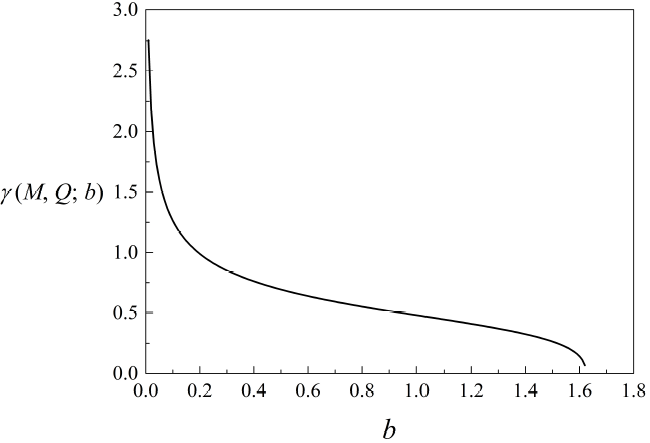}
\caption{The proportionality coefficient $\gamma (M, Q; b)$ as a function of $b$ as $M = 2$, $Q = 1$ and $R_0=-12$.}
\label{image-4.1}
\end{center}
\end{figure}

It is shown that $\gamma \left(M, Q; b \right)$ gradually decreases with increasing coefficient $b$. Since the solution will degenerate to standard Einstein charged AdS black holes as $b = 1$ and $R_0 = -\frac{12}{l^2} = -4 \Lambda$, the function $\gamma \left(M, Q; b \right)$ should be the evolution relation between the scalar field entropy and Bekenstien-Hawking entropy under Hawking radiation for usual charged AdS black holes. When $b < 1$, the value of the function $\gamma \left(M, Q; b \right)$ increases as $b$ decreases, and its value diverges at $b = 0$. It means that the variation of the scalar field entropy is greater than the variation of Bekenstein-Hawking entropy under an infinitesimal evaporation process. In other words, the growth rate of the scalar field entropy in the interior volume of the black hole is greater than the decreasing rate of Bekenstein-Hawking entropy under Hawking radiation. When $b > 1$, the value of the function $\gamma \left(M, Q; b \right)$ decreases, and its value will approach to zero at $b = 1.6$. It means that the variation of the scalar field entropy is much more less than the variation of Bekenstein-Hawking entropy. In other words, the growth rate of the scalar field entropy in the interior volume of the black hole is much more less than the decreasing rate of Bekenstein-Hawking entropy under Hawking radiation. Actually, the entropy of the scalar field in the interior volume always increases linearly with the advanced time. Therefore, we can conclude that as the modified coefficient $b$ increases, the decreasing rate of Bekenstein-Hawking entropy gradually increases, which means that the radiation rate of Hawking radiation can enhance with the increasing coefficient $b$.

Till now, the interior volume of a charged $f(R)$ black hole has been investigated. Considering the massless scalar field in the interior volume of the black hole, the entropy of the scalar field is calculated using the equilibrium statistical method. According to the two important assumptions which are the black-body radiation and the quasi-static process, the evolution of the scalar field entropy in an infinitesimal Hawking radiation has been investigated. In an infinitesimal evaporation process, the evolution of Bekenstein-Hawking entropy can be obtained naturally. The proportional relation between the entropy of the scalar field and the Bekenstein-Hawking entropy under Hawking radiation is given. Finally, the relation between the proportionality coefficient of the two types of entropy and the modified coefficient $b$ in the $f(R)$ gravity theory is investigated. It is shown that the radiation rate for the Hawking radiation of a charged $f(R)$ black hole can enhance with the increasing modified coefficient $b$.

\clearpage
\thispagestyle{empty}
\hfill
\clearpage
\newpage

\chapter{Configurational entropy and coexistence curves of the d-dimensional f(R) AdS black holes}

\section{Introduction}
Black hole Phase transition is the property of black hole to change its phase due to some thermal fluctuation, in which the zero-temperature degree of freedom is reorganized quantitatively in different form. Hence, from thermodynamical point of view the equilibrium state of the black holes will not be valid in phase transition phenomena \cite{Stephens:2001sd, Banerjee:2011raa, Banerjee:2011au, Banerjee:2011cz, Banerjee:2016nse, Quevedo:2016swn}. 
The study of black hole phase transition is naturally an important topic in this era. Since from the birth of black hole thermodynamics \cite{Hawking:1974sw, Bardeen:1973gs}, the issue of black holes phase transition has been found to playing a significant role in understanding several crucial and fundamental properties of black holes including the statistical and configuration origin of its entropy. The phase transition describes the complex interactions between the phases of a body. Typically, there three phases states of a body solid, liquid and gas. The laws of black hole mechanics become similar to the usual laws of thermodynamics after appropriate identifications between the black hole parameters and the thermodynamic variables \cite{Roychowdhury:2014cva}.  First time Davis \cite{Davies:1978zz} claimed a certain type of phase transition by observing the discontinuity of heat capicity as given by, $C=\frac{\partial M}{\partial T}$, which results to be negative but due to the lack of information this discontinuity was not treated as a phase transition. it means that an in-going will increase the the mass of black hole and the temperature will decreases because the rate of absorption will be more than rate of emission. At finite temperature at the spatial boundary the heat capacity comes out to be positive. defining the black hole partition function do not show any discontinuity. so studying the black hole will be meaningful in asymptotically curved space times (i.e. those space-times which admit cosmological constants $\Lambda$, whose description is also given in chapter 1) instead of flat spaces. It is a fascinating phenomenon after the Hawking-Page prediction \cite{Hawking:1982dh}. 

Hawking and Page discovered a first order phase transition Schwarzschild AdS black hole at high temperature. Note that the AdS nature is crucial because for asymptotically flat case the specific heat is negative and the black hole is thermodynamically unstable. Hawking-Page phase transition can be interpreted via the AdS/CFT correspondence, as a transition from a low-temperature confining phase to a high-temperature de-confining phase in the boundary field theory \cite{Hawking:1982dh, Witten:1998zw}. Gibbs proposed the thermodynamic potential, which can introduce the behaviour of a system at the point phase transition \cite{Banerjee:2016nse}. According to Gibbs if we know the temperature, pressure and the phase composition, then the equilibrium state of the system can be completely understood. 

The most prominent result of this prediction is the idea about charged AdS black hole is that they have act similar to liquid-gas phase transition. Chamblin et. al \cite{Chamblin:1999tk, Chamblin:1999hg} investigate the critical phenomenon of Reissner Nordstrom AdS black hole and showed that Phase transition associated with RN-AdS black hole is similar to the Van der Waal (VdW) like phase transition. Considering the cosmological constant as thermodynamic pressure, thermodynamic of black hole are discussed in an extended phase-space \cite{Caldarelli:1999xj, Dolan:2010ha, Dolan:2011xt, Dolan:2011jm, Cvetic:2010jb, Lu:2012xu}. Similarly, taking the thermodynamic volume as cosmological constant, the liquid-gas phase transition phenomenon is in-sighted via P-V criticality in \cite{Kubiznak:2012wp, Gunasekaran:2012dq}. The critical component was determined and are found to be similar as that of a Van der Waal Fluid. Following the analogy of Van der Waal’s like first order phase transition, the most interesting and meaningful Physics is revealed. Which extend the study to more deep and fundamental understanding about the black hole. For more detail, see \cite{Altamirano:2013ane, Wei:2014hba, Frassino:2014pha, Altamirano:2013uqa, Belhaj:2012bg, Hendi:2012um, Altamirano:2014tva, Dolan:2014jva}. 

Configuration entropy (information entropy) is the portion of a system's entropy, that is related to the position of its constituent particles, but not to their velocity or momentum. It is a useful tool for investigation of stability and relative dominance for the physical system under consideration. The configurational entropy has the roots in early studies of information entropy proposed by Shannon in \cite{shannon1948}, introduced in framework of mathematical communication theory. Shannon proposed that, for a uniformly equally probable system, the information entropy of an unknown $N$ memo,s is $log_2 N=k_BlogN$. Where $k_B$ is the Boltzmann constant. In a thermodynamic process variation of configurational entropy is equivalent to the variation of macroscopic entropy of the system. The Boltzmann-Gibbs entropy $S_BG $ and micro-states of the system in statistical thermodynamics are related as
\begin{equation}\label{GBent1}
S_{BG}=k_Blogw
\end{equation}
Where $w$ is the number of possible configurations at given state of energy. The probability of particle configuration over the micro-state has the form $w=\frac{1}{p_i}$ so, $S_{BG}$ can be described by a set of their microscopic state probabilities
\begin{equation}\label{GBent2}
S_{BG}=-k_B\sum p_i logp_i
\end{equation}
As the distribution of micro-states is considered as uniform, hence one can also consider the equal probability over the distribution. Where the total probability is $\sum p_i=1$. Let us consider the total number of molecule in system as $N_t$ consist of two type molecules $N_1$ and $N_2$, then the probability distribution over the micro-state is given by
\begin{equation}\label{GBent3}
p_i=\frac{N_t}{N_1!(N_t-N_1)!}
\end{equation}
Defining the effective number density $n$ in terms of Plank’s length $l_p ^2=\frac{Gh}{c^3}$ and thermodynamic volume $V$ as defined in \cite{Gunasekaran:2012dq, Lee:2017ero}
\begin{equation}\label{noDen}
n=\frac{N}{V}=\frac{1}{2l_p ^2 r_+}
\end{equation}
The total configurational entropy of the system in geometrical units can be calculated as 
\begin{equation}\label{GBent4}
S_{con}=-(n_1log(\frac{n_1}{n_1+n_2})+n_2log(\frac{n_2}{n_1+n_2}))
\end{equation}
Here, we involved the Sterling approximation with $n_1$  and $n_2$ as effective number densities resulting from first order phase transition.

The organization of this chapter is such that in next section we investigate the a general approach for the relations of coexistence curve and molecule number density of a d-dimensional $f(R)$ AdS black holes by reduced parameter space. We found that in reduced parameters the pressure and temperature are independent of parameter $b$ while the free energy is dependent on factor $b$. In section. $3$, using this general solution for specific dimensions, we evolve the difference in number densities and coexistence curves and configurational entropy. Finally, we also calculate the relation between difference in number densities and dimensional variation.

\section{Phase Transition in black holes}

\section{Coexistence Curve and Molecule Number Density of f(R) AdS Black Holes}
The Einstein-Maxwell action in higher dimensions can be read as
\begin{equation}\label{EMaction}
S=\frac{1}{2\kappa} \int{d^dx\sqrt{-g}(R-F_{{\mu}{\nu}}F^{{\mu}{\nu}}}-2)
\end{equation}
Here $\kappa=8\pi G$ and $\Lambda=\pm{\frac{(d-1)(d-2)}{2l_p ^2}}$. The positive-negative cosmological constants corresponding to the de sitter (dS) and anti-de sitter (AdS) space-times in cosmological scale $l_p$ respectively. $F_{{\mu}{\nu}}$ is the electromagnetic field tensor related to electromagnetic potential as $F_{{\mu}{\nu}}=\partial_{\mu}A_{\nu}-\partial_{\nu}A_{\mu}$. Accounting for a constant scalar curvature $(R=R_0)$ the d-dimensional metric for charged spherically symmetric $f(R)$ AdS black hole is given as \cite{Mo:2016sel, Chen:2013ce}
\begin{equation}\label{mettric}
ds^2=-f(r)dt^2+f(r)^{-1}dr^2+r^2d\Omega^2
\end{equation}
Where
$$f(r)=1-\frac{m}{r^{(d-3)}}+\frac{q^2}{br^{2(d-3)}}-\frac{R_0r^2}{12}$$
and 
$$b=f'(R)$$
This solution is asymptotically AdS, if $R_0=-4\Lambda$. The parameter $M$ and $Q$ are corresponding to the black hole ADM mass and electric charge can be expressed as 
\begin{equation}\label{masscharge}
m=\frac{16\pi M}{(d-2)\Omega_(d-2)} \qquad q=\frac{8{\pi} Q}{\Omega_{(d-2)}\sqrt{2(d-2)(d-3)}}
\end{equation}
Form reference \cite{Chen:2013ce} the relation between curvature scalar and thermodynamic pressure is $R_0=-\frac{64 \pi P}{(d-2)b}$ so, the horizon temperature $T$ of a $d$-dimensional $f(R$) black hole is given by
$$T=\frac{\partial_r f(r)}{4\pi}$$
\begin{equation}\label{temp}
=\frac{1}{12\pi r_+}\left (3(d-3)\left ( 1-\frac{q^2 r_+ ^{6-2d}}{b} \right )+\frac{16(d-1)\pi r_+ ^2 P}{(d-2)b}  \right )
\end{equation}
and the thermodynamic pressure is obtained from the above equation as 
$$P=\frac{3 b (d-2)}{d-1}\left ( \frac{T}{4 r_+} -\frac{d-3}{16 \pi r_+ ^2}+\frac{(d-3) q^2}{16 \pi  b r_+^{2 d-4}}\right )$$
or we can write as 

\begin{equation}
P=\frac{(3 b)}{d-1} \left(\frac{T}{\frac{4 r_+}{d-2}}-\frac{d-3}{\pi  (d-2) \left(\frac{4 r_+}{d-2}\right){}^2}+\frac{4^{2 d-5} (d-3) q^2}{4 \pi  b (d-2)^{2 d-5} \left(\frac{4 r_+}{d-2}\right){}^{2 d-4}}+\right)
\end{equation}
The Gibbs free energy can be calculated by following the expression $G=H-TS$ as
\begin{equation}\label{FE1}
G=\frac{\pi^{\frac{d-1}{2}}}{8\pi \Gamma({\frac{d-1}{2}})}\left ( br_+ ^{(d-3)}-\frac{16 \pi P r_+ ^{(d-1)}}{(d-1)(d-2)}+\frac{(2d-5)q^2}{r_+ ^{(d-3)}} \right )
\end{equation}

Now let us use the specific volume of black hole fluid $\nu=\frac{4l_p ^2 r_+}{(d-2}$, which is identified for comparison with Van der Waals equation. The above equation of state for pressure and Gibbs free energy becomes
\begin{equation}\label{P2}
P=\frac{3b}{(d-1)}\left (\frac{T}{\nu}-\frac{(d-3)}{\pi (d-2){\nu}^2}+\frac{4^{2d-5}(d-3)q^2}{4\pi b (d-2)^{2d-5}{\nu}^{2d-4}}\right )
\end{equation}
Recall the Legendre transformation of enthalpy $G=H-TS$, the Gibbs free energy $G$ is calculated as
\begin{equation}\label{FE2}
G=\frac{(d-2)^{d-3}\pi^{\frac{d-1}{2}}}{2^{2d-3}\pi \Gamma{(\frac{d-1}{2}})}\left (b\nu^{d-3}-\frac{(d-2)\pi P \nu^{d-1}}{d-1}+\frac{4^{2(d-3)}(2d-5)q^2}{(d-2)^{2(d-3)}\nu^{d-3}} \right )
\end{equation}
The critical point occurs at $P=P_c$, whose inflection point can be calculated satisfy the condition \cite{Kubiznak:2012wp, Chen:2013ce, Mo:2016sel}
\begin{equation}\label{CC}
\partial_\nu P_{r_h=r_\nu, T=T_c}=0 \qquad \partial_\nu ^2 P_{r_h=r_\nu, T=T_c}=0 
\end{equation}

\begin{equation}
\partial_\nu P=\frac{2 (d-3)}{\pi  (d-2) \nu ^3}-\frac{T}{\nu ^2}+\frac{4^{2 d-6} (d-3) (d-2)^{5-2 d} (4-2 d) q^2 \nu ^{3-2 d}}{\pi  b}=0
\end{equation}
Here we can make the conversion as 
$$\frac{6 (d-3)}{\pi  (d-2) \nu }=3T-\frac{3\ 4^{2 d-6} (4-2 d) (d-3) (d-2)^{5-2 d} q^2 \nu ^{5-2 d}}{\pi  b}$$
similarly 
\begin{equation}
\partial_\nu ^2 P=\left (- \frac{6 (d-3)}{\pi (d-2)\nu ^4}+\frac{2 T}{\nu ^3}+\frac{4^{2 d-6} (3-2 d) (4-2 d) (d-3) (d-2)^{5-2 d} q^2 \nu ^{2-2 d}}{\pi  b}\right )=0 
\end{equation}
$$\frac{6 (d-3)}{\pi  (d-2) \nu }=2T+\frac{4^{2 d-6} (3-2 d) (4-2 d) (d-3) (d-2)^{5-2 d} q^2 \nu ^{5-2 d}}{\pi  b}$$
Comparing these coefficients, we can calculate quantities at the critical quantities as
\begin{equation}\label{SVCP}
\nu_c =\frac{4}{d-2}\left(\frac{(d-2)(2d-5)q^2}{b}\right)^{\frac{1}{2(d-3)}}
\end{equation}
\begin{equation}\label{CPP}
P_c=\frac{3b(d-3)^2}{(d-2)^2(d-1)\pi \nu_c ^2}
\end{equation}
\begin{equation}\label{CPT}
T_c=\frac{4(d-3)^2}{(d-2)(2d-5)\pi \nu_c}
\end{equation}
The temperature of the critical point is independent on parameter b. Using the values of $v_c$ and $P_c$ in Eq. (\ref{FE2}), we get
\begin{equation}\label{CPFE}
G_c=\frac{\pi ^{\frac{d-3}{2}} \left(d^3-6 d^2+21 d-28\right)}{8 (d-1)^2 \Gamma \left(\frac{d-1}{2}\right)}\sqrt{\frac{b (2 d-5) q^2}{d-2}}
\end{equation}
 Eq. (\ref{SVCP}), (\ref{CPP}), (\ref{CPT}) and (\ref{CPFE}) are the thermodynamic parameters of critical point. Beyond this point the phase of small and large black hole can't be identified. For detail also see \cite{Chen:2013ce, Mo:2016sel}. From the above components at $d=4$, one can investigate that an $f(R)$ AdS black hole system have similarity with the Vander Walls fluid. Which also shows that first order phase transition between small/large black hole is of Van der Waal's type phase transition. The intersection relation is found to be
\begin{equation}\label{Pt.inter}
\rho=\frac{P_c\nu_c}{T_c}=\frac{3b(2d-5)}{4(d-2)(d-1)}
\end{equation}
Which is valid for a universal number that is predicted for all fluids in critical condition \cite{Kubiznak:2012wp, Mo:2016sel} in same way as in Van der Waals fluid. In reference \cite{Wei:2015iwa}, it is stated that if the microscopic structure of a thermodynamics system has discontinuous change during the during the phase transition then it shows the non-vanishing latent heat given by
\begin{equation}\label{LHeat}
L=T\Delta\nu\partial_T P=T\left(\frac{1}{n_1}-\frac{1}{n_2}\right)\partial_T P
\end{equation}
From this equation if $T=T_c$ the number densities of small and large black hole are equal i.e. $n_1=n_2$, so the latent heat vanishes and also one can’t distinguish between SBH/LBH. Here the coexistence curves are found that the slop of an $f(R)$ AdS black hole increase with increase in temperature as shown in Fig. (\ref{image-1}). To calculate the critical exponent, the reduced parameters are defined as
\begin{equation}\label{Red.Para}
\widetilde{p}=\frac{P}{P_c}\qquad \widetilde{\tau}=\frac{T}{T_c}\qquad \widetilde{\nu}=\frac{\nu}{\nu_c}\qquad\widetilde{G}=\frac{G}{G_c}
\end{equation}
So, the equation of state can be rewrite in reduced parameters. From Eq. (\ref{P2}), we have 
\begin{equation}\label{Red.P}
\widetilde{p}=\frac{4\widetilde{\tau}(d-2)}{(2d-5)\widetilde{\nu}}-\frac{(d-2)}{(d-3)\widetilde{\nu}^2}+\frac{1}{(d-3)(2d-5)\widetilde{\nu}^{2d-4}}
\end{equation}
from Eq. (\ref{P2}), we can also write as 
\begin{equation}\label{Red.T}
\widetilde{\tau}=\frac{1}{4} \left(\frac{(2 d-5) \tilde{\nu } \tilde{p}}{d-2}+\frac{2 d-5}{(d-3) \tilde{\nu }}-\frac{1}{(d-2) (d-3) \tilde{\nu }^{2 d-5}}\right)
\end{equation}
and the reduced Gibbs free energy from Eq. (\ref{FE2}) can be obtained as
\begin{equation}\label{Red.FE}
\widetilde{G}=\frac{\sqrt{b} \pi ^{\frac{d-1}{2}} q\sqrt{(d-2) (2 d-5)}}{8 \pi  \Gamma \left(\frac{d-1}{2}\right)}\left ( \tilde{\nu }^{d-3}+\frac{1}{(d-2) \tilde{\nu }^{d-3}}-\frac{3 (d-3)^2 \tilde{p} \tilde{\nu }^{d-1}}{(d-2) (d-1)^2} \right)
\end{equation}
The reduced parameters in Eq. (\ref{Red.P}), (\ref{Red.T}) are independent of factor $b$, whereas reduces free energy $\widetilde{G}$ is dependent of parameter $b$. The first order phase transition is found to occurs between small and large black hole after crossing the coexistence curve. It is found that the number density bears a sudden change, while crossing the coexistence curve \cite{Wei:2015iwa}. It is found that the number density and specific volume suffers a gap during this phenomenon. Below the two phases are denoted by script $(1, 2)$ for small and large numbers respectively. During this phase transition the specific volume and number density suffers a change. If the two states have the same Gibbs free energy then, we can write the Gibbs free energies of the two states as $\widetilde{G_1}=\widetilde{G_2}$
\begin{equation}\label{Comp.FE}
\tilde{\nu }_1^{d-3}+\frac{1}{(d-2) \tilde{\nu }_1^{d-3}}-\frac{3 (d-3)^2 \tilde{p} \tilde{\nu _1}^{d-1}}{(d-2) (d-1)^2}=\tilde{\nu _2}^{d-3}+\frac{1}{(d-2) \tilde{\nu _2}^{d-3}}-\frac{3 (d-3)^2 \tilde{p} \tilde{\nu _2}^{d-1}}{(d-2) (d-1)^2}
\end{equation}
And from the comparison of the critical temperature for the two phases 
\begin{equation}\label{Comp.T}
{\frac{2 d-5}{(d-3) \tilde{\nu }}-\frac{1}{(d-2) (d-3) \tilde{\nu }^{2 d-5}}+\frac{(2 d-5) \tilde{\nu } \tilde{p}}{d-2}=\frac{2 d-5}{(d-3) \tilde{\nu }}-\frac{1}{(d-2) (d-3) \tilde{\nu }^{2 d-5}}+\frac{(2 d-5) \tilde{\nu } \tilde{p}}{d-2}}
\end{equation}
and 
\begin{multline}\label{Add.T}
2 \tilde{\tau }=\frac{1}{4} \left(\frac{2 d-5}{(d-3) \tilde{\nu _1}}-\frac{1}{(d-2) (d-3) \tilde{\nu _1}{}^{2 d-5}}+\frac{(2 d-5) \tilde{\nu _1} \tilde{P}}{d-2}\right)+\\ \frac{1}{4} \left(\frac{2 d-5}{(d-3) \tilde{\nu _2}}-\frac{1}{(d-2) (d-3) \tilde{\nu _2}{}^{2 d-5}}+\frac{(2 d-5) \tilde{\nu _2} \tilde{P}}{d-2}\right)
\end{multline}
Consider Eq. (\ref{Comp.FE}), we can write as
\begin{equation}\label{A1}
\left ( \tilde{\nu }_1^{d-3}-\tilde{\nu _2}{}^{d-3} \right )-\frac{3 (d-3)^2 \tilde{p}}{(d-2) (d-1)^2}\left ( \tilde{\nu _1}{}^{d-1}-\tilde{\nu _2}{}^{d-1} \right )-\frac{1}{d-2}\left ( \frac{\tilde{\nu _1}{}^{d-3}-\tilde{\nu _2}{}^{d-3}}{\left(\tilde{\nu _1} \tilde{\nu _2}\right){}^{d-3}} \right )=0
\end{equation}
from Eq. (\ref{Comp.T}), we can write as 
\begin{equation}\label{A2}
\frac{(2 d-5) \tilde{p}}{d-2}\left ( \tilde{\nu }_1-\tilde{\nu }_2 \right )-\frac{2 d-5}{d-3}\left ( \frac{\tilde{\nu }_1-\tilde{\nu }_2}{\tilde{\nu }_1 \tilde{\nu }_2} \right )-\frac{1}{(d-2) (d-3)}\left ( \frac{1}{\tilde{\nu _1}{}^{2 d-5}}-\frac{1}{\tilde{\nu _2}{}^{2 d-5}} \right )=0
\end{equation}
Similarly, the sum of temperatures in reduced form for both phases, we have
\begin{equation}\label{A3}
8 \tilde{\tau }=\frac{(2 d-5) \tilde{p}}{d-2}\left ( \tilde{\nu _1}+\tilde{\nu _2} \right )+\frac{2 d-5}{d-3}\left ( \frac{\tilde{\nu _1}+\tilde{\nu _2}}{\tilde{\nu _2} \tilde{\nu _1}} \right )-\frac{1}{(d-2) (d-3)}\left ( \frac{\tilde{\nu _1}{}^{2 d-5}+\tilde{\nu _2}{}^{2 d-5}}{\left(\tilde{\nu _1} \tilde{\nu _2}\right){}^{2 d-5}} \right )
\end{equation}
Eq. (\ref{A1}) (\ref{A2}) (\ref{A3}) is a set of general solution for the reduced parameters in terms of effective volume for a d-dimensional f(R) AdS black hole. The main purpose of this solution is to find a relation for of reduced pressure and temperature at the coexistence state, which could lead us to the number densities and difference in number densities and finally to the configurational entropy using Eq. (\ref{GBent4}). We will consider the simple cases in next sections.
\section{Configuration entropy Framework}
\begin{figure}
\begin{center}
\includegraphics[width=0.6\textwidth]{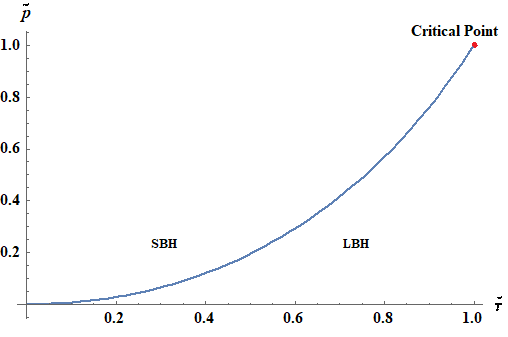}
\caption{Plot of reduced pressure $\tilde{p}$. vs reduced temperature $\tilde{\tau}$ for $d=4$ of f(R) AdS black hole. The coexistence curve (blue) and critical point (red small circle on the top of coexistence curve) is also shown for $(\tilde{p}, \tilde{\tau})$-plane}
\label{image-1}
\end{center}
\end{figure}

Taking $\tilde{\nu _1}+\tilde{\nu _2}=x$ and $\tilde{\nu _1} \tilde{\nu _2}=y$, the Eq. (\ref{A1}) (\ref{A2}) (\ref{A3}) can be written as 
\begin{equation}\label{A12}
y \tilde{p} \left(x^2-y\right)-6 y+3=0
\end{equation}
\begin{equation}\label{A22}
3 y^3 \tilde{p}+x^2-6 y^2-y=0
\end{equation}
\begin{equation}\label{A32}
16 y^3 \tilde{\tau }-3 x y^3 \tilde{p}+x \left(x^2-3 y\right)-6 x y^2=0
\end{equation}
Note that the Eq. (\ref{A12}), (\ref{A22}) and (\ref{A32}) are derived in situation where $\nu_1 \neq \nu_2$. At the critical point, this condition is not valid because at critical point $\nu_1 = \nu_2$ and  $\tilde{\tau}=1$. From Eq. (\ref{A22}), we can write as 
\begin{equation}\label{A23}
x^2=y-3 y^3 \tilde{p}+6 y^2
\end{equation}
So, from Eq. (\ref{A12}), we gets
\begin{equation}\label{A33}
3y^3 \tilde{p} \left(\tilde{p}y+2\right)-6 y+3=0
\end{equation}
Solving Eq. (\ref{A32}), (\ref{A23}) (\ref{A33}), we get the reduced pressure as
\begin{equation}\label{RPnRT}
\tilde{p}=\frac{2^{4/3} \tilde{\tau }^2 \left(\sqrt{\tilde{\tau }^2-2}-\tilde{\tau }\right)^{2/3}}{\left(\left(\sqrt{\tilde{\tau }^2-2}-\tilde{\tau }\right)^{2/3}+\sqrt[3]{2}\right)^2}
\end{equation}
This equation shows the reduced pressure as a function of reduced temperature and is independent of parameter $b$, as in Fig. (\ref{image-1}) above. It shows that the difference of the number densities between the small and large black hole decreases with the increase of reduced temperature and approaches to zero at the critical point. If we compare these results with references \cite{Wei:2015iwa, Moon:2011}, then it is clear that this is the generalized form.

Here $\tilde{p}$ depends on reduced temperature $\tilde{\tau}$. Form this, as we increase $\tilde{\tau}$ the value of $\tilde{p}$ increases. This coexistence curve of reduced pressure and temperature is independent of parameters like charge, Ricci scalar $R_0$ and parameter $b$ associated with $f'(R_0)$. Introducing the difference in number density of black hole molecule in the two phases by using $n=\frac{1}{\nu}$, we can write as
\begin{equation}\label{NDD1}
\frac{n_1-n_2}{n_c}=\frac{\sqrt{(\tilde{\nu _1}+\tilde{\nu _2})^2-4{\tilde{\nu _1}\tilde{\nu _2}}}}{\tilde{\nu _1}\tilde{\nu _2}}=\sqrt{6-6\sqrt{\tilde{p}}}
\end{equation}
Using the values of $x^2$ and $y$, we get the difference of number density as
\begin{equation}\label{NDD2}
\frac{n_1-n_2}{n_c}=\sqrt{6-\frac{6 \left(2^{2/3} \tau  \sqrt[3]{\sqrt{\tau ^2-2}-\tau }\right)}{\left(\sqrt{\tau ^2-2}-\tau \right)^{2/3}+\sqrt[3]{2}}}
\end{equation}
We see that the difference in number densities as a function of reduced temperature. This relation between the difference in number densities $\frac{n_1-n_2}{n_c}$ and reduced temperature $\tilde{\tau }$ is plotted numerically below.
\begin{figure}
\begin{center}
\includegraphics[width=0.6\textwidth]{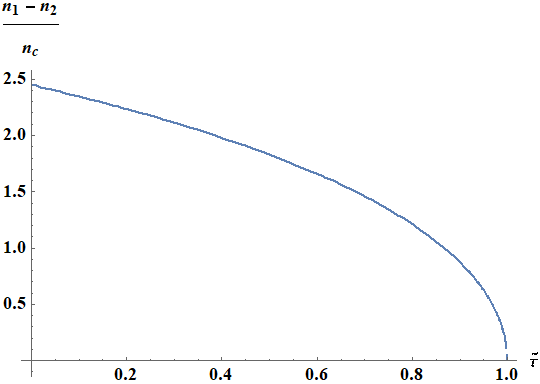}
\caption{Plot of difference between number densities of small and large molecules $\frac{n_1-n_2}{n_c}$ vs reduced temperature $\tilde{\tau }$ for $d=4$ of an f(R) AdS black hole.}
\label{image-2}
\end{center}
\end{figure}

Consider $n=\frac{1}{\nu}$, the number densities of each phase can be calculated.
From the above result, we see that the difference of number densities between small and large black hole decreases with the increase of temperature and these results are same as charged AdS black hole case. Now consider the number densities $n_1$ and $n_2$. As 
$$
n_{(1,2)}=\frac{\sqrt{(\tilde{\nu _1}+\tilde{\nu _2})^2-4{\tilde{\nu _1}\tilde{\nu _2}}}\pm(\tilde{\nu _1}+\tilde{\nu _2})}{\tilde{\nu _1}\tilde{\nu _2}}=\frac{\sqrt{x^2-4y}\pm x}{2y}
$$
\begin{equation}\label{NDDTZ}
=\frac{\left (\sqrt{3-\sqrt{\tilde{p}}}\pm \sqrt{3-3\sqrt{\tilde{p}}}  \right )\sqrt{\tilde{p}}}{\sqrt{2}}
\end{equation}
Note that here the $(+)$ sign is taken in correspondence to phase $1$ and $(-)$ sign for phase $2$. Using the value of $\tilde{p}$ from Eq. (\ref{RPnRT}), we gets
\begin{small}
\begin{equation}\label{n1n2}
n_{(1,2)}=\frac{1}{\sqrt{2}}\left ( \sqrt{3-\frac{ 2^{2/3} \tilde{\tau}  \sqrt[3]{\sqrt{\tilde{\tau} ^2-2}-\tilde{\tau}}}{\left(\sqrt{\tilde{\tau}^2-2}-\tilde{\tau}\right)^{2/3}+\sqrt[3]{2}}} \pm \sqrt{3-\frac{3\times 2^{2/3}\tilde{\tau}\sqrt[3]{\sqrt{\tilde{\tau}^2-2}-\tilde{\tau}}}{\left(\sqrt{\tilde{\tau}^2-2}-\tilde{\tau} \right)^{2/3}+\sqrt[3]{2}}} \right ){\frac{ 2^{2/3} \tilde{\tau}  \sqrt[3]{\sqrt{\tilde{\tau} ^2-2}-\tilde{\tau}}}{\left(\sqrt{\tilde{\tau}^2-2}-\tilde{\tau}\right)^{2/3}+\sqrt[3]{2}}}
\end{equation}
\end{small}

So, the configurational entropy from Eq. (\ref{GBent4}) can be written as 

\begin{align*}
-S_{con}=&\frac{\left(\sqrt{3-\sqrt{\tilde{p}}}+\sqrt{3-3 \sqrt{\tilde{p}}}\right) \sqrt{\tilde{p}}}{\sqrt{2}}\log\left(\frac{\left(\sqrt{3-\sqrt{\tilde{p}}}+\sqrt{3-3 \sqrt{\tilde{p}}}\right)}{2\sqrt{3-\sqrt{\tilde{p}}}} \right ) \\
 &+\frac{\left(\sqrt{3-\sqrt{\tilde{p}}}-\sqrt{3-3 \sqrt{\tilde{p}}}\right) \sqrt{\tilde{p}}}{\sqrt{2}}\log\left (\frac{\left(\sqrt{3-\sqrt{\tilde{p}}}-\sqrt{3-3 \sqrt{\tilde{p}}}\right)}{2 \sqrt{3-\sqrt{\tilde{p}}}} \right )
\end{align*}
or we can write as 
\begin{tiny}
\begin{equation}\label{conf-Ent}
S_{con}=-\sqrt{\frac{\tilde{p}}{2}} \left(\left(\sqrt{3-\sqrt{\tilde{p}}}+\sqrt{3-3 \sqrt{\tilde{p}}}\right)\text{log} \left(1+\frac{\sqrt{3-3 \sqrt{\tilde{p}}}}{2 \sqrt{3-\sqrt{\tilde{p}}}}\right)+\left(\sqrt{3-\sqrt{\tilde{p}}}-\sqrt{3-3 \sqrt{\tilde{p}}}\right)\text{log} \left(\frac{\sqrt{3-3 \sqrt{\tilde{p}}}}{2 \sqrt{3-\sqrt{\tilde{p}}}}-1\right) \right)
\end{equation}
\end{tiny}

Using the value of $\tilde{p}$ from Eq. (\ref{RPnRT}), we can get the result for higher dimensions $(d=4,5,6...)$ and plot the relation of $S_{con}$ vs reduce temperature $\tilde{\tau}$. From the above plot, it is clear, that for a $4$-dimensional case, the two quantities are linearly proportional to each other. Below are given for $d=4,5$ in Fig. (\ref{image-1}) and Fig. (\ref{image-2}), respectively.
\begin{figure}
\begin{center}
\includegraphics[width=0.6\textwidth]{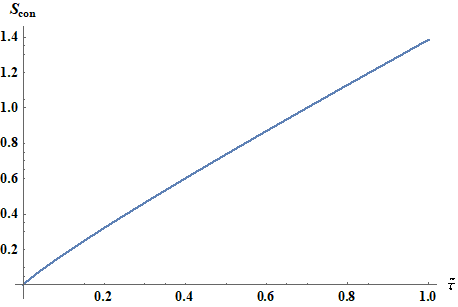}
\caption{Plot of configurational entropy vs reduced temperature for $d = 4$}
\label{image-3}
\end{center}
\end{figure}
Which shows the two quantities are approximately proportional to each other.

In such high order case, solving these equation is more complicated. So, we will consider the simple approach to avoid complexity in its solution. In previous case, we found that same to charged AdS black hole case, for $d$-dimensional f(R) black hole, reduced pressure, number density are functions of reduced temperature as shown in Fig. (\ref{image-1}) and Fig. (\ref{image-2}). The numerically calculated configurational entropy is plotted in Fig. (\ref{image-3}) below.
\begin{figure}
\begin{center}
\includegraphics[width=0.6\textwidth]{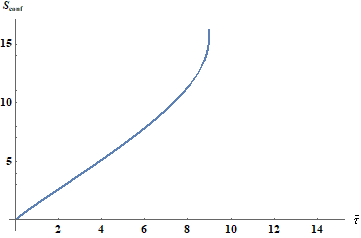}
\caption{Plot of configurational entropy vs reduced temperature for $d = 5$. This shows that as the dimension $d$ increases, the $(S_{con}-\tilde{\tau})$ plot becomes more and more concave.}
\label{image-4}
\end{center}
\end{figure}
As discussed before, that the relation between reduced pressure and reduced temperature is found to increases monotonically and the number density decreases with reduced temperature. From Eq. (\ref{A2}) and (\ref{A3}) at $\tilde{\tau}=0$, we can write the difference in number densities of SBH/LBH as
\begin{equation}\label{DD-d}
\frac{n_1-n_2}{n_c}=\left((d-3) (2 d-5) \tilde{p}\right)^{\frac{1}{2 (d-3)}}-((d-2) (2 d-5))^{\frac{1}{2 (d-3)}}
\end{equation}
\begin{figure}
\begin{center}
\includegraphics[width=0.6\textwidth]{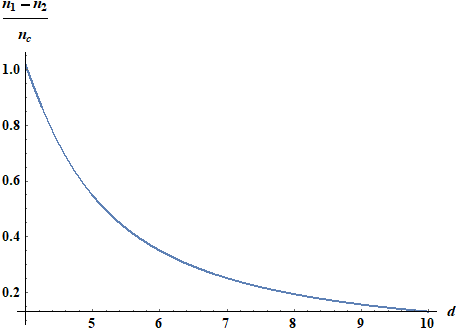}
\caption{The plot of difference in number densities $\frac{n_1-n_2}{n_c}$ of small and large f(R) AdS black hole vs dimension $d$.}
\label{image-5}
\end{center}
\end{figure}
This result show that change in number density of small and large black hole is not same.
\section{Discussions and conclusions}
In this chapter, we consider a d-dimensional f(R) AdS black hole for accounting the effects of modified gravity parameter b on the investigation of coexistence curves, difference of number densities for small and large f(R) black holes and configuration entropy.

We consider $d$-dimensional f(R) AdS black hole for investigating the thermodynamic quantities and reduced parameters. The reduced parameters of reduced pressure $\tilde{p}$ and temperature $\tilde{\tau}$ are independent of factor $b$, whereas reduces free energy $\tilde{G}$ is dependent of parameter $b$, that is also discussed in \cite{Mo:2016sel} for a $4$-dimensional f(R) AdS black hole and in \cite{Lee:2017ero} for a charged AdS black hole as well. From the expressions of reduced parameters $(\tilde{p}, \tilde{\tau}, \tilde{G})$. Employing these expressions, we found that for an f(R) AdS black hole the reduce pressure as a function of reduced temperature an independent of parameter $b$. So, the result obtained is consistent with those of Reissner Nordstrom and charged AdS black hole. We plot the coexistence curve of $(\tilde{p}-\tilde{\tau})$ in Fig. (\ref{image-1}). We investigate the difference in molecule number density of small and large black hole as a function of reduced temperature and is plotted in Fig. (\ref{image-2}), which shows that for an f(R) AdS black hole the difference in molecule number densities of SBH/LBH decreases with the increase in reduced temperature. Evaluating the individual number densities of SBH/LBH, we calculate the configurational entropy in Eq. (\ref{conf-Ent}). In addition to Eq. (\ref{RPnRT}), the configurational is also a function of $\tilde{\tau}$. The general result is plotted for five dimensions and it is found that as long as the dimension increases, the $(S_{con}-\tilde{\tau})$ plot becomes more and more concave. Finally, we consider the thermodynamic character and discussed the difference in number densities of small and large f(R) AdS black holes and plot the result in Fig. (\ref{image-5}). Which shows that as the dimension d increases the difference in number densities of small and large f(R) AdS black holes decreases.

\newpage
\clearpage
\thispagestyle{empty}
\hfill
\chapter{Summary and Outlook}
For summarizing  the work done in this thesis, I would like to present the detailed summary and the analysis carried out for the desired conclusions. Here, it is important to mention that the motivation of the present thesis is to analyze the volume and entropy evolution under Hawking radiation in the interior of black holes. The main work discussed in this thesis is based on references \cite{CR:2014, Zhang:2015, Yang:2018arj, Wang:2018dvo}. We also followed the work done in \cite{Majhi:2017tab} and found the corrected entropy, which is consistent with \cite{Zhang:2015}. We also extend the study to express the phase transition, the critical behaviour occurring in AdS black holes from various perspectives, as well as configuration entropy. Such analyses are important, since it could open a gateway to a new realm of black hole physics as well as the theory of quantum gravity and the solution to the problem of information paradox.
In classical physics, a black hole is compact object with strong gravitational attraction such that nothing can come out from it. Even though light also couldn't come out from it. Any black hole has 1-dimensional point like singularity in its center as the place where all the known laws of physics formulated on a classical background are  break downs. This was first proposed by Roger Penrose in 1965. Inside a black hole the space-time is coupled to infinity. Black holes are characterized by their boundary called event horizon. Which behaves as a trapped surface such that all future directed null geodesics orthogonal are converging to it. Actually, horizon is a null surface that separates the interior and exterior space times of black holes. In 1970 Brandon Carter and Stephen Hawking proved the No-Hair theorem, which states that all the black hole solutions of Einstein-Maxwell equation can be characterized by three physical quantities mass $M$, charge $Q$ and angular momentum $J$. all other information disappears behind the horizon.  In 1972, Hawking proposed, a black hole can emit radiations due to quantum effects so, It has temperature and entropy. The rate of emitting radiations is a quasi-static process, but with non-zero temperature. Bardeen et.al. derived the four laws of black hole thermodynamics, which are analogous to the common thermodynamic laws. The questions which are still need to be answer, are the black holes information paradox, the microscopic origin of entropy and final state of evaporation.

As no information can be retrieved form black hole due to existance of horizon and singularity. Which is a place for breakdown of physical laws inside it. So, in classical physics, it is not easy to find the volume enclosed by the black hole horizon. Inside a black hole the space-time is coupling to infinity and also the existence of event horizon as barrier between the two space times make the situation ambiguous . In curved space-time finding the interior volume first effort was made by M.K. Parikh. He proposed that the interior volume of black hole is independent of slicing. \cite{Parikh:2005qs} . He also added that if one could construct the bound area and unbound volume then on can find the interior volume of black hole. For a  black hole in four dimensions the interior volume turns out to be $\frac{4}{3}\pi R^3$. In Ref.  \cite{DiNunno:2008id},  B. DiNunno and R. A. Matzner proposed the black hole as null surface and investigate the interior volume of black holes. Recently CR reviewed these literature and proposed an advance definition for the black hole interior. According to them the interior volume of a sphere $S$ in Minkowski space-time of radius $R$, it is very convenient to compute the volume $V$ as the volume of  collapsed  spherically symmetric sphere inside a 2-sphere, which is given by $\frac{4}{3}\pi R^3$. If there are many hyper-surfaces, then for choosing the hyper-surface one need two equivalent statements as given below.
\begin{itemize}
\item{The hyper-surface $\sum$ chosen must be lie on the same simultaneity surface as $S$}
\item{The hyper-surface $\sum$ must be the largest spherically symmetric bounded surface of $S$.}
\end{itemize}
 
Such a hyper-surface will bound the largest possible volume $V$ as compared to any other hyper-surface in the same space-time.  
In CR work, following the proposed definition they obtained the interior volume as
\begin{equation}\label{CRvol}
V_{CR}=3\sqrt{3}\pi M^2 \nu
\end{equation}

Where $\nu$ is the advance time. The horizon is an out going null surface but the area remain constant. The special property of this equation is that the interior volume is linearly increases with advance time $\nu$. The interior of black hole don't lost long, because only time like geodesic can reaches to the singularity in proper time of order m and space-like region fits in it which is very large. it means that a black could have large space to store all in-going in formations. They also extend their result to the Reissner Nordstr$\ddot{o}$m black hole and found a consistent result. Following the CR work, Baocheng Zhang investigate the entropy of scalar modes for the interior quantum modes of Schwarzschild black hole using standard statistical method. They found, that the entropy of scalar modes is also proportional to surface area of black hole as given below
\begin{equation}\label{BZent}
S_{CR}={\frac{3\sqrt{3}\gamma}{(90\times8^4)\pi}}A
\end{equation}
Where $A=16\pi M^2$ is the surface area of black hole. The proportionality constant is much smallar then that Bekenstein entropy, which means that most of information from an in-going object resides the horizon of a black hole. The word information is used in context of all physical quantities, which are related to the specific quantum state of a single particle in a black hole. These quantities may be mass, of temperature, length, spin, position and so many other, which you could name it. The quantum field theory predicts that all the in-going information are conserved and are not lost inside a black hole, but this problem still has no proof in its physical shape.

Following the above discussion of CR volume and Baocheng Zhang entropy analysis, Here we investigate several problems regarding the interior entropy of black hole, and black hole evaporation due to Hawking Radiations. The main proposal of our investigations is based on the relation of interior volume and entropy with advance time $\nu$. This means that black hole could have a large volume to store information containing quantum modes. The entropy of these quantum modes is directly related to the Bekenstein Hawking entropy. So, if one could determine the entropy quantum modes in the interior of black hole and construct the evolution relation between the entropy of in the interior quantum modes and Bekenstein Hawking entropy under Hawking radiation, this may give a reasonable solution for the problem of information paradox. In the following discussion, we summarizes the structure of contents in this thesis:

In first chapter, we briefly discussed the concept of black holes. Starting from the Einstein theory of general relativity and field equation, we discussed different geometries as solution of field equation. We also discussed the Types of black holes and black hole thermodynamics and evaporation process.

In Second chapter, we followed the CR and Baocheng Zhang work using different black holes. The main proposal of our investigations is based on linear relations of interior volume and entropy with advance time as in Eq. (\ref{CRvol}) and (\ref{BZent}). This means that black hole could have a large volume containing quantum modes to store the lost information. The entropy of these quantum modes is directly related to the Bekenstein Hawking entropy with constant of proportionality less than one  (i.e. most of the information of an in-going objects are lost on the horizon of the black holes). If one could determine the entropy of these quantum modes in the interior of black hole and construct the evolution relation between the entropy of in the interior quantum modes with Bekenstein Hawking entropy under Hawking radiation, then this may give a reasonable solution for the problem of information paradox.

The main aim of second chapter is to revise the concept on which this thesis is based i.e. interior volume and entropy of different black holes as discussed in \cite{CR:2014, Zhang:2015}. We extend the concept to other types of black holes and also made comparison among them, in order to take out the effect on the interior entropy due to difference in geometry of black hole. Structure of our contents is such that, in second section we evaluate the interior volume of different black holes. For the entropy calculation of the scalar modes in the interior of black hole, we followed the analogy of Baocheng Zhang. We  also made discussion on the entropy correction by following the semi-classical way as followed by B. Majhi and S. Samanta. To discuss the black hole evaporation, we consider the two assumptions and calculate the differential form of entropy in different black holes and proportional relation between the entropy of the scalar modes and the Bekenstein Hawking Entropy is calculated.

In third chapter, we consider $d$-dimensional charged black hole to calculate the interior volume and entropy by following the analogy discussed in chapter. 2. Next, using the black hole emission rate as quasi-static process and black hole radiation as black body radiations to obtain the differential form of this entropy using Stefan Boltzmann law. Comparing the entropy of interior modes to the Bekenstein Hawking entropy, we found a proportional relation between the entropy's. The relation of proportionality relation vs dimensions is numerically analyzed, which shows that the proportional relation $(\frac{d S_{CR}}{dS_{BH}})$ decreases with the increase of dimensions $d$.

In fourth chapter Based on the 4-dimensional black hole solution in modified gravity theory coupled to a non-linear Maxwell field theory, we estimate the interior volume of a charged $f(R)$ black hole using the method proposed by Christodoulou and Rovelli. Next, considering massless scalar field in the interior volume and Hawking radiation carrying only energy, we calculate the entropy of the scalar field in the interior of charged $f(R)$ black hole and investigate the evolution of the entropy under Hawking radiation. In the meantime, the evolution of the Bekenstien-Hawking entropy under Hawking radiation has also been calculated. Based on these results, the proportional relation is obtained between the evolution of the scalar field entropy and the evolution of Bekenstein-Hawking entropy under Hawking radiation. According to the result, we investigate and discuss how the modified gravity coefficient $b$ in $f(R)$ gravity theory and its affects in the evolution relation between the two types of entropy. It is shown that the radiation rate for Hawking radiation of a charged $f(R)$ black hole can increase with the modified coefficient $b$.

In fifth chapter, considering phase transition of a $d$-dimensional $f(R)$ AdS black holes, we analyzed the effects of factor $b$ associated with $f'(R)$ on the phase transition and critical point parameters $(P_c, T_c, \nu_c, G_c)$, coexistence curves, difference in number densities of small and large black holes (SBH/LBH) $(\frac{n_1-n_2}{n_c})$ and configurational entropy $S_{con}$. At critical point, the pressure and temperature are found to be independent of parameter $b$, whereas the free energy $G$ is b dependent. A consistent result is also found for reduced parameters. Following the analogy of black hole first order phase transition, we found our results in agreement with Reissner Nordstrom and charge AdS black hole. For black hole crossing the critical point are found to have vanishing latent heat and become indistinguishable for and observer. Further analysis revealed that the parameter  $b$ doesn't have any effect on the coexistence curve, which shows that the our results are consistent with that of Reissner Nordstrom black hole and trick following is correct with the fact. Finally, we also investigate the relation between difference in number density and dimension $d$. Finally, as concluding remarks, we would like to make the following comments on the work done in this thesis.
\begin{itemize}
\item The volume inside a $2$-sphere say $S$ is the volume of the largest space-like spherically-symmetric surface $\sum$ bounded by $S$. A black hole may have many hyper-surfaces inside it so, the largest hyper-surface must be such that it surrounds the maximum interior volume.
\end{itemize}

\begin{itemize}
\item The interior volume and entropy of a black hole increases linearly with advance time. So, it will have large room for storing information inside it.  The connection between the black hole interior entropy and Bekenstein Hawking entropy always showed a constant relation in  them. Where the proportionality constant is always less than one. It is also found interior entropy to be less than that of horizon entropy. It means most of the black hole information reside on the horizon of black hole.
\item In case of higher dimensional black holes, the proportional relation between the two entropy's are found to be decreasing with the increase of dimensions and also the variation of entropy in the interior of black hole is much smaller as compared to the variation of Bekenstein-Hawking entropy. This means that the Bekenstien entropy rapidly decreases with the increase of dimensions .
\item A form Eq. (\ref{BZent}), the black hole entropy is proportional to its surface area, but the coefficient of area is much less then that of $\frac{1}{4}$. It seems that it will not satisfy the first law of black hole thermodynamics. For balancing the first law of black hole thermodynamics, it must have some extra term in it so that it will balance the first law of black hole thermodynamics
\item In the case of modified gravity theories, the proportional relation of the two entropy,s decreases with the increase of modified gravity factor b, as discussed above with detail in their numerical analysis along with possible results using b as$(\pm b,o)$. In other words, the radiation rate of Hawking radiation can enhance with the increasing coefficient b.
\item In modified gravity theories, the growth rate of the scalar field entropy in the interior volume of the black hole is much less than the decreasing rate of Bekenstein-Hawking entropy under Hawking radiation.
\item Considering the modified gravity theory, we extend the notion of phase transition to the d-dimensional f(R) black hole and found that it undergoes the same phase transition as that of Reissner Nordstr$\ddot{o}$m  black hole.
\item The reduced pressure and temperature are found to be independent of modified gravity coefficient b, while the free energy is its function. the relations are plotted. This context will explore the re-normalization group scheme to study the black hole phase transition via critical phenomena and configuration entropy.
\end{itemize}

\clearpage
\section*{Appendices}
\addcontentsline{toc}{section}{Appendices}
\section*{A}\label{app.A}
\addcontentsline{toc}{section}{A}

The Klein-Gordon equation in curved space-time is,
$$\frac{1}{\sqrt{-g}}{{\partial_ \mu }\left(\sqrt{-g} g^{\mu \nu } \partial_\nu \Phi \right)}=0$$
Expanding this equation in $(T,\lambda, \theta, \phi)$, we get the general equation as
$$\frac{1}{\sqrt{-g}}{{\partial_ T }\left(\sqrt{-g} g^{T \nu } \partial_\nu \Phi \right)}+\frac{1}{\sqrt{-g}}{{\partial_ \lambda }\left(\sqrt{-g} g^{\lambda  \nu } \partial_\nu \Phi \right)}+\frac{1}{\sqrt{-g}}{{\partial_ \theta }\left(\sqrt{-g} g^{\theta  \nu } \partial_\nu \Phi \right)}+\frac{1}{\sqrt{-g}}{{\partial_ \phi }\left(\sqrt{-g} g^{\phi  \nu } \partial_\nu \Phi \right)}=0$$
Where we used
$$\partial\Phi_\nu=\partial_ T  \Phi+\partial_{\lambda} \Phi+\partial_ {\theta} \Phi+\partial_ \phi \Phi$$
First solving $(\partial_ T  \Phi, \partial_{\lambda} \Phi, \partial_ {\theta} \Phi, \partial_ \phi \Phi)$, we get
$$\partial_ T  \Phi=\partial_ T (e^{\text{iET}}e^{\text{iI}(\lambda ,\theta ,\phi )})=-i E\Phi$$
$$\partial_{\lambda} \Phi=\partial_{\lambda} (e^{\text{iET}}e^{\text{iI}(\lambda ,\theta ,\phi )})=i I\partial_{\lambda} \Phi$$
$$\partial_ {\theta} \Phi=\partial_ {\theta} (e^{\text{iET}}e^{\text{iI}(\lambda ,\theta ,\phi )})=i I\partial_ {\theta} \Phi$$
$$\partial_ \phi \Phi)=\partial_ \phi (e^{\text{iET}}e^{\text{iI}(\lambda ,\theta ,\phi )})=i I \partial_ \phi \Phi$$
So, from above we can write Scalar field as 
$$\partial_\nu \Phi=i (E^2+\partial_{\lambda} I+\partial_ {\theta} I+\partial_ \phi I)\Phi$$
The first part of the above general equation can be written as 
$$\frac{1}{\sqrt{-g}}{{\partial_ T }\left(\sqrt{-g} g^{T \nu } \partial_\nu \Phi \right)}=\frac{1}{\sqrt{-g}}{{\partial_ T }\left ( \sqrt{-g}\left ( g^{TT} +g^{T\lambda}+g^{T\theta}+g^{T\phi}\right )\partial_\nu \Phi  \right )}$$
$$\frac{1}{\sqrt{-g}}{{\partial_ T }\left(\sqrt{-g} g^{T \nu } \partial_\nu \Phi \right)}=\frac{1}{\sqrt{-g}}{{\partial_ T }\left ( \sqrt{-g}\left ( g^{TT} +g^{T\lambda}+g^{T\theta}+g^{T\phi}\right )\times i (E^2+\partial_{\lambda} I+\partial_ {\theta} I+\partial_ \phi I)\Phi \right)}$$
Where for the metric Eq. (\ref{EDmet+T}), the coordinates are $g^{TT}=-1, g^{T\lambda}=g^{T\theta}=g^{T\phi}=0$
$$=\frac{1}{\sqrt{-g}}{{\partial_ T }\left ( \sqrt{-g}\left ( -1 +0+0+0\right )\times i (E^2+\partial_{\lambda} I+\partial_ {\theta} I+\partial_ \phi I)\Phi \right)}$$
or we can write as 
$$\frac{1}{\sqrt{-g}}{{\partial_ T }\left(\sqrt{-g} g^{T \nu } \partial_\nu \Phi \right)}=- E^2\Phi$$
Similarly, the other parts of the general equation are
$$\frac{1}{\sqrt{-g}}{{\partial_ \lambda }\left(\sqrt{-g} g^{\lambda  \nu } \partial_\nu \Phi \right)}=\frac{1}{\sqrt{-g}}{{\partial_ \lambda }\left ( \sqrt{-g}\left ( g^{\lambda T} +g^{\lambda\lambda}+g^{\lambda\theta}+g^{\lambda\phi}\right )\times i (E^2+\partial_{\lambda} I+\partial_ {\theta} I+\partial_ \phi I)\Phi \right)}$$
Here $g^{\lambda T}=0, g^{\lambda\lambda}=\frac{1}{\left(-f(r) \dot{\nu }^2+2 \dot{\nu } \dot{r}\right)} ,g^{\lambda\theta}=0, g^{\lambda\phi}=0$ so,
$$\frac{1}{\sqrt{-g}}{{\partial_ \lambda }\left(\sqrt{-g} g^{\lambda  \nu } \partial_\nu \Phi \right)}=\frac{1}{\left(-f(r) \dot{\nu }^2+2 \dot{\nu } \dot{r}\right)} \partial_ \lambda ^2 I \Phi$$

$$\frac{1}{\sqrt{-g}}{{\partial_ \theta }\left(\sqrt{-g} g^{\theta  \nu } \partial_\nu \Phi \right)}=\frac{1}{\sqrt{-g}}{{\partial_ \theta }\left ( \sqrt{-g}\left ( g^{\theta T} +g^{\theta\lambda}+g^{\theta\theta}+g^{\theta\phi}\right )\times i (E^2+\partial_{\lambda} I+\partial_ {\theta} I+\partial_ \phi I)\Phi \right)}$$
Here $g^{\theta T}=0, g^{\theta\lambda}=0 ,g^{\theta\theta}=\frac{1}{r^2}, g^{\theta\phi}=0$ so,
$$=\frac{1}{\sqrt{-g}}{{\partial_ \theta }\left ( \sqrt{-g}\left ( 0 +0+\frac{1}{r^2}+0\right )\times i (E^2+\partial_{\lambda} I+\partial_ {\theta} I+\partial_ \phi I)\Phi \right)}$$
$$\frac{1}{\sqrt{-g}}{{\partial_ \theta }\left(\sqrt{-g} g^{\theta  \nu } \partial_\nu \Phi \right)}=\frac{1}{r^2}\partial^2 _ {\theta}I\Phi$$

and 
$$\frac{1}{\sqrt{-g}}{{\partial_ \phi }\left(\sqrt{-g} g^{\phi  \nu } \partial_\nu \Phi \right)}=\frac{1}{\sqrt{-g}}{{\partial_ \phi }\left ( \sqrt{-g}\left ( g^{\phi T} +g^{\phi\lambda}+g^{\phi\theta}+g^{\phi\phi}\right )\times i (E^2+\partial_{\lambda} I+\partial_ {\theta} I+\partial_ \phi I)\Phi \right)}$$

Here $g^{\phi T}=0, g^{\phi\lambda}=0 ,g^{\phi\theta}=0, g^{\phi\phi}=\frac{1}{r^2sin^2 \theta}$ so,
$$=\frac{1}{\sqrt{-g}}{{\partial_ \phi }\left ( \sqrt{-g}\left ( 0 +0+0+\frac{1}{r^2 sin^2 \theta}\right )\times i (E^2+\partial_{\lambda} I+\partial_ {\theta} I+\partial_ \phi I)\Phi \right)}$$

$$\frac{1}{\sqrt{-g}}{{\partial_ \phi }\left(\sqrt{-g} g^{\phi \nu } \partial_\nu \Phi \right)}=\frac{1}{r^2 sin^2 \theta}\partial^2 _ {\phi}I\Phi$$
using the above results, the general solution of Klein-Gordon equation in curved space-time for $(T,\lambda, \theta, \phi)$ can be written as 
$$- E^2\Phi+\frac{1}{\left(-f(r) \dot{\nu }^2+2 \dot{\nu } \dot{r}\right)} \partial_ \lambda ^2 I \Phi+\frac{1}{r^2}\partial^2 _ {\theta}I\Phi+\frac{1}{r^2 sin^2 \theta}\partial^2 _ {\phi}I\Phi=0$$
using $\partial_ \lambda ^2 I=p_\lambda ^2, \partial^2 _ {\theta}I=p_\theta ^2, \partial^2 _ {\phi}I=p_\phi ^2$, we gets 
$$-E^2+\frac{1}{\left(-f(r) \dot{\nu }^2+2 \dot{\nu } \dot{r}\right)} p_\lambda ^2 +\frac{1}{r^2}p_\theta ^2+\frac{1}{r^2 sin^2 \theta}p_\phi ^2=0$$
Which is the Eq. (\ref{eqofenergy})

\section*{B}\label{app.B}
\addcontentsline{toc}{section}{B}
$$\dot{x}^a=\frac{dx^a}{d\tau}\Rightarrow u^a=\{x^a, H_T\}$$
Here $H_T=\zeta(\tau)(P^2-M^2)$
$$\dot{x}^a=\{x^a, \zeta(\tau)(P^2-M^2)\}$$
$$=\{x^a, \zeta(\tau)P^2-0\}$$
$$=\{x^a, \zeta(\tau)P^2-0\}$$
Using the property
$$\{x^a, H_T\}=\sum \left(\frac{\partial x^a}{\partial x^a}\frac{\partial H_T}{\partial P^a}-\frac{\partial x^a}{\partial P^a}\frac{\partial H_T}{\partial x^a} \right)$$
Using this the above expression can be write as 
$$\dot{x}^a=2\zeta\{x^a, P_a\}P^a=2\zeta P^a$$
$$\dot{P}^a=\{P^a, H_T\}=-\frac{\partial H_T}{\partial x^a}$$
using the valve of $H$, we gets
$$\dot{P}^a=-\zeta \frac{\partial}{\partial x^a}  \left(g^{bc}P_b P_c\right)=-\zeta \frac{\partial g^{bc}}{\partial x^a}  P_b P_c$$
\section*{C}\label{app.C}
\addcontentsline{toc}{section}{C}
$$\dot{P}^0=\{P^0, H_T\}=-\zeta \frac{\partial}{\partial x^a} g^{bc} P_b P_c$$

$$=-\zeta \frac{\partial g^{bc}}{\partial x^a}  P_b P_c-2\zeta g_{bc}\frac{\partial P_b}{\partial x^a}  P_b P_c$$
Here, we used the product rule for the second term as
$$g_{bc}\frac{\partial P_b}{\partial x^a}  P_c=\frac{\partial g^{bc} P_b}{\partial x^a}  P_c-\frac{\partial g^{bc}}{\partial x^a} P_b P_c$$

$$\dot{P}^0=\{P^0, H_T\}=\zeta \left( \frac{\partial g^{bc}}{\partial x^a}  P_b P_c-2 g_{bc}\frac{\partial P_b}{\partial x^a}  P_b P_c \right)$$
\section*{D}\label{app.D}
\addcontentsline{toc}{section}{D}
As the general equation for the constraint metric in Eq. (\ref{constmetgenrleq}) is
$$C_{AB}=\left\{\phi _A,\phi _B\right\}, \qquad (A,B)=(1,2)$$
so, 
$$C_{11}=\left\{\phi _1,\phi _1\right\}=0$$

$$C_{12}=\left\{\phi _1,\phi _2\right\}=\{P^2-M^2, \frac{P^0}{M}\tau-x^0\}$$

$$=\{P^2, \frac{P^0}{M}\tau\}-\{P^2,x^0\}$$
$$=\frac{\tau}{M}\{P^2, P^0\}+\frac{P^0}{M}\{P^2, \tau\}-\{P^2,x^0\}=-2P^0$$

Similarly as $C_{12}=-C_{21}$ and 
$$C_{22}=\left\{\phi _2,\phi _1\right\}=0$$
So, the constraint metric is given in Eq.(\ref{constmet}).

\clearpage
\thispagestyle{empty}
\hfill
\clearpage
\newpage

\thispagestyle{empty}


\begin{thebibliography}{99}
\addcontentsline{toc}{section}{References}



\bibitem{Weber}   
   J.~Weber,
   ``General Relativity and gravitational waves,''
   Courier Corporation. (2004).
   
\bibitem{Einstein}
   A.~Einstein, N.~Rosen, 
   ``The particle problem in the general theory of relativity,''   
   Physical Review, 48(1), 73. (1935). 
   
\bibitem{John}
    John Lighton Synge
    ``Relativity: the special theory'' 
    Prabhat Prakash, (1964).
 
\bibitem{rindler1991}
  Rindler, Wolfgang
  ''Introduction to special relativity. 2,''
  ERIC, (1991).
 
\bibitem{Wald:1984rga} 
  R.~M.~Wald,
  ``General Relativity,''
  doi:10.7208/chicago/9780226870373.001.0001, (1984).
  
\bibitem{Carroll:2004st} 
  S.~M.~Carroll,
  ``Spacetime and geometry: An introduction to general relativity,''
  San Francisco, USA: Addison-Wesley, 513 p, (2004).
 
\bibitem{Stephani:2004ud} 
  H.~Stephani,
  ``Relativity: An introduction to special and general relativity,''
  Cambridge, UK: Univ. Pr. 396p, (2004)

\bibitem{Misner:1974qy} 
  C.~W.~Misner, K.~S.~Thorne and J.~A.~Wheeler,
  ``Gravitation,''
  San Francisco, 1279 p, (1973).

\bibitem{Gary}
  G.~T.~Horowitz,
  ''Black Holes in Higher Dimensions''
  ISBN(1107013453,9781107013452)
  Cambridge University Press, (2012)
  
\bibitem{Visser:2007fj} 
  M.~Visser,
  arXiv:0706.0622 [gr-qc].
  
\bibitem{Heinicke:2015iva} 
  C.~Heinicke and F.~W.~Hehl,
  Int.\ J.\ Mod.\ Phys.\ D {\bf 24}, no. 02, 1530006 (2014).
  doi:10.1142/S0218271815300062
  [arXiv:1503.02172 [gr-qc]].
  
\bibitem{ruffini1971}
  Remo Ruffini and John A.~ Wheeler
  Physics Today,
  ERIC, (1971).

\bibitem{Novikov:1989sz} 
  I.~D.~Novikov and V.~P.~Frolov,
  Fundam.\ Theor.\ Phys.\  {\bf 27} (1989).
  doi:10.1007/978-94-017-2651-1

\bibitem{Hawking:1974sw} 
  S.~W.~Hawking,
  Commun.\ Math.\ Phys.\  {\bf 43}, 199 (1975)
  Erratum: [Commun.\ Math.\ Phys.\  {\bf 46}, 206 (1976)].
  doi:10.1007/BF02345020, 10.1007/BF01608497
  
\bibitem{Hawking:1971tu} 
  S.~W.~Hawking,
  Phys.\ Rev.\ Lett.\  {\bf 26}, 1344 (1971).
  doi:10.1103/PhysRevLett.26.1344
  
\bibitem{Hawking:1976de} 
  S.~W.~Hawking,
  Phys.\ Rev.\ D {\bf 13}, 191 (1976).
  doi:10.1103/PhysRevD.13.191
  
\bibitem{Li:2018bny} 
  C.~Li, C.~Fang, M.~He, J.~Ding, P.~Li and J.~Deng,
  arXiv:1812.02567 [hep-th].

\bibitem{Bekenstein:1972tm}
J.~D.~Bekenstein,
  Lett.\ Nuovo Cim.\  {\bf 4}, 737 (1972).
  doi:10.1007/BF02757029

\bibitem{Wald:1999vt} 
  R.~M.~Wald,
  Living Rev.\ Rel.\  {\bf 4}, 6 (2001)
  doi:10.12942/lrr-2001-6
  [gr-qc/9912119].

\bibitem{Ross:2005sc} 
  S.~F.~Ross,
  hep-th/0502195, (2005)
  
\bibitem{Davies:1978mf} 
  P.~C.~W.~Davies,
  Proc.\ Roy.\ Soc.\ Lond.\ A {\bf 353}, 499 (1977).
  doi:10.1098/rspa.1977.0047

\bibitem{Rovelli:1997yv} 
  C.~Rovelli,
  Living Rev.\ Rel.\  {\bf 1}, 1 (1998)
  doi:10.12942/lrr-1998-1
  [gr-qc/9710008].

\bibitem{Smolin:2004xb} 
  L.~Smolin,
  In *Barrow, J.D. (ed.) et al.: Science and ultimate reality* 492-527, (2004).

\bibitem{Emparan:2008eg} 
  R.~Emparan and H.~S.~Reall,
  Living Rev.\ Rel.\  {\bf 11}, 6 (2008)
  doi:10.12942/lrr-2008-6
  [arXiv:0801.3471 [hep-th]].

\bibitem{Reall:2015esa} 
  H.~S.~Reall,
  Int.\ J.\ Mod.\ Phys.\ D {\bf 21}, 1230001 (2012)
  doi:10.1142/S0218271812300017
  [arXiv:1210.1402 [gr-qc]].
  
\bibitem{Strominger:1996sh} 
  A Strominger and C Vafa,
  Phys.\ Lett.\ B {\bf 379}, 99 (1996)
  doi:10.1016/0370-2693(96)00345-0
  [hep-th/9601029].

\bibitem{dehghani:2009gu} 
   M.~Dehghani and A.~Farmany,
  Brazilian Journal of Physics, {\bf 39}, (3), pp. 570-573,(2009).
  
\bibitem{Hawking:1974rv} 
  S.~W.~Hawking,
  Nature {\bf 248}, 30 (1974).
  doi:10.1038/248030a0
  
\bibitem{jacobson1996}  
  Ted Jacobson,
  Given at Utrecht U. in, {\bf 26},, (1996).
 
\bibitem{Hartman:2008b} 
  T.~Hartman, K.~Murata, T.~Nishioka and A.~Strominger,
  JHEP {\bf 0904}, 019 (2009)
  doi:10.1088/1126-6708/2009/04/019
  [arXiv:0811.4393 [hep-th]].

\bibitem{Page:1977um} 
  D.~N.~Page,
  Phys.\ Rev.\ D {\bf 16}, 2402 (1977).
  doi:10.1103/PhysRevD.16.2402
  
\bibitem{Zhang:2019pzd} 
  M.~Zhang,
  Phys.\ Lett.\ B {\bf 790}, 205 (2019)
  doi:10.1016/j.physletb.2019.01.032
  [arXiv:1901.04128 [gr-qc]].

\bibitem{CR:2014}
  M.~Christodoulou and C.~Rovelli,
  Phys.\ Rev.\ D {\bf 91}, no. 6, 064046 (2015)
  doi:10.1103/PhysRevD.91.064046
  [arXiv:1411.2854 [gr-qc]].

\bibitem{Ong:2015tua} 
  Y.~C.~Ong,
  JCAP {\bf 1504}, no. 04, 003 (2015)
  doi:10.1088/1475-7516/2015/04/003
  [arXiv:1503.01092 [gr-qc]].

\bibitem{Bengtsson:2015zdaa} 
  I.~Bengtsson and E.~Jakobsson,
  Mod.\ Phys.\ Lett.\ A {\bf 30}, no. 21, 1550103 (2015)
  doi:10.1142/S0217732315501035
  [arXiv:1502.01907 [gr-qc]].

\bibitem{Ong:2015}
  Y.~C.~Ong,
  Gen.\ Rel.\ Grav.\  {\bf 47}, no. 8, 88 (2015)
  doi:10.1007/s10714-015-1929-x
  [arXiv:1503.08245 [gr-qc]].

\bibitem{DiNunno:2008id} 
  B.~S.~DiNunno and R.~A.~Matzner,
  Gen.\ Rel.\ Grav.\  {\bf 42}, 63 (2010)
  doi:10.1007/s10714-009-0814-x
  [arXiv:0801.1734 [gr-qc]].

\bibitem{Bhaumik:2016sav} 
  N.~Bhaumik and B.~R.~Majhi,
  Int.\ J.\ Mod.\ Phys.\ A {\bf 33}, no. 02, 1850011 (2018)
  doi: 10.1142/S0217751X18500112
  [arXiv:1607.03704 [gr-qc]].

\bibitem{Yang:2018arj} 
  J.~Z.~Yang and W.~B.~Liu,
  Phys.\ Lett.\ B {\bf 782}, 372 (2018).
  doi:10.1016/j.physletb.2018.05.050

\bibitem{Zhang:2015}
  B.~Zhang,
  Phys.\ Rev.\ D {\bf 92}, no. 8, 081501 (2015)
  doi:10.1103/PhysRevD.92.081501
  [arXiv:1510.02182 [gr-qc]].

\bibitem{Majhi:2017tab} 
  B.~R.~Majhi and S.~Samanta,
  Phys.\ Lett.\ B {\bf 770}, 314 (2017)
  doi: 10.1016/j.physletb.2017.05.003
  [arXiv:1703.00142 [gr-qc]].

\bibitem{Hossenfelder:2012jw} 
  S.~Hossenfelder,
  Living Rev.\ Rel.\  {\bf 16}, 2 (2013)
  doi:10.12942/lrr-2013-2
  [arXiv:1203.6191 [gr-qc]].

\bibitem{Parikh:2005qs} 
  M.~K.~Parikh,
  Phys.\ Rev.\ D {\bf 73}, 124021 (2006)
  doi:10.1103/PhysRevD.73.124021
  [hep-th/0508108].

\bibitem{Han:2018}
  S.~Z.~Han, J.~Z.~Yang, X.~Y.~Wang and W.~B.~Liu,
  Int.\ J.\ Theor.\ Phys.\  {\bf 57}, no. 11, 3429 (2018).
  doi:10.1007/s10773-018-3856-6

\bibitem{Wang:2018dvo} 
  X.~Y.~Wang, J.~Jiang and W.~B.~Liu,
  Class.\ Quant.\ Grav.\  {\bf 35}, no. 21, 215002 (2018)
  doi:10.1088/1361-6382/aae276
  [arXiv:1803.09649 [gr-qc]].
  
\bibitem{Wang:2018txl} 
  X.~Y.~Wang, S.~Z.~Han and W.~B.~Liu,
  Phys.\ Lett.\ B {\bf 787}, 64 (2018).
  doi:10.1016/j.physletb.2018.10.033

\bibitem{Christodoulou:2016tuua} 
  M.~Christodoulou and T.~De Lorenzo,
  Phys.\ Rev.\ D {\bf 94}, no. 10, 104002 (2016)
  doi:10.1103/PhysRevD.94.104002
  [arXiv:1604.07222 [gr-qc]].

\bibitem{Wang:2017zfn} 
  S.~J.~Wang, X.~X.~Guo and T.~Wang,
  Phys.\ Rev.\ D {\bf 97}, no. 2, 024039 (2018)
  doi:10.1103/PhysRevD.97.024039
  [arXiv:1702.05246 [gr-qc]].

\bibitem{Kerr:1963ud} 
  R.~P.~Kerr,
  Phys.\ Rev.\ Lett.\  {\bf 11}, 237 (1963).
  doi:10.1103/PhysRevLett.11.237
  
\bibitem{Zhang:2017aqf} 
  B.~Zhang,
  Phys.\ Lett.\ B {\bf 773}, 644 (2017)
  doi:10.1016/j.physletb.2017.09.035
  [arXiv:1709.07275 [gr-qc]].

\bibitem{Page:1976ki} 
  D.~N.~Page,
  Phys.\ Rev.\ D {\bf 14}, 3260 (1976).
  doi:10.1103/PhysRevD.14.3260

\bibitem{Bekenstein:1973ur} 
  J.~D.~Bekenstein,
  Phys.\ Rev.\ D {\bf 7}, 2333 (1973).
  doi:10.1103/PhysRevD.7.2333

\bibitem{Bardeen:1973gs} 
  J.~M.~Bardeen, B.~Carter and S.~W.~Hawking,
  Commun.\ Math.\ Phys.\  {\bf 31}, 161 (1973).
  doi:10.1007/BF01645742

\bibitem{Davies:1978zz} 
  P.~C.~W.~Davies,
  Rept.\ Prog.\ Phys.\  {\bf 41}, 1313 (1978).
  doi:10.1088/0034-4885/41/8/004

\bibitem{Hawking:1982dh} 
  S.~W.~Hawking and D.~N.~Page,
  Commun.\ Math.\ Phys.\  {\bf 87}, 577 (1983).
  doi:10.1007/BF01208266

\bibitem{Wald:2002mm} 
  R.~M.~Wald,
  (2002)

\bibitem{Natsuume:2014sfa} 
  M.~Natsuume,
  Lect.\ Notes Phys.\  {\bf 903}, pp.1 (2015)
  doi:10.1007/978-4-431-55441-7
  [arXiv:1409.3575 [hep-th]].

\bibitem{Grumiller:2005zk} 
  D.~Grumiller,
  J.\ Phys.\ Conf.\ Ser.\  {\bf 33}, 361 (2006)
  doi:10.1088/1742-6596/33/1/044
  [gr-qc/0509077].
  
\bibitem{Ballik:2010rx} 
  W.~Ballik and K.~Lake,
  arXiv:1005.1116 [gr-qc].
  
\bibitem{Ballik:2013uia} 
  W.~Ballik and K.~Lake,
  Phys.\ Rev.\ D {\bf 88}, no. 10, 104038 (2013)
  doi:10.1103/PhysRevD.88.104038
  [arXiv:1310.1935 [gr-qc]].

\bibitem{Finch:2012fq} 
  T.~K.~Finch,
  Gen.\ Rel.\ Grav.\  {\bf 47}, no. 5, 56 (2015)
  doi:10.1007/s10714-015-1891-7
  [arXiv:1211.4337 [gr-qc]].
  
\bibitem{Cvetic:2010jb} 
  M.~Cvetic, G.~W.~Gibbons, D.~Kubiznak and C.~N.~Pope,
  Phys.\ Rev.\ D {\bf 84}, 024037 (2011)
  doi:10.1103/PhysRevD.84.024037
  [arXiv:1012.2888 [hep-th]].
  
\bibitem{Gibbons:2012ac} 
  G.~W.~Gibbons,
  AIP Conf.\ Proc.\  {\bf 1460}, 90 (2012)
  doi:10.1063/1.4733363
  [arXiv:1201.2340 [gr-qc]].

\bibitem{Gunasekaran:2012dq} 
  S.~Gunasekaran, R.~B.~Mann and D.~Kubiznak,
Born-Infeld vacuum polarization,''
  JHEP {\bf 1211}, 110 (2012)
  doi:10.1007/JHEP11(2012)110
 [arXiv:1208.6251 [hep-th]].

\bibitem{Montvay:1981jj}
  I.~Montvay and E.~Pietarinen,
  Phys.\ Lett.\  {\bf 110B}, 148 (1982).
  doi:10.1016/0370-2693(82)91024-3

\bibitem{Page:1993}
  D.~N.~Page,
  Phys.\ Rev.\ Lett.\  {\bf 71}, 3743 (1993)
  doi:10.1103/PhysRevLett.71.3743
  [hep-th/9306083].

\bibitem{Marolf:2017}
  D.~Marolf,
  Rept.\ Prog.\ Phys.\  {\bf 80}, no. 9, 092001 (2017)
  doi:10.1088/1361-6633/aa77cc
  [arXiv:1703.02143 [gr-qc]].

\bibitem{Jacob:2005}
  T.~Jacobson, D.~Marolf and C.~Rovelli,
  Int.\ J.\ Theor.\ Phys.\  {\bf 44}, 1807 (2005)
  doi:10.1007/s10773-005-8896-z
  [hep-th/0501103].

\bibitem{Ali:2018}
  S.~Ali, X.~Y.~Wang and W.~B.~Liu,
  Int.\ J.\ Mod.\ Phys.\ A {\bf 33}, no. 27, 1850159 (2018).
  doi:10.1142/S0217751X18501592

\bibitem{Sheykhi:2012}
  A.~Sheykhi,
  Phys.\ Rev.\ D {\bf 86}, 024013 (2012)
  doi:10.1103/PhysRevD.86.024013
  [arXiv:1209.2960 [hep-th]].

\bibitem{Nojiri:2003}
  S.~Nojiri and S.~D.~Odintsov,
  Phys.\ Rev.\ D {\bf 68}, 123512 (2003)
  doi:10.1103/PhysRevD.68.123512
  [hep-th/0307288].

\bibitem{Atazadeh:2008}
  K.~Atazadeh, M.~Farhoudi and H.~R.~Sepangi,
  Phys.\ Lett.\ B {\bf 660}, 275 (2008)
  doi:10.1016/j.physletb.2007.12.057
  [arXiv:0801.1398 [gr-qc]].

\bibitem{Moon:2011}
  T.~Moon, Y.~S.~Myung and E.~J.~Son,
  Gen.\ Rel.\ Grav.\  {\bf 43}, 3079 (2011)
  doi:10.1007/s10714-011-1225-3
  [arXiv:1101.1153 [gr-qc]].

\bibitem{LEP:2009}
L. E. Parker and D. Toms, 
''Quantum Field Theory in Curved Spacetime: Quantized Field and Gravity'', 
Cambridge University Press (2009)


\bibitem{Chamblin:1999tk} 
  A.~Chamblin, R.~Emparan, C.~V.~Johnson and R.~C.~Myers,
  Phys.\ Rev.\ D {\bf 60}, 064018 (1999)
  doi:10.1103/PhysRevD.60.064018
  [hep-th/9902170].
 
\bibitem{Chamblin:1999hg} 
  A.~Chamblin, R.~Emparan, C.~V.~Johnson and R.~C.~Myers,
  Phys.\ Rev.\ D {\bf 60}, 104026 (1999)
  doi:10.1103/PhysRevD.60.104026
  [hep-th/9904197].
  
\bibitem{Caldarelli:1999xj} 
  M.~M.~Caldarelli, G.~Cognola and D.~Klemm,
  Class.\ Quant.\ Grav.\  {\bf 17}, 399 (2000)
  doi:10.1088/0264-9381/17/2/310
  [hep-th/9908022].
  
\bibitem{Dolan:2010ha} 
  B.~P.~Dolan,
  Class.\ Quant.\ Grav.\  {\bf 28}, 125020 (2011)
  doi:10.1088/0264-9381/28/12/125020
  [arXiv:1008.5023 [gr-qc]].
  
\bibitem{Dolan:2011xt} 
  B.~P.~Dolan,
  Class.\ Quant.\ Grav.\  {\bf 28}, 235017 (2011)
  doi:10.1088/0264-9381/28/23/235017
  [arXiv:1106.6260 [gr-qc]].
  
\bibitem{Dolan:2011jm} 
  B.~P.~Dolan,
  Phys.\ Rev.\ D {\bf 84}, 127503 (2011)
  doi:10.1103/PhysRevD.84.127503
  [arXiv:1109.0198 [gr-qc]].

\bibitem{Lu:2012xu} 
  H.~Lu, Y.~Pang, C.~N.~Pope and J.~F.~Vazquez-Poritz,
  Phys.\ Rev.\ D {\bf 86}, 044011 (2012)
  doi:10.1103/PhysRevD.86.044011
  [arXiv:1204.1062 [hep-th]].
  
\bibitem{Kubiznak:2012wp} 
  D.~Kubiznak and R.~B.~Mann,
  JHEP {\bf 1207}, 033 (2012)
  doi:10.1007/JHEP07(2012)033
  [arXiv:1205.0559 [hep-th]].
 
\bibitem{Altamirano:2013ane} 
  N.~Altamirano, D.~Kubiznak and R.~B.~Mann,
  Phys.\ Rev.\ D {\bf 88}, no. 10, 101502 (2013)
  doi:10.1103/PhysRevD.88.101502
  [arXiv:1306.5756 [hep-th]].
  
\bibitem{Wei:2014hba} 
  S.~W.~Wei and Y.~X.~Liu,
  Phys.\ Rev.\ D {\bf 90}, no. 4, 044057 (2014)
  doi:10.1103/PhysRevD.90.044057
  [arXiv:1402.2837 [hep-th]].
  
\bibitem{Frassino:2014pha} 
  A.~M.~Frassino, D.~Kubiznak, R.~B.~Mann and F.~Simovic,
  JHEP {\bf 1409}, 080 (2014)
  doi:10.1007/JHEP09(2014)080
  [arXiv:1406.7015 [hep-th]].
  
\bibitem{Altamirano:2013uqa} 
  N.~Altamirano, D.~Kubizňák, R.~B.~Mann and Z.~Sherkatghanad,
  Class.\ Quant.\ Grav.\  {\bf 31}, 042001 (2014)
  doi:10.1088/0264-9381/31/4/042001
  [arXiv:1308.2672 [hep-th]].
  
\bibitem{Belhaj:2012bg} 
  A.~Belhaj, M.~Chabab, H.~El Moumni and M.~B.~Sedra,
  Chin.\ Phys.\ Lett.\  {\bf 29}, 100401 (2012)
  doi:10.1088/0256-307X/29/10/100401
  [arXiv:1210.4617 [hep-th]].
  
\bibitem{Hendi:2012um} 
  S.~H.~Hendi and M.~H.~Vahidinia,
  Phys.\ Rev.\ D {\bf 88}, no. 8, 084045 (2013)
  doi:10.1103/PhysRevD.88.084045
  [arXiv:1212.6128 [hep-th]].
  
\bibitem{Altamirano:2014tva} 
  N.~Altamirano, D.~Kubiznak, R.~B.~Mann and Z.~Sherkatghanad,
  Galaxies {\bf 2}, 89 (2014)
  doi:10.3390/galaxies2010089
  [arXiv:1401.2586 [hep-th]].
  
\bibitem{Dolan:2014jva} 
  B.~P.~Dolan,
  Mod.\ Phys.\ Lett.\ A {\bf 30}, no. 03n04, 1540002 (2015)
  doi:10.1142/S0217732315400027
  [arXiv:1408.4023 [gr-qc]].
  
\bibitem{shannon1948}
  C.~E.~Shannon
  Bell system technical journal 27.3 (1948): 379-423.Wiley Online Lib.
  doi:10.1145/584091.584093
  
\bibitem{Stephens:2001sd} 
  G.~J.~Stephens and B.~L.~Hu,
  Int.\ J.\ Theor.\ Phys.\  {\bf 40}, 2183 (2001)
  doi:10.1023/A:1012930019453
  [gr-qc/0102052].
  
  
\bibitem{Banerjee:2011raa} 
  R.~Banerjee, S.~K.~Modak and D.~Roychowdhury,
  JHEP {\bf 1210}, 125 (2012)
  doi:10.1007/JHEP10(2012)125
  [arXiv:1106.3877 [gr-qc]].
  
\bibitem{Banerjee:2011au} 
  R.~Banerjee and D.~Roychowdhury,
  JHEP {\bf 1111}, 004 (2011)
  doi:10.1007/JHEP11(2011)004
  [arXiv:1109.2433 [gr-qc]].
  
\bibitem{Banerjee:2011cz} 
  R.~Banerjee and D.~Roychowdhury,
  Phys.\ Rev.\ D {\bf 85}, 044040 (2012)
  doi:10.1103/PhysRevD.85.044040
  [arXiv:1111.0147 [gr-qc]].
  
\bibitem{Banerjee:2016nse} 
  R.~Banerjee, B.~R.~Majhi and S.~Samanta,
  Phys.\ Lett.\ B {\bf 767}, 25 (2017)
  doi:10.1016/j.physletb.2017.01.040
  [arXiv:1611.06701 [gr-qc]].
  
\bibitem{Quevedo:2016swn} 
  H.~Quevedo, M.~N.~Quevedo and A.~Sánchez,
  Phys.\ Rev.\ D {\bf 94}, no. 2, 024057 (2016)
  doi:10.1103/PhysRevD.94.024057
  [arXiv:1606.02048 [gr-qc]].
  
\bibitem{Roychowdhury:2014cva} 
  D.~Roychowdhury,
  arXiv:1403.4356 [gr-qc].
  
\bibitem{Witten:1998zw} 
  E.~Witten,
  Adv.\ Theor.\ Math.\ Phys.\  {\bf 2}, 505 (1998)
  doi:10.4310/ATMP.1998.v2.n3.a3
  [hep-th/9803131].
  
\bibitem{Mo:2016sel} 
  J.~X.~Mo and G.~Q.~Li,
  Phys.\ Rev.\ D {\bf 92}, no. 2, 024055 (2015)
  doi:10.1103/PhysRevD.92.024055
  [arXiv:1604.07931 [gr-qc]].
  
\bibitem{Chen:2013ce} 
  S.~Chen, X.~Liu, C.~Liu and J.~Jing,
  Chin.\ Phys.\ Lett.\  {\bf 30}, 060401 (2013)
  doi:10.1088/0256-307X/30/6/060401
  [arXiv:1301.3234 [gr-qc]].
  
\bibitem{Wei:2015iwa} 
  S.~W.~Wei and Y.~X.~Liu,
  Phys.\ Rev.\ Lett.\  {\bf 115}, no. 11, 111302 (2015)
  Erratum: [Phys.\ Rev.\ Lett.\  {\bf 116}, no. 16, 169903 (2016)]
  doi:10.1103/PhysRevLett.116.169903, 10.1103/PhysRevLett.115.111302
  [arXiv:1502.00386 [gr-qc]].

\bibitem{Lee:2017ero} 
  C.~O.~Lee,
  Phys.\ Lett.\ B {\bf 772}, 471 (2017)
  doi:10.1016/j.physletb.2017.07.013
  [arXiv:1705.09047 [gr-qc]].

\end{thebibliography}
\end{document}